\documentclass[3p,times]{elsarticle}

\usepackage{ecrc}

\volume{00}

\firstpage{1}

\journalname{Physics Reports}

\runauth{}

\CopyrightLine{2011}{Published by Elsevier Ltd.}

\usepackage{amssymb}

\usepackage[pagewise]{lineno}

\usepackage[figuresright]{rotating}

\usepackage{amsthm} 

\usepackage{amsmath,amssymb,graphicx,bm,color,mathrsfs,verbatim,epstopdf,dcolumn,cancel}
\usepackage{esint}

\usepackage{ulem}

\usepackage[colorlinks=true]{hyperref}

\definecolor{violet}{rgb}{0.58, 0.0, 0.83}
\definecolor{brown}{rgb}{0.82,0.41,0.12}

\newcommand{\be}{\begin{equation}}
\newcommand{\ee}{\end{equation}}

\begin{document}
	

\begin{frontmatter}



\dochead{}

\title{Adiabatic Perturbation Theory and Geometry \\ of Periodically-Driven Systems}


\author[BostonU]{Phillip Weinberg}\corref{correspondingauthor}
\cortext[correspondingauthor]{Corresponding author}
\ead{weinbe58@bu.edu}

\author[BostonU]{Marin Bukov}

\author[BostonU,UPenn]{Luca D'Alessio}

\author[BostonU]{Anatoli Polkovnikov}

\author[BostonU,BudapestU]{Szabolcs Vajna}

\author[BostonU,UCBerkley,LBNL]{Michael Kolodrubetz}

\address[BostonU]{Department of Physics, Boston University, 590 Commonwealth Ave., Boston, MA 02215, USA}
\address[BudapestU]{Department of Physics and BME-MTA Exotic Quantum Phases Research Group, Budapest University of Technology and Economics, 1521 Budapest, Hungary}
\address[UPenn]{Department of Physics, The Pennsylvania State University, University Park, PA 16802, USA}
\address[UCBerkley]{Department of Physics, University of California, Berkeley, CA 94720, USA}
\address[LBNL]{Materials Sciences Division, Lawrence Berkeley National Laboratory, Berkeley, CA 94720, USA}

\begin{abstract}
We give a systematic review of the adiabatic theorem and the leading non-adiabatic corrections in periodically-driven (Floquet) systems. These corrections have a two-fold origin: (i) conventional ones originating from the gradually changing Floquet Hamiltonian and (ii) corrections originating from changing the micro-motion operator. These corrections conspire to give a Hall-type linear response for non-stroboscopic (time-averaged) observables allowing one to measure the Berry curvature and the Chern number related to the Floquet Hamiltonian, thus extending these concepts to periodically-driven many-body systems. The non-zero Floquet Chern number allows one to realize a Thouless energy pump, where one can adiabatically add energy to the system in discrete units of the driving frequency. We discuss the validity of Floquet Adiabatic Perturbation Theory (FAPT) using five different models covering linear and non-linear few and many-particle systems. We argue that in interacting systems, even in the stable high-frequency regimes, FAPT breaks down at ultra slow ramp rates due to avoided crossings of photon resonances, not captured by the inverse-frequency expansion, leading to a counter-intuitive stronger heating at slower ramp rates. Nevertheless, large windows in the ramp rate are shown to exist for which the physics of interacting driven systems is well captured by FAPT.  
\end{abstract}


\end{frontmatter}


\section{\label{sec:intro} Introduction}

	The concept of adiabaticity in equilibrium systems has profound importance of both a fundamental and practical nature. Fundamentally, it allows one to identify and label families of adiabatically connected microscopic states and macroscopic phases. The existence of the adiabatic limit is a cornerstone of equilibrium thermodynamics, as it allows one to calculate thermodynamic forces, formulate the notion of reversibility, define the laws of thermodynamics, and put restrictions on possible outcomes of macroscopic processes, such as efficiency bounds of heat engines and refrigerators~\cite{book_LL_Stat, book_balian}. Practically, the existence of an adiabatic limit allows for the preparation of complex ground or excited states in interacting isolated systems by slowly changing the couplings of the Hamiltonian. This idea, for instance, underlies adiabatic quantum computation and quantum annealing~\cite{Farhi_00, santoro_06, das_08}.
	
	Recently, the adiabatic control of model parameters, such as the drive coupling strength, has become relevant in analyzing the response of systems subject to periodic modulation. For instance, one may wish to prepare a target state by gradual change of a rapidly oscillating Hamiltonian. Such periodic systems, realized in a variety of settings from irradiation by lasers to application of periodic mechanical kicks, has been the subject of extensive experimental and theoretical study throughout the modern history of physics~\cite{shirley_65,sambe_73,breuer_90,breuer_91}. Prominent examples in mechanics include the Kapitza pendulum and the closely related kicked rotor, whose dynamics feature tantalising integrability-to-chaos transitions as a function of the drive parameters. These and similar models also feature dynamical stabilization~\cite{kapitza_51,broer_04, book_LL} and localization~\cite{chirikov_81,rahav_03_pra,rahav_03,reichl_04,grossmann_91,grossmann_92,gavrila_02,grifoni_98,bavli_92,neu_96}, among a variety of counter-intuitive effects induced by periodic modulations. In atomic physics, driving leads to reduced ionisation rates in systems irradiated by electromagnetic fields at high frequency and intensity~\cite{su_90,bialynicki-birula_94,eberly_93,piraux_98,pont_90,zakrzewski_95}, which can be traced back to decreased spreading of wave packets reported in periodically-driven systems~\cite{holthaus_95,buchleitner_02}.  Last but not least, the effects of periodic drives on transport  has recently become an active field of study, predicting non-trivial behavior in more traditional condensed matter settings~\cite{platero_04,kohler_05,denisov_07,auerbach_07,poletti_09,salger_13,denisov_14,grossert_16,gu_11,foa-torres_14}.

	Adiabatic protocols in the presence of periodic drive have been extensively studied in few-body problems, such as NMR and qubit experiments. Adiabatic passages are robust protocols based on the Floquet adiabatic principle to prepare excited states with a high tolerance to the inhomogeneity of the applied radio-frequency (RF) field \cite{tannus_97, garwood_01}. Different driving protocols have been used successfully to reach higher speed and to increase robustness \cite{silver_84, staewen_90,garwood_91,norris_89}. Adiabatic protocols are also used to enhance sensitivity of spins with low gyromagnetic ratio \cite{hediger_95}, for spin-decoupling \cite{bendall_95}, and for refocusing \cite{hwang_97}. Adiabatic passages for population transfer between quantum states are also applied in the optical domain\cite{vitanov_01,desbuquois_17}.

	However, much remains to be understood in transferring these ideas to the many-body domain. The recent surge of activity in applying periodic drives to many body systems has spawned a new branch of quantum physics known as Floquet engineering~\cite{goldman_14,bukov_14}, i.e., the synthetic generation of novel effective Hamiltonians that either manipulate the properties of the un-driven system or produce Hamiltonians which are hard to realize in static condensed matter systems. For instance, periodic modulations have been reported to change the critical properties of systems, by inducing critical points not present without the drive, allowing a controllable dependence of the critical point on the drive parameters~\cite{oka_09,bastidas_Dicke_12,bastidas_Ising_12, engelhardt_13, bastidas_14,kitagawa_11,fregoso_14} or even enable new phases which cannot exist in equilibrium~\cite{khemani_16,bastidas_Dicke_12,bastidas_Ising_12,oka_09,kitagawa_11,fregoso_14}. Similarly, cold atom experiments in `shaken' optical lattices have progressed to realise phenomena such as dynamical localisation and stabilisation\cite{dunlap_86,dunlap_88,lignier_07,sias_07,eckardt_09,zenesini_09,creffield_10}, artificial gauge fields~\cite{struck_11,struck_12,hauke_12,struck_13,aidelsburger_13,miyake_13,atala_14,kennedy_15}, topological~\cite{oka_09,kitagawa_11,jotzu_14,aidelsburger_14,flaeschner_15,mikami_15} and spin-dependent~\cite{jotzu_15} bands, topological pumps~\cite{rudner_13,lindner_16}, and spin-orbit coupling~\cite{jimenez-garcia_15,galitski_13}. These ideas are not restricted to cold atoms, as evidenced by Floquet topological insulators~\cite{wang_13} and photonic topological insulators~\cite{rechtsman_13,hafezi_14,mittal_14}, the latter effectively obeying the Schr\"odinger equation with the additional spatial dimension playing the role of periodic time modulation. Future similar experiments in this vein are expected to produce synthetic Hamiltonians realizing Laughlin states, fractional topological insulators~\cite{grushin_14}, Weyl points~\cite{dubcek_14,kennedy_15},  quantum motors similar to quantum ratchets~\cite{ponomarev_10,ponomarev_09}, as well as other systems hard to create statically.

    Much of the interest in driving these systems is inspired by Floquet's theorem, which states that the dynamics of a periodically driven quantum system is stroboscopically (i.e.~at times $t=lT$ with $l\in\mathbb{Z}$) governed by the time-independent Floquet Hamiltonian~\cite{goldman_14,bukov_14,mikami_15,goldman_14_res,eckardt_15} $H_F$. More generally, Floquet's theorem states that the unitary time evolution $U(t, 0)$ from time $0$ to $t$ can be represented as 
	\begin{eqnarray}
	U(t,0) = P(t)\exp\left(-it H_F\right) P^\dagger (0),
	\label{eq:Floquet_thm_general}
	\end{eqnarray}
	where $P(t)=P(T+t)$ is a time-periodic, unitary micromotion operator, which is related to the commonly used kick operator $K(t)$ via $P(t)=\exp[-i K(t)]$. The micromotion operator and the Floquet Hamiltonian are defined up to a global unitary transformation, which defines the Floquet gauge choice~\cite{bukov_14}, i.e.~the initial phase of the drive. A particularly simple choice is the stroboscopic one, where the micromotion operator $P(0)={\bf 1}$ equals the identity operator at the initial time, such that Floquet's theorem reduces to
	\begin{eqnarray}
	U(t,0) = P(t)\exp\left(-i t H_F\right) \implies  U(t=lT,0) =  \exp\left(-i H_F lT\right) ~.
	\label{eq:Floquet_thm}
	\end{eqnarray} 
	In the following, we shall always work in this stroboscopic Floquet gauge, unless explicitly stated otherwise.
	
	Despite this striking similarity with their static counterparts, periodically driven systems are a priori out-of-equilibrium. In systems with an unbounded spectrum, e.g.~in the thermodynamic or classical limits, the Floquet Hamiltonian, as a local operator, is not even guaranteed to exist~\cite{bukov_14,bukov_15_erg}. Therefore, the question as to \emph{how to prepare the system in a desired state} is of equal importance as engineering the effective parent Hamiltonian~\cite{goldman_14,bukov_14}. Adiabatic preparation of Floquet states in certain quantum many-body systems has been reported both numerically and experimentally. For instance, in Ref.~\cite{poletti_11}, the authors studied a large but finite periodically-driven Bose-Hubbard chain using DMRG. They found an adiabatic regime for intermediate velocities, which enabled the adiabatic transfer of a superfluid from the zero-momentum to the $\pi$-momentum mode at high driving frequencies. Slowly turning on the amplitude of a periodic drive has also recently been reported to enable successful preparation of Floquet condensates~\cite{heinisch_16}. At the same time, a study on periodically-driven Luttinger liquids reported that the momentum distribution of fermions changes immediately after the drive is turned on, due to enhanced photon-assisted scattering near the Fermi edge, and concluded that the existence of an adiabatic limit is not possible at low drive frequencies~\cite{bukov_12}. More recently, a number of exciting experiments~\cite{aidelsburger_13,atala_14,jotzu_14} with cold atoms slowly turned on the amplitude of the drive to come sufficiently close to the desired ground state of a carefully engineered topological Floquet Hamiltonian. It must be noted, though, that another experiment, which managed to prepare the ground state of the $\pi$-flux Hofstadter model~\cite{kennedy_15}, reported lower fidelities when adiabatically ramping the drive compared to a sudden switch on of the periodic modulation. 
	
	Conventional adiabatic perturbation theory (APT) predicts that for very slow and smooth ramps, during which the system remains gapped, the excitations accumulated during the ramp are small, and thus the systems follows the adiabatically connected eigenstates of the Hamiltonian without transitions~\cite{rigolin_08,degrandi_10}. The leading non-adiabatic corrections to observables, such as the energy or various generalised forces, are analytic functions of the ramp rate. These corrections give rise to various velocity-dependent forces, such as the Lorentz force, the Magnus force, or the Coriolis force, as well as to various inertia-type forces proportional to the acceleration of the system~\cite{kolodrubetz_16}. In gapless or open systems, finite ramp rates result in additional dissipative forces, such as friction~\cite{kirkwood_46, zwanzig_64, berry_93, sivak_12}. These forces, however, also vanish in the adiabatic limit and can be captured within APT~\cite{sivak_12, dalessio_14_mass}.
	
	The idea of an adiabatic passage between continuously connected Floquet eigenstates was introduced to study the behaviour of single particle quantum systems in the presence of intense radiation fields. It was soon afterwards argued that resonant transitions can be understood as Landau-Zener (LZ) processes between Floquet levels. For these few-level systems, APT was successfully extended to incorporate Floquet theory, and produced accurate estimates of the ionization rates in various single-atomic systems~\cite{zakrzewski_95,young_70,breuer_89,viennot_05,dietz_96}. Besides contributing to the understanding of the physical processes involved, APT has also lead to the development of dynamic control over the population of single-particle states in strongly-driven atoms~\cite{breuer_93,breuer_91_2,gabriel_93,jakubetz_90,gu_11,foa-torres_14,usaj_14}. Beyond few-level systems, it was conjectured that generic periodically-driven many-body Hamiltonians do not possess a well-defined adiabatic limit due to the exponentially large number of interacting states in the thermodynamic limit~\cite{hone_97,drese_99,eckardt_08}.
		
	As we discuss in this review, APT can be extended to Floquet systems essentially retaining the form of leading non-adiabatic corrections. However, there is a crucial caveat for generic Floquet systems: one must in addition avoid photon resonances, which correspond to the closing of effective gaps in the Floquet spectrum due to hybridisation of (nearly) resonant states~\cite{shirley_65,sambe_73,hone_97,drese_99}. This is only possible if the ramp rate is not too slow. In this sense, one can anticipate a finite window of rates where Floquet adiabatic perturbation theory (FAPT) is applicable: the rates should be sufficiently fast that the photon resonances are passed diabatically, but also sufficiently slow that non-adiabatic processes which do not involve photon absorption remain suppressed. Intuitively, one expects that this window can exist only for special protocol conditions, typically at fast driving frequencies -- larger than the natural energy scales of the non-driven system or, more accurately, away from single-particle resonances.

	In this work, we present an extensive overview of the problem of slowly changing the parameters in a periodically driven system. We illustrate the main ideas discussed here using various different models which cover a range of single-particle and many-body condensed matter systems. At the same time, we complement the general theory with new, previously unpublished results, by pointing out new constraints on adiabaticity imposed by the presence of the micromotion. Intuitively, as a consequence of the ramp, the $P(t)$ operator never comes back exactly to itself after one period, which induces additional non-adiabatic corrections to the wave function. Hence, in systems where the Floquet Hamiltonian vanishes identically, such as some topological pumps~\cite{nakajima_16,lohse_16,rudner_13}, the dynamics is governed entirely by the micromotion operator and understanding its contribution to  non-adiabatic corrections is crucial. Since Floquet engineering often requires one to scale the driving amplitude with the driving frequency~\cite{bukov_14}, the effects of micromotion can remain finite even in the infinite-frequency limit. As we discuss in detail below, it is only the sum of the contributions to FAPT coming from micromotion and the Floquet Hamiltonian, which leads to a unique, Floquet gauge-invariant result which is insensitive to the choice of folding. 
	
	This review is organised as follows: 
	\begin{itemize}
		\item In Sec.~\ref{sec:general} we briefly review adiabatic perturbation theory (APT) for non-driven systems before proceeding to discuss Floquet adiabatic perturbation theory (FAPT) for Floquet systems with a well-defined adiabatic limit. Using this formalism, we derive the leading non-adiabatic response of observables such as the excitation probability and the Floquet diagonal entropy, which constitute convenient measures of adiabaticity. Finally, similar to conventional linear response theory, we exemplify the crucial connections between leading corrections to adiabaticity and natural Floquet generalizations of quantum geometry. In particular, we describe the Floquet gauge potential, i.e., the generator of adiabatic transformations in Floquet systems, the Floquet Berry curvature, and the Floquet Chern number, showing that these objects are naturally measurable in the language of FAPT.

		\item Section~\ref{sec:examples_SP} demonstrates the applicability of FAPT using three single-particle examples. We begin with the exactly solvable model of a quantum harmonic oscillator whose potential is displaced periodically, and find perfect agreement between FAPT and the exact analytical predictions. This model allows us to highlight and isolate the individual terms giving rise to non-adiabatic effects. Then, in Sec.~\ref{subsec:kapitza} we study a nonlinear driven oscillator, the quantum Kapitza pendulum, for which we discuss the applicability and breakdown of FAPT due to photon absorption resonances. Last, in Sec.~\ref{subsec:qubit_topology} we illustrate the connection between leading non-adiabatic corrections to observables and quantum geometry using a driven two-level system (qubit). We show how one may use this to measure frequency-dependent topological transitions of the Floquet Chern number via dynamical response.	
		
		\item Section~\ref{sec:examples_MB} is devoted to the applicability of FAPT to many-body models. In particular, we simulate integrable and non-integrable versions of the one-dimensional transverse field Ising model, which demonstrates the validity of these methods for generic quantum many-body systems.
		
		\item In Sec.~\ref{sec:HFE} we briefly introduce the widely-used inverse-frequency expansion and discuss the relation of a variant thereof -- the van Vleck expansion~\cite{goldman_14,bukov_14,eckardt_15,mikami_15} -- to FAPT. We comment on the general convergence properties of the inverse-frequency expansion and examine its relevance to adiabaticity.
		
		\item Finally, a discussion of the main conclusions reached in this review with an outlook to future studies is presented in Sec.~\ref{sec:outro}.
	\end{itemize}

	\section{\label{sec:general} Floquet Adiabatic Perturbation Theory}
	
	We open up the discussion by briefly recapitulating the main results of conventional adiabatic perturbation theory (APT) for non-driven systems. We then proceed to generalise this formalism to periodically-driven systems, which we shall refer to as Floquet adiabatic perturbation theory (FAPT).

	\subsection{\label{subsec:APT}Adiabatic Perturbation Theory (APT).} 
	
	Let us first outline some key results of quantum adiabatic perturbation theory; for more details see Refs.~\cite{rigolin_08,degrandi_10,dalessio_15_neg_mass}. Consider a Hamiltonian $H(\lambda)$ which depends on some parameter $\lambda$ that slowly changes in time. For simplicity, we assume that the Hamiltonian has a discrete energy spectrum with no degeneracies so that the adiabatic limit is well defined. Furthermore, we assume that the system is prepared in the ground state of the initial Hamiltonian and thus, in the adiabatic limit, it remains in the instantaneous ground state as $\lambda$ is ramped\footnote{Notice that this discussion applies as well to any excited state.}.  
	
	Suppose that $V(\lambda)$ is a unitary transformation which diagonalizes the Hamiltonian, i.e.,~$\tilde H(\lambda)=V^\dagger(\lambda) H(\lambda) V(\lambda)$ is a diagonal matrix whose entries are the eigenenergies of $H(\lambda)$. It is convenient to go to a moving frame with respect to the instantaneous Hamiltonian by defining $|\tilde \psi\rangle= V^\dagger(\lambda) |\psi\rangle$. Substituting this into the time-dependent Schr\"odinger equation,  the time evolution $i\mathrm{d}_t |\tilde \psi\rangle=\tilde H_m |\tilde \psi\rangle$ of $|\tilde \psi\rangle$ is governed by the moving-frame Hamiltonian
	\[
	\tilde H_\mathrm{m}(\lambda)=\tilde H(\lambda)-\dot \lambda \tilde{\mathcal A_\lambda},
	\]
	where $\tilde{\mathcal{A}}_\lambda=i V^\dagger(\lambda)\partial_\lambda V(\lambda)$ is the adiabatic gauge potential in the moving-frame, i.e., the generator of translations of the energy eigenstates w.r.t.~$\lambda$~\cite{kolodrubetz_16}. This gauge potential is a Hermitian operator whose diagonal elements are the Berry connections of the energy eigenstates. Moreover, it follows from the above definition that the unitary $V(\lambda)$ also describes the basis transformation of the instantaneous energy eigenstates $|n(\lambda)\rangle$ to a $\lambda$-independent basis $|e_n\rangle$:
	\[
	|n(\lambda)\rangle = V(\lambda) |e_n\rangle.
	\]
	This implies that $\mathcal A_\lambda\equiv V(\lambda) \tilde {\mathcal A}_\lambda V^\dagger(\lambda)=i [\partial_\lambda V(\lambda)] V^\dagger(\lambda)$ acts as $i \partial_\lambda$ in the energy basis:
		\begin{equation}
			\langle m(\lambda)| \mathcal A_\lambda |n(\lambda)\rangle=\langle e_m|\tilde{\mathcal A}_\lambda | e_n \rangle=
			\langle m(\lambda)| i \partial_\lambda |n(\lambda) \rangle.
			\label{eq:matr_elem_A_lam_apt}
		\end{equation}
	
	Since in the moving frame the Hamiltonian $\tilde H(\lambda)$ is diagonal, it does not lead to transitions between the instantaneous levels. Consequently, all the transitions are due to the Galilean term $\dot \lambda \tilde {\mathcal A}_\lambda$ \footnote{The term $\dot \lambda \tilde {\mathcal A}_\lambda$ is called Galilean in analogy with the $V P$ term appearing in the Hamiltonian if we do the Galilean transformation into the moving frame. The adiabatic gauge potential plays the role analogous to the momentum $P$ generating the adiabatic transformations with respect to the parameter $\lambda$ and $\dot\lambda$ plays the role of the velocity $V$, see Ref.~\cite{kolodrubetz_16} for more details.}. As this term is suppressed at slow ramp rates (a.k.a.~velocities) $\dot\lambda$, the system approximately (i.e., up to order $\dot\lambda^2$) follows the instantaneous ground state of $H_\mathrm{m}(\lambda)$. In order to obtain the transition amplitudes in the instantaneous basis, one uses first-order static perturbation theory with respect to the Galilean term. Then, expanding in the instantaneous basis $|\psi(t)\rangle =\sum_n c_n(t) |n(\lambda)\rangle$, we find
	\begin{eqnarray}
	c_0(t)&\approx& \exp[i \Phi_0]=\exp\left[-i\int_{t_0}^t \mathrm{d}t' \left(\epsilon_0(\lambda(t'))-\dot \lambda (t')  A_\lambda (\lambda(t'))\right)\right]\nonumber
	\\c_{n \neq 0}(t) &\approx& \mathrm e^{i\Phi_0}\dot \lambda {\langle n(\lambda) | \mathcal A_\lambda| 0(\lambda)\rangle\over \epsilon_n -\epsilon_0}=i \mathrm e^{i\Phi_0} \dot{\lambda}  {\langle n(\lambda) |\partial_\lambda |0(\lambda)\rangle \over \epsilon_n-\epsilon_0}=-i \mathrm e^{i\Phi_0} \dot \lambda  {\langle n(\lambda) |(\partial_\lambda H) |0(\lambda)\rangle \over (\epsilon_n-\epsilon_0)^2}~,
	\label{eq:APT_cn}
	\end{eqnarray}
	where $A_\lambda(\lambda')=\langle 0(\lambda')| \mathcal A_\lambda(\lambda')| 0(\lambda')\rangle$ is the ground state Berry connection. Note that, to obtain the above lab-frame expressions, one rotates $|\psi(t)\rangle$ into the moving basis $|e_n\rangle$, applies perturbation theory, and then rotates back to the lab frame via Eq.~\eqref{eq:matr_elem_A_lam_apt}.One might recognize that this expression is nothing but the result of the static first order perturbation theory applied to the Galilean term in the moving Hamiltonian. The reason the static perturbation theory applies in the first order is that any retardation effect will only manifest themselves as higher order derivatives in $\dot\lambda$ and hence will affect only higher order terms in the perturbation theory (see Ref.~\cite{degrandi_10} for more details).
	
	Using these expressions for the transition amplitudes, one can go one step further and find the leading non-adiabatic correction to various observables. It is convenient to represent such observables $\mathcal M_\mu$ as conjugate to the parameters of the Hamiltonian: $\mathcal M_\mu=-\partial_\mu H$, where $\mu$ can coincide with $\lambda$ or be any other parameter. Then we find
	\be
	M_\mu(t)\equiv\langle \psi(t) |\mathcal M_\mu|\psi(t)\rangle =  M_\mu^{(0)} + F_{\mu\lambda} \dot\lambda + \mathcal{O}(\dot\lambda^2,\ddot{\lambda}), 
	\ee
	where $F_{\mu\lambda}=i\langle 0| [\mathcal A_\mu,\mathcal A_\lambda]|0\rangle \equiv i\langle [\mathcal A_\mu,\mathcal A_\lambda]\rangle_0$ is the Berry curvature evaluated in the instantaneous ground state and $M_{\mu}^{(0)}=\langle 0| \mathcal M_\mu |0\rangle$ is the instantaneous ground state expectation of the generalized force. This force reduces to the Born-Oppenheimer force for heavy nuclei interacting with fast electrons~\cite{book_balian}, or to the Casimir force for macroscopic objects interacting with fast photon modes~\cite{kardar_07}. For the special class of observables which commute with the instantaneous Hamiltonian, e.g. the Hamiltonian itself, the leading non-adiabatic contribution is quadratic in the ramp speed $\dot\lambda$. For example, we find for the energy and the energy variance 
	\begin{eqnarray*}
		\langle H \rangle&\approx& \epsilon_{0}+\dot \lambda^2 \sum_{n\neq 0} {\left|\langle n(\lambda) |(\partial_\lambda H) |0(\lambda)\right |^2 \over (\epsilon_n-\epsilon_0)^3} + \mathcal{O}\left(\dot\lambda^4,\ddot\lambda^2\right),
		\\ \langle H^2\rangle-\langle H \rangle^2 & \approx & \dot\lambda^2 \sum_{n\neq 0} {\left|\langle n(\lambda) |(\partial_\lambda H) |0(\lambda)\rangle\right|^2 \over (\epsilon_n-\epsilon_0)^2} + \mathcal O(\dot\lambda^4,\ddot\lambda^2)=\dot\lambda^2 g_{\lambda\lambda} + \mathcal O(\dot\lambda^4,\ddot\lambda^2),
	\end{eqnarray*}
	where $g_{\lambda\lambda}=\langle \mathcal A_\lambda^2\rangle_0-(\langle \mathcal A_\lambda\rangle_0)^2$ is the fidelity susceptibility~\cite{yang_08, damski_13} or equivalently the diagonal component of the Fubini-Study metric tensor~\cite{provost_vallee_80, zanardi_07, kolodrubetz_13, kolodrubetz_16}.

	\subsection{\label{subsec:FAPT}Floquet Adiabatic Perturbation Theory (FAPT).} 
	
	After this brief introduction to conventional adiabatic perturbation theory, we proceed with a similar approach to Floquet systems. As before, we assume that the Floquet Hamiltonian and the adiabatic limit are well defined. In particular, we assume that the Floquet ``ground state'' (or more accurately the Floquet state we target) is non-degenerate. These conditions can be realized, for instance, in a driven system with a finite-dimensional Hilbert space. They are also realized in special classes of ``Floquet integrable'' systems~\cite{liao_15,chandran_15}. As we show later in Sec.~\ref{subsec:kapitza}, the situation becomes much more interesting and complex in Floquet systems whose time-averaged Hamiltonian features an unbounded spectrum, where these assumptions may break down in a fascinating and physically important way. 
	
	We now consider a system described by the Hamiltonian $H(\lambda,t)$, which is periodic in time with period $T=2\pi/\Omega$ at any fixed $\lambda$. The parameter $\lambda$ is arbitrary; for example, it can be the amplitude, phase, or frequency of the drive, or some other parameter which is not directly coupled to the drive. We mostly focus on the situations where the Floquet Hamiltonian is adiabatically connected to some static non-driven Hamiltonian in the sense that Floquet eigenstates may be continuously tracked as the drive is turned off, in which case it is often convenient to think of $\lambda$ as the driving amplitude. However, this assumption is not essential in the general discussion presented below. Also let us point out that any smooth time dependence of the driving frequency can be eliminated by rescaling time $t\to\tau=\Omega(t) t$ in Schr\"odinger's equation, effectively resulting in the smooth time dependence of the other coupling parameters~\cite{drese_99}.
	
	Since the Hamiltonian is time-periodic, for fixed $\lambda$ it satisfies Floquet's theorem (Eq.~\ref{eq:Floquet_thm}). It is useful to go to a preliminary rotating frame with respect to the micromotion operator $P(\lambda,t)$ to define the instantaneous Floquet Hamiltonian,
	\be
	H_F(\lambda)=P^\dagger(\lambda, t) H(\lambda,t) P(\lambda,t)-i P^\dagger(\lambda,t) \partial_t P(\lambda,t)~,
	\ee
	where $\partial_t$ is used to emphasize that these expressions are for fixed $\lambda$\footnote{Note that the Floquet Hamiltonian defined in this way generally depends on the choice of the initial time $t_0$. This defines the choice of the Floquet gauge mentioned in the introduction~\cite{bukov_14}.}. Let us denote by $|n_F(\lambda)\rangle$ the eigenbasis of this Floquet Hamiltonian, for which $H_F(\lambda) |n_F(\lambda)\rangle = \epsilon_n^F(\lambda) | n_F(\lambda) \rangle$. By our assumption regarding the absence of level crossings\footnote{The importance of the level crossings is discussed in detail starting from Sec.~\ref{subsec:kapitza}.}, the basis states $|n_F(\lambda)\rangle$ and the Floquet Hamiltonian $H_F(\lambda)$ are smooth functions of $\lambda$. Note that this generally implies that we are dealing with a Floquet Hamiltonian whose spectrum is unfolded, or otherwise, if the Floquet energy crosses the edge of the Floquet zone, we would have to introduce a discontinuity into the Floquet spectrum and the $P$ operator. The final expressions for observables, however, will be insensitive to the choice of folding.
	
	Similarly to the stationary case, let us denote by $V(\lambda)$ the unitary transformation which diagonalizes the Floquet Hamiltonian such that $\tilde H_F(\lambda)=V^\dagger(\lambda) H_F(\lambda) V(\lambda)$ is diagonal and $|n_F(\lambda)\rangle=V(\lambda) |e_n\rangle$. Now the moving frame for this Floquet Hamiltonian is defined by two consecutive unitary transformations,
	\[
	|\tilde \psi\rangle = V^\dagger(\lambda) P^\dagger(\lambda,t) |\psi\rangle~,
	\]
	yielding the effective moving-frame Floquet Hamiltonian
	\be
	i \mathrm{d}_t |\tilde \psi\rangle = (\tilde H_F- \dot \lambda \tilde{\mathcal A}^{F}_\lambda) |\tilde \psi\rangle,
	\ee
	where
	\be
	\label{eq:Floquet_gauge_pot}
	\tilde{ \mathcal A}^{F}_\lambda(t)=i V^\dagger(\lambda) \partial_\lambda V(\lambda)+i V^\dagger(\lambda) P^\dagger(\lambda,t) [\partial_\lambda P(\lambda,t)] V(\lambda)
	\ee
	is the Floquet generalization of the adiabatic gauge potential. Unlike in conventional APT, the gauge potential naturally splits into two contributions: the first one describes the adiabatic changes of the instantaneous eigenstates of the Floquet Hamiltonian and thus only depends on $V$, while the second one describes transitions due to the micromotion $P$. Intuitively, this new contribution can be understood by noticing that during the ramp the Hamiltonian is not strictly periodic, and thus there are corrections induced when the $P$ operator does not come back to itself after one cycle. In the Floquet stationary frame, obtained by removing the $V$-rotation, $\mathcal A^F_\lambda \equiv V \tilde{ \mathcal A}^F_\lambda V^\dagger$ is given by
	\begin{eqnarray}
	\mathcal A^F_\lambda &=& \mathcal A_\lambda^V(\lambda)+\mathcal A_\lambda^P(\lambda,t)~, \nonumber\\
	\mathcal A_\lambda^V(\lambda) &\equiv& -iV(\lambda)\partial_\lambda V^\dagger(\lambda),\nonumber\\
	\mathcal A_\lambda^P(\lambda,t) &\equiv& i P^\dagger(\lambda,t) \partial_\lambda P(\lambda,t),\nonumber\\
	\langle m_F | \mathcal A^F_\lambda | n_F \rangle &=& i \langle  m_F | \partial_\lambda| n_F \rangle +  \langle  m_F | \mathcal A_\lambda^P | n_F \rangle~
	\label{eq:AFdef}
	\end{eqnarray}
	While the first term here does not explicitly depend on time, the second one depends on time both implicitly via the slowly changing $\lambda$ and explicitly through the oscillating in time terms in $P(\lambda, t)$. Similarly to APT, the matrix elements of $i\partial_\lambda$ are related to the matrix elements of the Floquet generalized forces and the Floquet energies via
	\be
	\langle m_F(\lambda)| i \partial_\lambda | n_F(\lambda)\rangle=-i {\langle m_F(\lambda) |\partial_\lambda H_F(\lambda) |n_F(\lambda)\rangle\over \epsilon_m^F(\lambda)-\epsilon_n^F(\lambda)}~;~m\neq n,
	\ee
	which can be obtained by differentiating $\langle m_F | H_F | n_F \rangle = 0$ with respect to $\lambda$. The $\mathcal A_\lambda^P$ part of the gauge potential describes adiabatic changes in the micromotion operator and is unrelated to the Floquet Hamiltonian. It therefore does not have a simple equilibrium analogue.
	
	It bears mention that the Floquet gauge potential takes on an even simpler form when written out in the basis $|n_F(\lambda,t)\rangle \equiv P(\lambda,t) |n_F(\lambda)\rangle$. One can think of these states as the natural basis in the absence of ramping because if one starts in the state $|n_F(\lambda,t)\rangle$ at time $t$, then at later time $t'$ for fixed $\lambda$ one will end up in $|n_F(\lambda,t')\rangle$. Then defining $\mathcal A^{F \prime}_\lambda \equiv P \mathcal A^F_\lambda P^\dagger = (P V) \tilde{\mathcal A}^F_\lambda (PV)^\dagger$, it has matrix elements $\langle m_F(\lambda, t) | \mathcal A^{F\prime}_\lambda | n_F(\lambda,t) \rangle = i \langle m_F(\lambda, t) | \partial_\lambda | n_F(\lambda,t) \rangle$. Our results can be easily re-expressed in this basis, but doing so makes it harder to distinguish between micromotion and non-micromotion effects of $\mathcal A^F_\lambda$. Therefore, for the remainder of this article we work in the ``stroboscopic'' basis $|n_F(\lambda)\rangle$.
	
	Combining all these transformations together we see that the exact time evolution of the amplitude $c_n$ in the instantaneous Floquet basis
	\[
	|\psi(t)\rangle=\sum_n c_n(t) P(\lambda,t) |n_F(\lambda)\rangle
	\]
	reads:
	\be
	i \dot c_n=\left(\epsilon^F_n(\lambda) -\dot\lambda \langle n_F(\lambda) |\mathcal A^F_\lambda(\lambda,t))|n_F(\lambda)\rangle\right) c_n-\dot\lambda\sum_{m\neq n} \langle n_F(\lambda) | \mathcal A^F_\lambda(\lambda,t) |m_F(\lambda)\rangle c_m.
	\label{eq:floquet_moving_frame}
	\ee
	In general, we will not be able to solve these equations analytically, but we will show how FAPT allows us to solve them to a good approximation in the limit of slow ramps.
	
	{\em Adiabatic limit.}  Assume that the system is initially prepared in the (Floquet) ground state at time $t=t_i$, i.e., $c_0(t_i)=1$ and $c_n(t_i)=0$  for $n\neq 0$\footnote{Of course a word of caution is needed here as the notion of the Floquet ground state is usually ill defined. As in this section we consider Floquet Hamiltonians adiabatically connected to static Hamiltonians we can define the ground state as simply an adiabatic continuation of the static ground state. In more general situations one can understand by the ground state some target state, e.g. the state with lowest entanglement or the state with the lowest mean energy of an approximate local Floquet Hamiltonian.}. Then we see that, similarly to the non-driven case, the system follows the instantaneous Floquet ground state and the wave function acquires a phase:
	\be
	\Phi_0^F(t)=-\int_{t_i}^t \epsilon^F_0(\lambda(t')) \mathrm{d}t' +\int_{t_i}^t \mathrm{d}t' \dot\lambda(t') \langle 0_F(\lambda(t')) | \mathcal A^F_\lambda(\lambda(t'),t') |0_F(\lambda(t'))\rangle.
	\label{eq:phi(t)}
	\ee
	The first term here is the usual dynamic phase. The second term gives both the Berry phase associated with the Floquet Hamiltonian (coming from $\mathcal A_\lambda^V$) and an additional contribution due to the $P$ operator, which explicitly depends on time. The expression for the phase $\Phi_0^F$ greatly simplifies if we ramp over many periods, such that the ramp effectively becomes slow and there is little change on the scale of a single driving period. Then  only its period-averaged value contributes:
	\begin{eqnarray}
	\gamma &\equiv& \int_{t_i}^t \mathrm{d}t' \dot\lambda(t') \langle 0_F(\lambda(t')) | \mathcal A^F_\lambda(\lambda(t'),t') |0_F(\lambda(t'))\rangle\nonumber
	\\ & \approx & \int_{t_i}^t \mathrm{d}t' \dot\lambda(t') \overline{\langle 0_F(\lambda(t')) | \mathcal A^F_\lambda(\lambda(t'),t') |0_F(\lambda(t'))\rangle} =  \int_{\lambda(t_i)}^{\lambda(t)} \mathrm{d}\lambda' (\langle 0_F(\lambda') |\overline{\mathcal A^F_\lambda}(\lambda')|0_F(\lambda')\rangle,
	\end{eqnarray}
	where $\overline{(\cdot)} = T^{-1} \int_0^T (\cdot) \mathrm{d}t$ is the average over a cycle at fixed $\lambda$.

	{\em Leading non-adiabatic response.} In order to approximate Eq.~\eqref{eq:floquet_moving_frame} beyond the adiabatic limit, it is convenient to go to the interaction picture with respect to the diagonal term:
	\[
	c_n=c_n' \exp[-i \Phi_n^F(t)],
	\]
	where the phase $\Phi_n^F(t)$ is defined similarly to Eq.~\eqref{eq:phi(t)}. Then Eq.~\eqref{eq:floquet_moving_frame} becomes
	\be
	i\dot c_n'=-\dot\lambda\sum_{m\neq n} \langle n_F(\lambda) | \mathcal A^F_\lambda(\lambda,t) |m_F(\lambda)\rangle \mathrm e^{i (\Phi_n^F(t)-\Phi_m^F(t))} c_m'.
	\label{eq:FAPT_exact}
	\ee
	To leading order in $\dot\lambda$ we thus find for $n\neq 0$
	\be
	c_n'(t)=  i\int_{t_i}^t \mathrm{d}t'\dot\lambda(t') \langle n_F(\lambda) | \mathcal A^F_\lambda(\lambda,t) |0_F(\lambda)\rangle \mathrm e^{i (\Phi_n^F(t')-\Phi_0^F(t'))} + \mathcal{O}(\dot\lambda^2).
	\label{eq:cnprime_leading}
	\ee
	To evaluate this integral it is convenient to decompose the gauge potential into Fourier harmonics:
	\[
	\mathcal A^F_\lambda(\lambda,t)=\sum_{\ell=-\infty}^\infty \mathrm e^{i\Omega \ell t} \mathcal A^{F,\ell}_\lambda(\lambda)
	\]
	Assuming that the protocol starts smoothly ($\dot\lambda(t_i)=0$) such that transients can be neglected, we can approximately evaluate the integrals by expanding around $t'=t$. This procedure is similar to what is done in standard APT (cf. Ref.~\cite{degrandi_10}), and is detailed in~\ref{app:general_thy}. Returning to the Floquet stationary frame, we obtain to the leading order in $\dot\lambda$: 
	\begin{equation}
	c_n(t)= \mathrm e^{-i\Phi_0^F(t)} \dot \lambda(t) \sum_{\ell=-\infty}^\infty \mathrm e^{i \ell \Omega t} {\langle n_F(\lambda)|\mathcal A^{F,\ell}_\lambda|0_F(\lambda)\rangle\over \epsilon^F_n-\epsilon^F_0+\ell\Omega}+\mathcal{O}(\ddot \lambda, \dot\lambda^2).
	\label{eq:c_n_FAPT}
	\end{equation}
	We note that this expression for the transition amplitudes has previously been derived by different means in Ref.~\cite{drese_99} in the context of quantum chemistry. In the following, we discuss the implications of this result to various physical observables. Examining this expression, we see that unlike the non-driven case, the leading non-adiabatic response in Floquet systems generates additional oscillating terms which can be interpreted as non-adiabatic corrections to the $P$-operator. As we shall see, in a wide class of problems these oscillating terms are equally or sometimes even more important than the non-adiabatic corrections due to the slowly changing Floquet Hamiltonian.
	
	To measure the deviations from the adiabatic limit, we consider the probability $p_n^F=|c_n(t)|^2$ of being in the Floquet state $|n_F \rangle$. Using Eq.~\eqref{eq:c_n_FAPT} one finds that the probabilities $p_n^F$ are given by:
	\begin{equation}
	\label{eq:FAPT_pF}
	p_n^F= \dot \lambda(t)^2\sum_{\ell,\ell'}\mathrm e^{i(\ell-\ell')\Omega t}\frac{\langle n_F(\lambda)|\mathcal A^{F,\ell}_\lambda|0_F(\lambda)\rangle\langle 0_F(\lambda)|\mathcal A^{F,\ell'}_\lambda|n_F(\lambda)\rangle}{(\epsilon^F_n-\epsilon^F_0+\ell\Omega)(\epsilon^F_n-\epsilon^F_0+\ell'\Omega)}
	\end{equation}
	From these probabilities, we can define the log-fidelity $f_d$ and the associated Floquet diagonal entropy $S_d^F$ as
	\begin{equation}
	f_d=-\log|c_0|^2=-\log\left(1-\sum_{n>0}|c_n|^2\right),~S_d^F = - \sum_n p^F_n \log p^F_n.
	\label{eq:FAPT_entropy}
	\end{equation}	
	Since $c_{n \neq 0} \sim v = \dot{\lambda}(t_f)$ for small velocities, both $f_d$ and $S_d^F$ scale as $v^2$ in the low velocity limit, up to a small log correction in $S_d^F$. We shall use this characteristic feature as a benchmark of adiabaticity in various models. While one can use either $f_d$ or $S_d^F$ to measure the magnitude of the non-adiabatic corrections, notice that the former requires the identification of the adiabatically connected Floquet state, while the latter does not. Hence, in complicated models, the entropy often constitutes a simpler measure of adiabaticity. As in non-Floquet systems, the diagonal entropy is simply a measure of delocalisation of the wave-function (or more generally density matrix) among the eigenstates of the instantaneous Floquet Hamiltonian.

	It is useful to understand Eq.~\eqref{eq:c_n_FAPT} in two important limits. First, in the limit of vanishing driving amplitude, there is no micromotion, and therefore all terms with $\ell\neq 0$ may be neglected. Furthermore in this limit $\epsilon_n^F(\lambda)\to\epsilon_n(\lambda)$ and $\mathcal{A}^{F}(\lambda)\to\mathcal{A}(\lambda)$. As a result, Eq.~\eqref{eq:c_n_FAPT} reduces to Eq.~\eqref{eq:APT_cn}, reproducing conventional APT as expected. 
	
	A similar situation occurs in the infinite-frequency limit, where all the Fourier modes in~\eqref{eq:c_n_FAPT} disappear leaving only the $\ell=0$ component\footnote{When the amplitude of the drive scales with the frequency, which is the relevant case for Floquet engineering, $\mathcal A^{F,\ell}$ acquires $\Omega$ dependence, and $\ell\neq 0$ terms may also survive the infinite frequency limit of~\eqref{eq:c_n_FAPT}}:
	\begin{equation}
	c_n(t)\approx \mathrm e^{-i\Phi_0^F(t)} \dot \lambda(t)  {\langle n_F(\lambda)|\mathcal A^{F,0}_\lambda|0_F(\lambda)\rangle\over \epsilon^F_n-\epsilon^F_0}+\mathcal{O}(\ddot \lambda, \dot\lambda^2).
	\label{eq:cn_Omega_infty}
	\end{equation}
	This is equivalent to assuming time-scale separation and averaging over the fast time variable (cf.~Ref.~\cite{kitagawa_11}), in which case  one loses information about the higher Fourier modes. While the probability amplitudes in the limits of vanishing drive amplitude, Eq.~\eqref{eq:c_n_FAPT}, and infinite frequency, Eq.~\eqref{eq:cn_Omega_infty}, look deceptively similar, there exists a subtle difference: the physics in the two limits could be governed by two Hamiltonians with completely different properties. This is particularly relevant when Floquet engineering methods are applied, which requires that the driving amplitude is of the order of the driving frequency~\cite{bukov_14}.

	\subsection{\label{subsec:observables} Observables.}
	
	Using the transition amplitudes it is straightforward to find the leading non-adiabatic corrections to the expectation values of observables. As in the the non-driven case, it is convenient to represent observables in terms of generalized forces, $\mathcal M_\mu(t)=-\partial_\mu H(t)$. Using Eq.~\eqref{eq:c_n_FAPT} we find
	\begin{multline}
	M_\mu(t) \equiv \langle \psi(t)| \mathcal M_\mu(t) | \psi(t)\rangle  \approx  \langle 0_F| P^\dagger(t) \mathcal M_\mu(t) P(t) |0_F\rangle+
	\\  \dot \lambda \sum_{n\neq 0} \sum_{\ell=-\infty}^\infty \left( \mathrm e^{i\ell\Omega t} {\langle 0_F| P^\dagger(t) \mathcal M_\mu(t) P(t) |n_F\rangle\langle n_F|\mathcal A_\lambda^{F,\ell} |0_F\rangle\over \epsilon_n^F-\epsilon_0^F+\ell\Omega}+ \mathrm{c.c.}\right)
	\label{eq:observable_fapt}
	\end{multline}
	Here we have dropped the argument $\lambda$ in the $P$ operator to simplify the notation. The result above can be simplified further by expressing it through the Floquet generalized force, $\mathcal M_\mu^F=-\partial_\mu H_F$. In order to do this we note that
	\begin{eqnarray}
	\label{eq:PdH_FP}
	P^\dagger(t) \partial_\mu H(t) P(t)&=& \partial_\mu [P^\dagger(t) H(t) P(t) -iP^\dagger(t) \partial_t P(t) + i P^\dagger(t)\partial_t P(t)] \nonumber\\
	&& - [\partial_\mu P(t)^\dagger] H(t) P(t) - P^\dagger(t) H(t) \partial_\mu P(t)\nonumber\\
	&=& \partial_\mu H_F+\partial_t \mathcal A_\mu^P(t)+i[H_F, \mathcal A_\mu^P(t)],
	\end{eqnarray}
	where we separated out the $P$-component of the gauge potential $\mathcal A^P_\mu(t)=i P^\dagger(t) \partial_\mu P(t)$ as in Eq.~\eqref{eq:AFdef}. Recall that $\mathcal A_\mu^V=-iV\partial_\mu V^\dagger$ does not explicitly depend on time. Then, using the right-hand side of Eq.~\eqref{eq:PdH_FP}, we have 
	\begin{eqnarray}
	\langle 0_F | \partial_t \mathcal A_\mu^P(t)| n_F\rangle &=& \langle 0_F | \partial_t \mathcal A_\mu^F(t)| n_F\rangle = -i\sum_{\ell'=
		-\infty}^\infty \mathrm e^{-i\ell'\Omega t }
	\ell'\Omega \langle 0_F |\mathcal A_\mu^{F,-\ell'} | n_F\rangle,\nonumber\\
	\langle 0_F | \partial_\mu H_F +i [H_F,\mathcal A_\mu^P(t)]| n_F\rangle &=& -i(\epsilon_n^F-\epsilon_0^F)  \langle 0_F |\mathcal A_\mu^{V} | n_F\rangle -i\sum_{\ell'=-\infty}^\infty \mathrm e^{-i\ell'\Omega t }
	(\epsilon_n^F-\epsilon_0^F)  \langle 0_F |\mathcal A_\mu^{P,-\ell'} | n_F\rangle \nonumber\\
	&=& -i\sum_{\ell'=-\infty}^\infty \mathrm e^{-i\ell'\Omega t }
	(\epsilon_n^F-\epsilon_0^F)  \langle 0_F |\mathcal A_\mu^{V,-\ell'} + \mathcal A_\mu^{P,-\ell'} | n_F\rangle \nonumber\\
	&=& -i\sum_{\ell'=-\infty}^\infty \mathrm e^{-i\ell'\Omega t }
	(\epsilon_n^F-\epsilon_0^F)  \langle 0_F |\mathcal A_\mu^{F,-\ell'} | n_F\rangle .
	\end{eqnarray}
	To see the last equality note that that by constructions all non-zero harmonics of $\mathcal A_\mu^{V,\ell\neq 0} = 0$ vanish identically, while $\mathcal A_\mu^{V,\ell= 0} = \mathcal A_\mu^{V}$. We then combined the two gauge potentials into the single Floquet gauge potential using Eq.~\eqref{eq:AFdef}.
	
	Adding the expression above in the right-hand side of Eqs.~\eqref{eq:PdH_FP} and substituting the result in Eq.~\eqref{eq:observable_fapt}, we find that the generalized force reads
	\begin{eqnarray}
	M_\mu(t) &\approx&\langle 0_F|\mathcal M_\mu^F |0_F\rangle-i  \sum_{\ell\neq 0} \Omega \ell \mathrm e^{i\ell \Omega t} \langle 0_F|\mathcal A_\mu^{F,\ell}|0_F\rangle + 
	\nonumber
	\\ & & i \dot \lambda \sum_{n \neq 0}\sum_{\ell,\ell'=-\infty}^\infty \left(\mathrm e^{i(\ell-\ell')\Omega t} \frac{\epsilon_n^F - \epsilon_0^F + \ell' \Omega}{\epsilon_n^F - \epsilon_0^F + \ell \Omega}\langle 0_F | \mathcal A_\mu^{F,-\ell'} | n_F \rangle \langle n_F | \mathcal A_\lambda^{F,\ell} | 0_F\rangle  - \mathrm{c.c.}\right)
	\label{eq:observable_fapt_fs}
	\end{eqnarray}
	
	There are two types of measurements one usually applies to periodically driven systems. Floquet stroboscopic (FS) measurements are performed at integer multiples of the driving period and are given by the general expression in Eq.~\eqref{eq:observable_fapt_fs}. Floquet non-stroboscopic (FNS) measurements are averaged over many cycles, or equivalently averaged over the driving phase $\varphi_0$~\cite{bukov_14}. We thus refer to FNS measurements as ``phase-averaged'' throughout the course of this paper. The choice of driving phase is often uncontrolled in experiments and, thus, its fluctuations from shot to shot effectively lead to phase-averaged measurements. The expressions for observables in FAPT greatly simplify for the phase-averaged measurement protocol as all non-zero harmonics average to zero. Then the generalized force becomes the Floquet generalized force, as anticipated:
	\be
	\overline{M_{\mu}^{(0)}}=\langle 0_{ F}|\mathcal M_\mu^F |0_{ F}\rangle =-\partial_\mu \epsilon_0^F~,
	\label{eq:Floquet_gen_force}
	\ee
	where in the second equality we have used the (Floquet) Feynman-Hellmann theorem~\cite{book_sakurai}. Meanwhile, the leading non-adiabatic correction becomes
	\be
	\overline {M_\mu^{(1)}}=i \dot \lambda  \sum_{\ell} \left\langle 0_F\left| \left[\mathcal A_\mu^{F,\ell}, \mathcal A_\lambda^{F,-\ell}\right]\right|0_F\right\rangle=i\dot \lambda \overline{\left\langle0_F\left|  \left[\mathcal A_\mu^F(t), \mathcal A_\lambda^F(t)\right]\right|0_F\right\rangle \phantom{^X\!\!\!\!}}~,
	\label{eq:linear_respose_FNS}
	\ee
	where as before
	\begin{equation}
	\overline{\mathcal O} = \frac{1}{T} \int_0^T \mathcal O(t) \mathrm{d}t
	\label{eq:fns_definition}
	\end{equation}
	denotes period (or equivalently phase) averaging over the cycle at fixed $\lambda$.

	\subsection{\label{subsec:Ftopology} Floquet Berry Curvature and Floquet Chern Number.}
	
	In APT, the leading-order correction to $M_\mu$ for a ramp of the parameter $\lambda$ is related to the Berry curvature~\cite{gritsev_12} $F_{\lambda\mu}$(see Sec.~\ref{subsec:APT}). Thus, it is natural to ask in which sense this generalises to Floquet systems. If we consider the state $|0_F(\lambda,t)\rangle = P(\lambda, t) |0_F(\lambda)\rangle$ introduced earlier, then the natural extension of the Berry curvature to Floquet systems is
	\begin{equation}
	\label{eq:Floquet_Berry_curvature}
	F^F_{\lambda\mu}(t) = i \langle \partial_\lambda 0_F (t) | \partial_\mu 0_F(t) \rangle+\mathrm{h.c.} = i\left \langle 0_F\left|  \left[\mathcal A_\lambda^F(t), \mathcal A_\mu^F(t)\right]\right|0_F\right \rangle~.
	\end{equation}
	This Floquet Berry curvature can also be expressed through the derivatives of the instantaneous Berry connection $A^F_\lambda(t)=i \langle 0_F(t)|\partial_\lambda| 0_F(t)\rangle$ in a standard fashion: $F_{\lambda\mu}^F(t)=\partial_\lambda A_\mu^F(t)-\partial_\mu A_\lambda^F(t)$.  While the instantaneous non-adiabatic response of observables in Floquet systems is not directly related to the Berry curvature (c.f. Eqs.~\eqref{eq:observable_fapt_fs} and \eqref{eq:Floquet_Berry_curvature}), the leading non-adiabatic correction is proportional to the period (phase) averaged Floquet Berry curvature. Indeed, comparing Eqs.~\eqref{eq:linear_respose_FNS} and \eqref{eq:Floquet_Berry_curvature} we see that:
	\be
	\overline{M_\mu} \approx \overline{M_\mu^{(0)}} + \dot \lambda \overline{F_{\mu\lambda}^F} ~.
	\ee
	
	Whereas $\overline{F_{\lambda \mu}^F}$ is an interesting curvature form in its own right, one may ask how to use it to obtain geometric and topological properties of the time-dependent system. One nice topological invariant which is unaffected by this time averaging is the Floquet Chern number~\cite{lindner_11,oka_09,kitagawa_11,jotzu_14,aidelsburger_14,flaeschner_15}, which is defined on a closed manifold for $(\lambda,\mu)$ and for any given time $t$ during the cycle as $C_1(t) = \frac{1}{2\pi}\int \mathrm{d} \lambda \mathrm{d}\mu F^F_{\lambda\mu}(\lambda,\mu,t)$. Notice that the Floquet eigenstates corresponding to different times within the period are connected by a continuous unitary gauge transformation, which does not change the energy spectrum and cannot lead to gap closings. Therefore, the corresponding Floquet Chern number~\cite{dalessio_15_Chern} is independent of the time within the period, and hence also independent of the driving phase. Thus, $C_1(t) = C_1(t^\prime) = C_1^F$ defines the Floquet Chern number, which can be found by measuring $\overline{F_{\lambda \mu}^F}$ and integrating:
	\be
	\label{eq:floquet_chern_number}
	C_1^F=\frac{1}{T} \int \mathrm{d} t C_1(t) = \frac{1}{2\pi T} \oiint \mathrm{d} \lambda \mathrm{d}\mu \mathrm{d}t F^F_{\lambda\mu} (\lambda,\mu;t) = \frac{1}{2 \pi} \oiint \mathrm{d} \lambda \mathrm{d}\mu \overline{F^F_{\lambda\mu} (\lambda,\mu)}.
	\ee
	This important result tells us that one can engineer, at least in principle, Floquet systems with quantized Hall-type response. In order to do this, one has to be in a position to prepare these systems sufficiently close to the corresponding Floquet ground state and perform phase-averaged measurements of the  current~\cite{bukov_14_pra,atala_14} or other related observables.
	
	\section{\label{sec:examples_SP} Single-Particle Examples.}
	
	Having introduced the FAPT formalism and used it to discuss non-adiabatic corrections and their connection to the Berry curvature, we now move on to illustrate these ideas with a variety of examples of increasing complexity. We start with the simplest case of the exactly-solvable single particle in a periodically-displaced harmonic potential, where corrections due to FAPT may be cleanly isolated and analyzed. We then move on to the quantum Kapitza pendulum -- a non-linear single-particle system --  where we delineate the role of photon absorption resonances. Finally, we study linear response and non-adiabatic corrections to observables in a linearly-driven qubit system, where a simple measurable connection to the topological Floquet Chern number is demonstrated through the Thouless energy pump.
	
	\subsection{\label{subsec:HO}The Periodically Driven Harmonic Oscillator.}
	
	Let us now consider the quantum harmonic oscillator with a periodically-displaced confining potential. This model is exactly solvable and shall therefore prove useful as a first check of the FAPT expressions derived in Sec.~\ref{subsec:FAPT}.  We consider the case where the drive consists of an oscillating force with frequency $\Omega$ whose amplitude is ramped according to some slow parameter $\lambda(t)$:
	\begin{equation}
	H(t)=\frac{p^2}{2m}+\frac{1}{2}m\omega_0^2 x^2 - A_f\Omega^2\lambda(t)\cos(\Omega t+\varphi_0)x~.
	\label{eq:H0_H}
	\end{equation}
	We pick units with $m\omega_0=1$ and explicitly introduce the driving phase $\varphi_0$, such that one can easily distinguish between phase-averaged and non-averaged protocols. Further, we chose the amplitude of the drive to scale quadratically with the driving frequency, which leads to a non-trivial high-frequency regime.
	
	We consider slow ramps, such that the oscillator starts in its ground state with the drive off at some $t_i<0$ [$\lambda(t_i)=\dot\lambda(t_i)=0$], and then smoothly ramp up the drive amplitude $\lambda(t)$ to the final value. It is convenient to set the final measurement time where we evaluate all observables to $t_f=0$ such that $|t_i|$ is equal to the ramp time. To simplify notations we define $v$ as the instantaneous velocity at the final time, i.e.~$v\equiv \dot \lambda(0)$ and we set the final value of $\lambda(0)$ to unity. The simplest protocol which satisfies these constraints is a quadratic ramp:
	\begin{equation}
	\lambda(t)=\left(\frac{t-t_i}{|t_i|}\right)^2\label{quad_ramp}
	\end{equation}
	with $v=2/|t_i|$. As a consequence of the analysis in Sec.~\ref{subsec:FAPT}, the FAPT expansion is independent of the protocol used for the ramp. We confirmed this by comparing the exact dynamics numerically for various ramping protocols, and found an excellent agreement in the small $\dot{\lambda}$ limit. 
	
	One advantage of choosing this Hamiltonian is that it is exactly solvable. Namely for any driving protocol $f(t)$ 
	\[
	H(t)={p^2\over 2m}+{m\omega^2 x^2\over 2}-f(t) x
	\] 
	one can reduce the time-dependent problem to an effective static harmonic oscillator by transforming into a moving frame with respect to the classical trajectory $\eta(t)$, whose equation of motion is $\ddot{\eta}(t)+\omega_0^2\eta(t)=f(t)/m$. The eigenstates in this frame are given in terms of the static harmonic oscillator states $\phi_n(x)$ and eigenenergies $E_n=\omega_0(n+1/2)$ as:
	\begin{equation}
	\chi_n(x,t)=\phi_n(x-\eta(t))\exp\left[i\left(m\dot{\eta}(t)\left(x-\eta(t)\right)-E_n t+\int_{t_i}^t  \mathrm{d}t' L(\eta,\dot{\eta},t')\right)\right]
	\end{equation}
	where $L(\eta,\dot{\eta},t)$ is the classical Lagrangian for a driven oscillator:
	\[
	L(\eta,\dot{\eta},t)=\frac{1}{2}m\dot{\eta}^2(t)-\frac{1}{2}m\omega_0^2\eta^2(t)+\eta(t)f(t).
	\]
	Any solution to the time-dependent Schr\"odinger equation with Hamiltonian \eqref{eq:H0_H} is given by a linear combination of $\chi_n(x,t)$ with \emph{time independent} coefficients. Using this method, we obtain not only the exact solution of the dynamics for arbitrary $\lambda(t)$ protocols, but also the exact Floquet eigenstates at any fixed $\lambda$. For more details on the exact solution used to compare this model to FAPT see~\ref{app:HO_thy}.
	
	To see why this model is interesting in the context of FAPT, let us start by solving it in the infinite-frequency limit. This is trivially done by a pair of unitary rotations, $|\psi^\mathrm{rot}(t)\rangle = V_2^\dagger(t) V_1^\dagger(t) |\psi(t) \rangle$, where 
	\be
	V_1(t)=\mathrm e^{i A_f \Omega \lambda \sin (\Omega t + \varphi_0) x}~,~V_2(t)=\mathrm e^{i A_f \lambda \cos (\Omega t + \varphi_0) p/m}~.\label{eq:HO_rot_frame}
	\ee
	One may readily confirm that this gives the rotating frame Hamiltonian
	\be
	\label{eq:H0_H_rot}
	H^\mathrm{rot}(t)=\frac{p^2}{2m}+\frac{1}{2}m\omega_0^2 x^2 - A_f\omega_0^2\lambda(t)\cos(\Omega t + \varphi_0)\; x~.
	\ee
	up to an irrelevant constant. This looks identical to the original Hamiltonian, except with $\Omega \to \omega_0$ in the drive amplitude. But now the $\Omega \to \infty$ limit is trivial because the drive strength remains finite, and thus the infinite frequency Floquet Hamiltonian is simply the time average of $H_{\mathrm{rot}}$~\cite{bukov_14}. This Floquet Hamiltonian is obviously $\lambda$-independent, and so are its eigenstates. However, as we shall see shortly, there is an important difference in the micromotion between the time-evolution due to Eq.~\eqref{eq:H0_H} and Eq.~\eqref{eq:H0_H_rot}.  Hence, any non-adiabatic effects can occur solely due to the $\lambda$-dependence of the micromotion operator $P(\lambda,t)$. In~\ref{app:HO_thy} we explicitly demonstrate that the contribution due to this operator to the transition probabilities and observables has a well-defined infinite-frequency limit. Thus, this example serves as a direct proof that neglecting the effects of the micromotion operator on the non-adiabatic response can lead to erroneous conclusions. 
	
	\begin{figure}[ht]
		\includegraphics[width=\columnwidth]{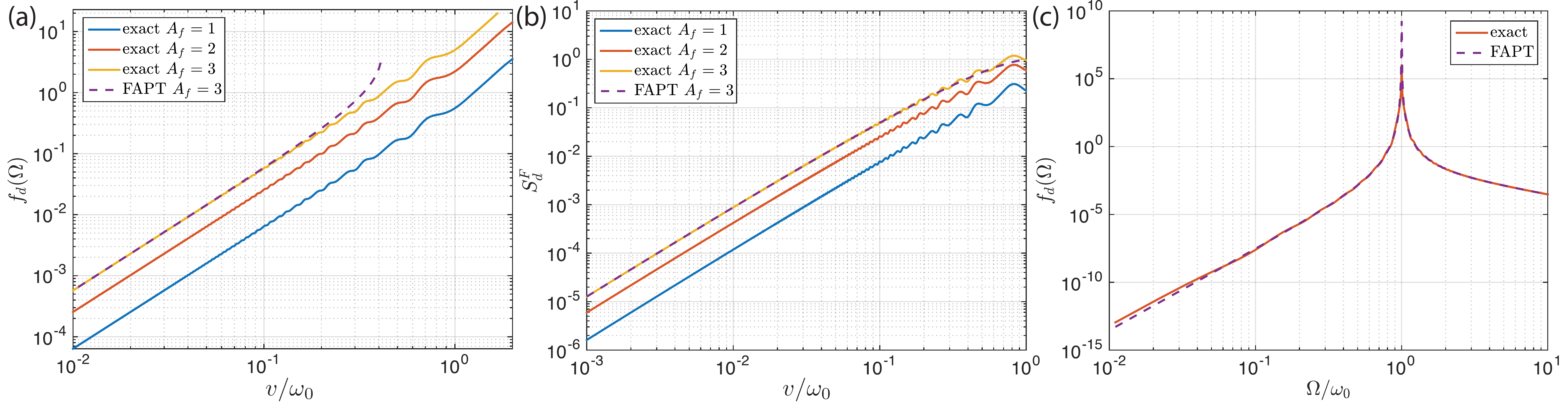}
		\caption{\label{fig:HO_Pexc}(Color online)[Driven harmonic oscillator]. Non-adiabatic transitions: exact results for the log-fidelity (a) and Floquet diagonal entropy (b) at $\Omega/\omega_0=5$ as a function of the driving amplitude and the ramp velocity. The FAPT prediction is shown for $A_f=3$ as a dashed line. (c) Log-fidelity as a function of $\Omega/\omega_0$ for $A_f=3$, $m\omega_0=1$, and $v/\omega_0=0.04$. The dashed line shows the FAPT prediction. All excitations in the model are solely due to the micromotion (see text). The data is shown for the driving phase $\varphi_0=0$. } 
	\end{figure}
	
	Next, let us discuss the exact results for this setup and compare them with the predictions of the FAPT. In Fig.~\ref{fig:HO_Pexc} (a) and (b) we show the log-fidelity and the Floquet diagonal entropy versus the ramp rate $v$. As we discussed in the previous section, these are the observable-independent measures of the non-adiabatic corrections. We also show a comparison of the exact results with the predictions of FAPT, and find an excellent agreement at small values of $v$. At this point we should briefly highlight a few important features of this model: first, it is interesting to note that the only source of excitations is the micromotion, not just at infinite frequency, but at any finite frequency as well; see~\ref{app:HO_thy} for details. One readily can check that in the leading order of FAPT and in the infinite frequency limit the system can only undergo the transitions to the first excited state, yielding:
	\begin{equation}
	p_1^F\stackrel{\Omega\to\infty}{\longrightarrow}\frac{A_f^2v^2}{2m\omega_0}\cos^2(\varphi_0).
	\end{equation} 
	In turn, this transition probability defines the log-fidelity and the Floquet diagonal entropy:
	\be
	f_d=-\log(1-p_1^F)\approx p_1^F,\; S_d^F\approx p_1^F (1-\log(p_1^F)),
	\nonumber
	\ee
	which match well the numerical results plotted in Figs.~\ref{fig:HO_Pexc}(a) and (b).

\begin{figure*}
		\begin{minipage}{\textwidth}
			\includegraphics[width=\columnwidth]{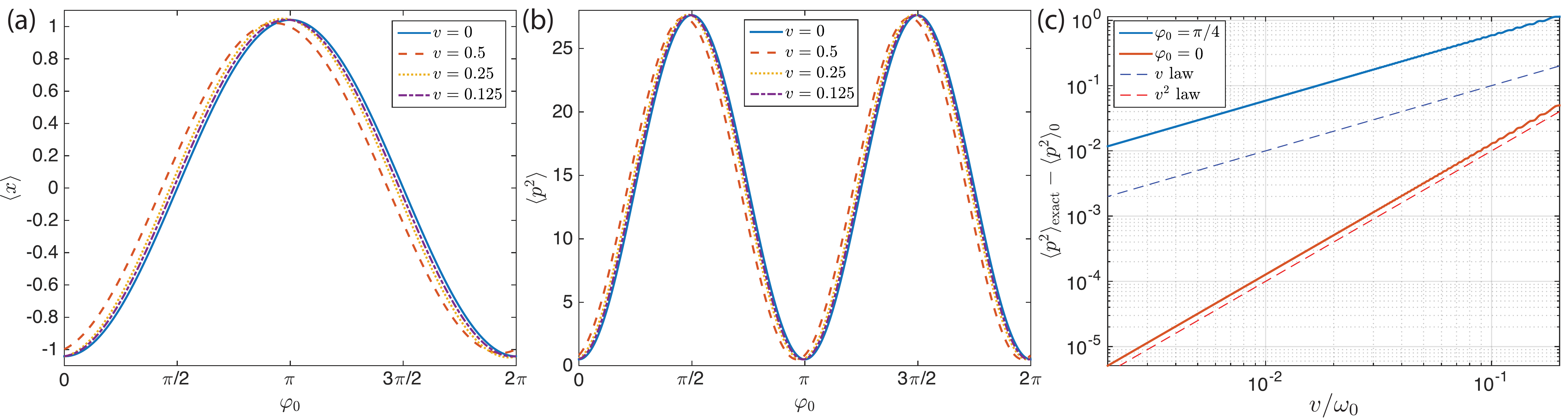}
			\caption{\label{fig:HO_obs}(Color online)[Driven harmonic oscillator]. Observables for $A_f=1$ and $\Omega/\omega_0=5$. Position operator $\langle x\rangle$ (a) and momentum-squared $\langle p^2 \rangle$ (b) as a function of the driving phase $\varphi_0$, showing agreement in the adiabatic limit $v\to 0$. (c) Difference between the exact value of $\langle p^2 \rangle$ and the value predicted by FAPT for  $\varphi_0=0$ (red) and $\varphi_0=\pi/4$ (blue). The blue and red dashed lines show a $v$ and a $v^2$--power law for comparison. }
		\end{minipage}
	\end{figure*}

	The above story is also supported by the behaviour of observables. For large $v$, the expectation values $\langle p^2\rangle$ and $\langle x\rangle$ are misaligned from the corresponding expectations in the Floquet ground state, as seen in~Fig.~\ref{fig:HO_obs}(a) and (b). As the velocity approaches zero, they converge to the ground state expectation values. One caveat when scaling the amplitude of the drive quadratically with the driving frequency is that observables, which do not commute with the driving, may not have a well-defined behaviour in the strict $\Omega\rightarrow\infty$ limit due to a divergent amplitude of the micromotion in the lab frame. In this example, the adiabatic expectation value $\langle p^2\rangle$ diverges as $\Omega^2$, while the non-adiabatic correction remains finite. For the expectation value $\langle x\rangle$, on the other hand, the ground state converges to a finite value in the infinite-frequency limit, but the non-adiabatic corrections vanishes as $\Omega^{-1}$.
	Figure~\ref{fig:HO_obs} (c) shows the difference between the exact expectation value of $p^2$ at the measurement point and the corresponding FAPT prediction to order $\mathcal{O}(v)$. Whenever the measurement is taken at a time-reversal symmetric point with $H(t)=H(-t)$, i.e., for $\varphi_0=0$ or $\pi$, there is no linear non-adiabatic correction to the observables [cf. Sec.~\ref{subsec:observables}] and the leading non-adiabatic contribution scales as $v^2$. However, if the measurement breaks time-reversal symmetry, then a linear correction appears. This situation is very reminiscent to that in non-driven systems, where time-reversal symmetry (specifically real Hamiltonians) leads to zero Berry curvature and hence vanishing linear non-adiabatic corrections to generalized forces~\cite{degrandi_13, kolodrubetz_16}.

	\subsection{\label{subsec:kapitza} The Quantum Kapitza Pendulum.}
	
	We now turn our attention to a more complicated single-particle model -- the quantum Kapitza pendulum. In the classical limit, this model is the prototype to study dynamical localisation~\cite{kapitza_51}, since strong fast shaking of the pivot point of a pendulum bob leads to stabilisation of the originally unstable inverted equilibrium ($\theta = \pi$). In the classical limit, this problem is also known to feature coexisting regions of regular and chaotic behavior suggesting that the Floquet Hamiltonian as a local operator is ill-defined. Nevertheless, we shall see that quantum Floquet theory can be defined practically by first introducing an ultraviolet (UV) cutoff, which makes the Hilbert space finite-dimensional, and then identifying the states which are insensitive to the cutoff. The Kapitza pendulum differs from the harmonic oscillator due to the non-linearity of the confining potential. We shall also shortly see that this non-linearity is, in fact, responsible for the existence of photon resonances, which result in new non-adiabatic effects absent in non-driven systems or integrable Floquet systems. 
	
	The Hamiltonian of the quantum Kapitza pendulum reads
	\begin{eqnarray}
	H(t) = \frac{p_\theta^2}{2m}  - m\omega_0^2\cos\theta -m A_f\lambda(t) \Omega\cos\Omega t\cos\theta,
	\label{eq:H_Kapitza}
	\end{eqnarray}
	where $p_\theta$ is the angular momentum operator and $m$ is the pendulum's momentum of inertia. As in the previous example we scale the driving amplitude to have non-trivial infinite-frequency limit.  For practical purposes, we work in the angular momentum basis\footnote{Since the Kapitza pendulum is a one-dimensional system, the angular momentum has only a $z$-component, and thus the quantum number $l$ is understood as the eigenvalue of the operator $p_\theta=L_z$.},  $p_\theta|l\rangle = l |l\rangle$, such that the operator $\mathrm{exp}(i\theta)|l\rangle  = |l+1\rangle$ shifts the angular momentum by one quantum. Consequently, the Hamiltonian in this basis assumes the form
	\begin{equation}
	H(t) = \frac{1}{2m}\sum_{l=-\infty}^{\infty} l^2|l\rangle\langle l| - \frac{m}{2}\left(\omega_0^2 + A_f\lambda(t)\Omega\cos\Omega t\right)\sum_{ l=-\infty}^{\infty}( | l+1\rangle\langle l|\;\; +\;\; | l\rangle\langle l+1|  ) ~,
	\label{eq:kapiza_mapping}
	\end{equation}
	which is equivalent to a free particle hopping on a lattice in the presence of a harmonic trap. The drive translates to periodically-modulated hopping. We note that this Hamiltonian could be readily realised, for instance, with non-interacting ultracold atoms in a harmonic trap.
	
	The angular momentum basis is particularly convenient to simulate the time-dependence of the system because it discretizes the Hilbert space. To deal with the unbounded spectrum, we impose a high-frequency cut-off by keeping only a finite number of angular momentum states: $l = -M,-M+1,\dots,0,\dots,M-1,M$. We made sure that the results presented here do not change with $M$. Further, we use parity symmetry ($l \to -l$) to divide the total Hilbert space into an even-parity subspace, containing $M+1$ states (including the $l=0$ state), and an odd subspace containing the remaining $M$ states.

	We are now interested in slowly ramping up the driving amplitude according to a smooth protocol, which we choose to be slightly different than the quadratic protocol used in the harmonic oscillator example:
	\be
	\lambda(t)=2\sin^2\left( \frac{\pi}{2}\frac{t-t_i}{2|t_i|}\right)
	\label{eq:kapitz a_ramp}
	\ee
	from time $t_i<0$ to the final time at $t=0$. This protocol slowly ramps the driving amplitude from zero to a finite value $\lambda(0)=1$ with the final velocity $v\equiv \dot\lambda(0)=\pi/2|t_i|$. We start in the ground state of the non-driven Hamiltonian $H(t_i)$. Due to the unbounded character of the Kapitza spectrum, the numerical simulations necessarily produce a folded quasienergy spectrum at any fixed driving frequency. This poses the fundamental problem of identifying the adiabatic state in the first place. As we shall see shortly, this is not a mere mathematical difficulty but rather a genuine physical problem. Obviously, if we cannot identify the proper adiabatic state the whole concept of FAPT is meaningless.

\begin{figure*}[ht]
		\begin{minipage}{\textwidth}
			\includegraphics[width=\columnwidth]{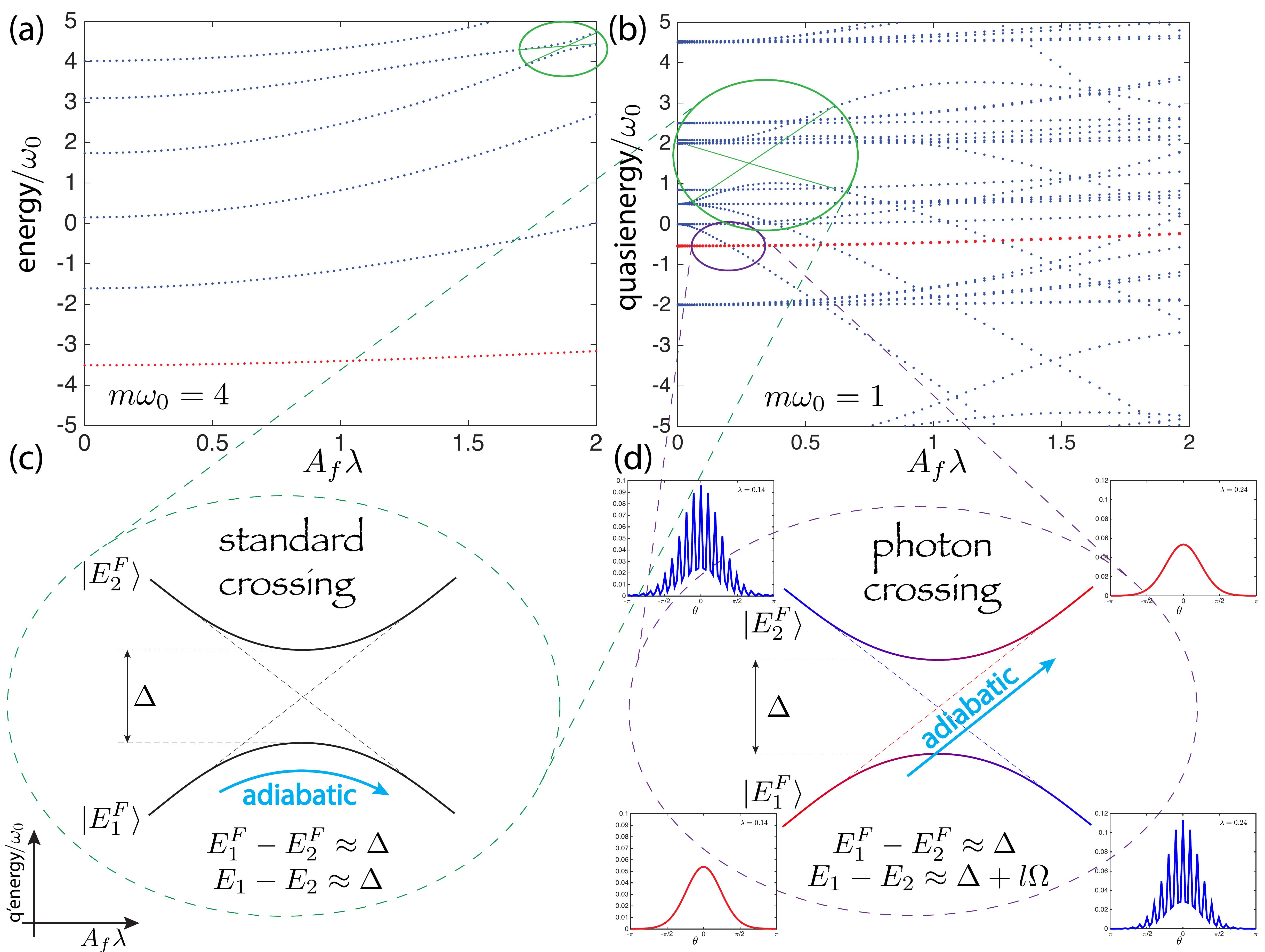}
			\caption{\label{fig:avoided_crossings}(Color online)[Kapitza pendulum]. (a) The low-energy spectrum of the Hamiltonian~\eqref{eq:H_Kapitza_nodrive} for $m\omega_0=4$ showing the standard avoided crossings (within green ellipse), which impose the limits for the validity of conventional adiabatic perturbation theory. (b) The Floquet spectrum of the Kapitza Hamiltonian~\eqref{eq:H_Kapitza} for $m\omega_0=1$. Apart from the standard avoided crossings (green ellipse), new avoided crossings appear due to photon absorption processes (purple ellipse). The Floquet ground state is shown in red dots. The insets show the wave function right before and after such a photon-absorption avoided crossing. We see that standard gaps must be crossed adiabatically (c), while photon absorption gaps need to be crossed \emph{diabatically} for FAPT to hold (d). The parameters are $\Omega/\omega_0 = 10$, and the cut-off parameter is $M=20$.}
		\end{minipage}
	\end{figure*}

	However, the situation is not as bad as it seems. It is intuitively clear that at high driving frequencies the Kapitza pendulum should remain stable against small perturbations, at least near the equilibrium positions. One can readily observe this stability numerically as well. To find the adiabatically-connected state, we start from the non-driven Hamiltonian at $\lambda=0$ and continuously follow it as we gradually increase the drive amplitude (see red dots in Fig.~\ref{fig:avoided_crossings}). We refer to this state of smallest angular momentum spread as the Floquet ground state, and this is the state we choose to target. We checked numerically that this procedure is reliable for frequencies $\Omega/\omega_0\gtrsim 7$, but it eventually fails once the Floquet operator becomes nonlocal and then there is no natural state to call the Floquet ground state. We note that, while this procedure seems somewhat ad-hoc, it may be systematically extended to arbitrary systems using the inverse-frequency expansion. In particular, one can identify the Floquet ground state as the eigenstate that has the largest overlap with the effective static ground state obtained via the first few orders of the inverse-frequency expansion. We will explore this connection to the high-frequency expansion in more detail in Sec.~\ref{sec:HFE}. Instead of the high-frequency expansion one can use some other expansion producing a local Floquet Hamiltonian or even use some variational approach giving a local Hamiltonian having the highest overlap  of eigenstates with the eigenstates of the Floquet Hamiltonian. Alternatively, in Ref.~\cite{russomanno_15} the Floquet ground state for an extended spin system was defined as the lowest entanglement state. Generically this identification is also only applicable to high driving frequencies. At low frequencies all Floquet eigenstates are expected to become maximally entangled infinite-temperature states~\cite{dalessio_14,bukov_15_erg}, and therefore the very notion of adiabaticity becomes ill defined. 
	
\begin{figure}
		\includegraphics[width=\columnwidth]{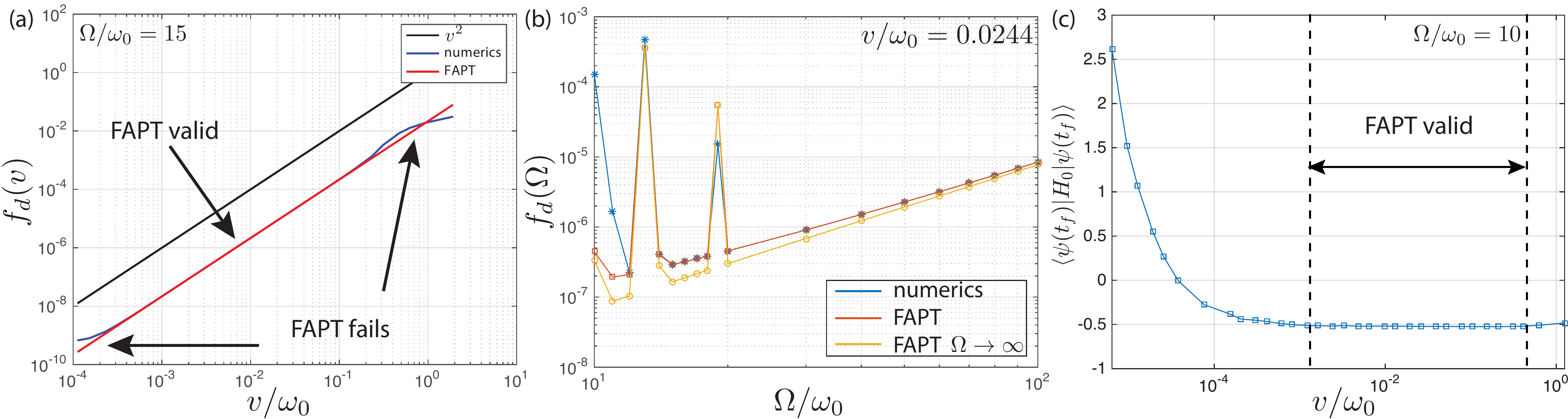}
		\caption{\label{fig:Kapitza_Pexc}(Color online)[Kapitza pendulum]. (a) Comparison between FAPT and exact numerics shows the existence of an adiabatic regime in the excitation probability (similar to log-fidelity) in for $\Omega/\omega_0 = 15$. (b) Frequency-dependence of the excitation probability (log fidelity) displaying the exact numerical data (blue stars), FAPT (red squares), and FAPT in the limit $\Omega\rightarrow\infty$ (yellow circles), for $v = 0.0244$. (c) Velocity dependence of the energy absorbed from the drive at the measurement time for $\Omega/\omega_0=10$. The model parameters are $m\omega_0 = 1$, $A_f = 2$. The infinite-frequency FAPT is obtained by keeping the $\ell=0=\ell'$ term from the sum in Eq.~\eqref{eq:FAPT_pF}. }
	\end{figure}

	Figure~\ref{fig:Kapitza_Pexc} gives a confirmation that there exists an adiabatic regime at large frequencies in which the Floquet ground state can be prepared with high fidelity. Not only does the excitation probability scale quadratically with the ramp speed, but there also exists a large interval of velocities for which the FAPT formula in Eq.~\eqref{eq:FAPT_entropy} quantitatively reproduces the correct results for the leading non-adiabatic correction\footnote{We note that in the figures comparing the FAPT formula with the numerical solution, we evaluated \eqref{eq:c_n_FAPT} numerically, and did not use any high frequency approximations discussed in later sections.}. Interestingly, lowering the ramp speed too far leads to an increase of the excitations in the system. This increase in excitations and Floquet diagonal entropy (Fig.~\ref{fig:Kapitza_entropy}) with decreasing velocity is also associated with a stronger heating of the system at slower ramp rates, as seen in Fig.~\ref{fig:Kapitza_Pexc}c. This is clearly inconsistent with the expectations from equilibrium thermodynamics, representing an important fundamental consideration for Floquet thermodynamics\footnote{Similar non-adiabatic effects can be anticipated in disordered systems with localized excitations, both single-particle and MBL, see Ref.~\cite{khemani_15} for discussion.}. Moreover such a non-monotonic increase in entropy and energy implies that there is no local differentiable Floquet Hamiltonian even in the high frequency regime.

	In Fig.~\ref{fig:Kapitza_Pexc}(b) we show the frequency dependence at fixed ramp rate of the log-fidelity $f_d$, see Eq.~\eqref{eq:FAPT_entropy}. In the infinite-frequency limit, we find an agreement between the exact numerical curve (blue stars) with both the finite-frequency (red squares) and infinite-frequency (yellow circles) stroboscopic FAPT predictions.  However, at finite frequencies the two deviate, with the difference reaching up to a factor of $2$ at low frequencies. Notice that finite-frequency FAPT is significantly more accurate than its infinite-frequency counterpart, as it includes the crucial contributions due to $\mathcal A_\lambda^P$, i.e., due to the kick operator. The exact numerical curve features isolated peaks at specific values of $\Omega$, which we will see correspond to strong resonances encountered during the ramp. Figure~\ref{fig:Kapitza_Pexc}(c), shows the energy at the measurement time as a function of the ramp velocity. There exists a large plateau at intermediate velocities which is described by FAPT. However, at smaller ramp speeds the excitations appear in the energy as well. We thus see that the failure of FAPT for small frequencies and velocities is related to physically-observable heating.  

	\subsubsection{The Role of the Level Crossings.}
	
	We will now show that these excitations at low ramp rate are due to the existence of photon absorption avoided crossings in the Floquet spectrum\cite{hone_97,eckardt_08}. The basic idea is that energy in Floquet systems is only defined modulo $\Omega$. Then as the UV cutoff $M$ is taken to infinity, the quasienergy spectrum becomes increasingly dense. As the density goes to infinity, one will find many accidental crossings between quasienergies (cf. Fig.~\ref{fig:avoided_crossings}b), which in turn have very small gaps opened up at high order in perturbation theory by multi-photon processes. So as the UV and/or thermodynamic limit is taken, the Floquet spectrum approaches an infinitely-dense set of infinitely-weak avoided crossings. We refer to these as photon-absorption avoided crossings or resonances.
	It is precisely these resonances that led Hone et al.~in Ref.~\cite{hone_97} to conclude that no adiabatic limit exists for Floquet systems, but as our numerics have demonstrated, there is still a wide range of ramp velocities for which these resonances are unimportant and FAPT provides a good description of the excitations in the system. This is especially relevant for experiments which need to target the correct parameter regime. We will now examine the effect of these resonances on adiabaticity, using the Kapitza pendulum as an example.

	To better elucidate the role of these photon absorption avoided crossings, consider first ramping the non-driven Hamiltonian
	\begin{eqnarray}
	H_\mathrm{ave}(\lambda(t)) = \frac{p_\theta^2}{2m}  - m\omega_0^2\cos\theta + \frac{m\lambda^2(t)A_f^2}{4}\sin^2\theta,
	\label{eq:H_Kapitza_nodrive}
	\end{eqnarray}
	which is nothing but the Floquet Hamiltonian for the Kapitza pendulum in the infinite frequency limit. Since this system is not periodically-driven, the conventional quantum adiabatic theorem applies. As the spectrum exhibits avoided level crossings upon tuning $\lambda$ the adiabatic theorem requires that the velocity be small enough so that one remains in the same \emph{energy} manifold while passing through the avoided crossing. An example of such crossing, which should be avoided in adiabatic limit is shown in Fig.~\ref{fig:avoided_crossings}(a). These crossings also occur in the finite frequency Floquet Hamiltonian identified by a green circle in Fig.~\ref{fig:avoided_crossings}(b). One can numerically identify these crossings by comparing the spectra of the Floquet Hamiltonian and of the approximate unfolded Floquet Hamiltonian obtained e.g.~within the high frequency expansion. Physically these crossings occur between Floquet eigenstates with small difference in both Floquet energies (Fig.~\ref{fig:avoided_crossings}(b)) and the energies defined as expectation values of the infinite frequency Hamiltonian (Fig.~\ref{fig:avoided_crossings}(a)). In the following, we refer to this type of avoided crossings as `standard'. 

	At any fixed frequency one can also identify additional avoided crossings, which do not show up in the infinite frequency Hamiltonian and in fact in any order in high frequency expansion. These ``photon-absorption crossings'' or ``photon resonances'' only show up in the \emph{quasienergy} spectrum (Fig.~\ref{fig:avoided_crossings}(b)) and appear as a result of strong hybridization between energy levels and the photon field. Such crossings correspond to a small difference between the Floquet quasienergies but large, of the order of $n\Omega$ with $n=1,2,\dots$, difference between the energies of the infinite frequency Floquet Hamiltonian. Adiabatic transition through such photon resonances should be understood as in fact a diabatic crossing of these levels as shown in  Fig.~\ref{fig:avoided_crossings}(d).  Indeed these crossings arise due to a finite matrix element with a highly excited folded state such that the wave functions of the two participating states exhibit a very different behaviour. For instance, in the case of the Kapitza pendulum, the GS is a smooth non-oscillatory function resembling a Gaussian, while a highly excited scattering state typically has many oscillations corresponding to its large kinetic energy, cf.~Fig.~\ref{fig:avoided_crossings}(d). When passing a photon-absorption avoided crossing, the two states hybridize strongly and amplitude may be transferred to the high-energy state, depending on the crossing speed. If one goes too slowly the wave function changes \emph{drastically} after the crossing and we find the system in the highly excited state instead. Hence, we are lead to the conclusion that photon-absorption avoided crossings should be passed \emph{diabatically}, so that the system will remain in the appropriately connected \emph{energy} manifold. Therefore, when we speak of ``adiabaticity'' in the context of FAPT, we keep in mind that this truly involves adiabatic ramping across standard avoided crossings (Fig. \ref{fig:avoided_crossings}c) and diabatic ramping across photon absorption crossings (Fig. \ref{fig:avoided_crossings}d), meaning that adiabaticity in the conventional sense is not adiabaticity in the Floquet sense. 
	
\begin{figure}[ht]
		\includegraphics[width=\columnwidth]{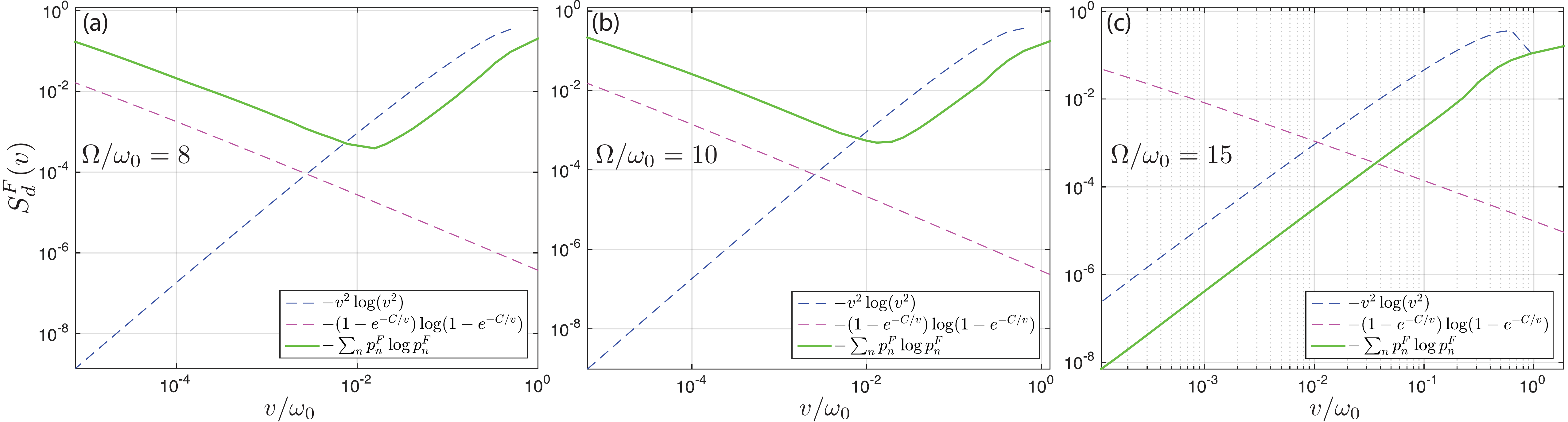}
		\caption{\label{fig:Kapitza_entropy}(Color online)[Kapitza pendulum]. Numerical results for the Floquet diagonal entropy $S^F_d=-\sum_n p^F_n\log p^F_n$, $p^F_n = |\langle n_F|\psi(t_f)\rangle|^2$ in the basis of the final Floquet Hamiltonian as a function of the ramp speed $v$. The three panels correspond to $\Omega/\omega_0=8$ (a), $\Omega/\omega_0=10$ (b), and $\Omega/\omega_0=15$ (c). The regime of validity of FAPT is determined by the velocity range for which the entropy approximately follows the $v^2$--law. The breakdown of FAPT is well-captured by Landau-Zener transitions between resonantly coupled states (cf. Fig.~\ref{fig:avoided_crossings}). The model parameters are $m\omega_0 = 1$, $A_f = 2$. The constant $C$ in low velocity fitting was chosen arbitrary.}
	\end{figure}

	The physics at ultra small velocities beyond FAPT can be understood as a cascade of Landau-Zener (LZ) transitions due to resonances with higher-energy states induced
	by the drive~\cite{eckardt_08}. To test this idea heuristically, we compare the prediction of LZ theory and find a reasonable agreement,
	cf.~Fig.~\ref{fig:Kapitza_entropy}(a),(b). While we fit to a single LZ avoided crossing, note that in general as one ramps slower one expects a cascade of such avoided crossings on different energy scales \cite{hone_97} and the simple LZ formula is expected to break down. Finding the sweet spot in velocity between standard and photon-absorption avoided crossings becomes increasingly difficult to achieve as the driving frequency decreases (or the driving amplitude increases) and eventually at low frequencies this window disappears and adiabaticity is completely lost, as seen in Fig.~\ref{fig:Kapitza_entropy}. Indeed, while this distinction between the two types of avoided crossings is quite sharp for the data shown, at lower driving frequencies it will be lost, and the choice of targeted state will require more care. We also note that the scenario described above generalizes to other nonintegrable models. In the generic situation one cannot always classify all crossings as being 'standard' or originating from photon-absorption resonances.  In Sec.~\ref{sec:HFE}, we shall argue that these photon absorption level crossings are absent not only in the spectrum of the infinite frequency Floquet Hamiltonian, but are in fact beyond any order of the inverse-frequency expansion.

	\subsection{\label{subsec:qubit_topology} A Nonequilibrium Topological Transition and the Thouless Energy Pump in the Linearly Driven Qubit.}
	
	Having demonstrated that the adiabatic theorem for periodically driven systems agrees well with numerical simulations, we now elaborate on the previously-stated connection between non-adiabatic corrections and Berry curvature, cf.~Sec.~\ref{subsec:Ftopology}. Previous work generalised and studied the Kubo response of noninteracting systems with electron conduction using the example of driven graphene~\cite{oka_09,dehghani_15,dehghani_15_2}, and derived expressions for the Floquet Berry curvature and the Chern number (a.k.a.~the quantised conductivity) in the limits of weak probe coupling. These papers tacitly assumed the presence of an adiabatic limit, which we have seen does not always exist for Floquet systems. Furthermore, a number of cold-atoms experiments reported successful measurements of the Berry curvature and the associated Chern number of a topological Floquet band~\cite{aidelsburger_14,jotzu_14,flaeschner_15} using linear response techniques. The experiments involved high-frequency driving, relative to the bare energy scales. Hence, it is natural to expect that for $\Omega\to\infty$, this procedure allows one to measure the Chern number of the bands associated with the Floquet Hamiltonian $H_F$. However, as experiments are performed at finite frequencies, where non-equilibrium effects become important, one might wonder how this simple picture acquires modification.
	
	In Sec.~\ref{subsec:Ftopology}, we generalised these results to arbitrary interacting systems and drive strengths, and demonstrated that they hold true only as long as FAPT holds. One important point that we have seen in the Kapitza pendulum is that FAPT is not generally valid for all ramp velocities, which is intricately related to the absence of a generic Floquet adiabatic limit~\cite{hone_97}. Therefore, the discussion of Floquet geometry and topology holds exclusively in the regime of validity of FAPT, and is expected to fail when the effect of photon absorption resonances becomes sizable and FAPT fails. This is an important result of our theory, suggesting that care must be taken in measuring Floquet geometry and topology using these linear response techniques.
	
\begin{figure}
		\includegraphics[width=\columnwidth]{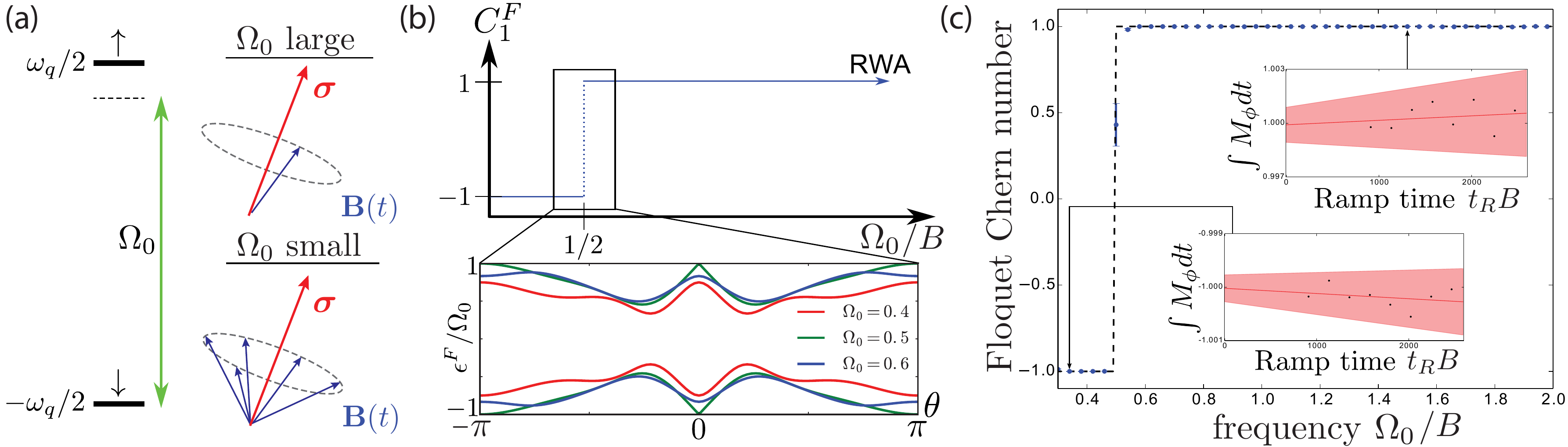}
		\caption{
		(Color online)[Driven qubit]. Floquet topology of a strongly driven two-level system. (a) Illustration of the setup. Two levels are split by a bare frequency $\omega_q$ and driven near resonance at frequency $\Omega_0$. In the rotating frame the system looks like a spin-1/2 in a rotating magnetic field ${\bf B}(t)$ with frequency $\Omega=2\Omega_0$. At large $\Omega_0$ compared to the Rabi frequency set by the driving strength, the rotating wave approximation (RWA) applies and ${\bf B}$ can be averaged. At low $\Omega_0$ it cannot. (b) By sweeping $B(t)$ over a sphere in parameter space, one may define the Floquet Chern number $C_1^F$. At high frequency the Chern number is $1$, like that of a spin-1/2 in a magnetic field. However, at $\Omega_0=1/2$ the system undergoes a topological transition where a gap closes at the edge of the Floquet zone, as shown in the bottom panel. (c) The topological transition can be measured by linear response as described in the text. Using the protocol $\theta(t)=\pi \sin^2\left( \frac{\pi(t+t_R)}{2 t_R} \right)$, a linear fit to the time-integrated generalized force $M_\phi$ gives the Floquet Chern number as the y-intercept (insets). The main figure shows the frequency-driven topological Chern transition from these fits. All energies are in units of $B=1$, the magnitude of the effective magnetic field. \label{fig:qubit_chern}
	}
	\end{figure}

	We now illustrate these ideas using an example of a driven two-level system, a.k.a.~a qubit. In Sec.~\ref{subsec:Ftopology} we showed that the leading correction to the phase average of the expectation value of the generalized force $M_\mu(t) =-\langle \psi(t)| \partial_\mu H(t)|\psi(t) \rangle$ is proportional to the phase-averaged Berry curvature. This is very similar to the conventional APT case, where leading corrections to adiabaticity have been used to measure the Berry curvature of one and two-qubit systems and subsequently integrated over a closed manifold to give their topologically-invariant Chern number~\cite{schroer_14,roushan_14}. However, a detailed look at these particular superconducting qubits shows that they are actually Floquet systems. In particular, at first approximation, they consist of two far-detuned bare levels $|\uparrow \rangle$ and $|\downarrow \rangle$ whose splitting $\omega_q$ is much larger than the desired qubit operation frequency. Then microwave fields are shone on this system at frequency $\Omega_0$ that is nearly resonant with the qubit transition, which is able to couple these levels (see Fig.~\ref{fig:qubit_chern}a). 
	
	In the lab frame, this system is described by the Hamiltonian
	\be
	H_\mathrm{lab}=\frac{\omega_q}{2} \sigma^z + g \cos(\Omega_0 t + \phi) \sigma^x~,
	\ee
	where $g$ is proportional to the strength of the driving field. In general, the drive is controllable such that $\Omega_0$, $g$, and $\phi$ are arbitrary functions of time. Going to the rotating frame via the unitary $V(t) = \mathrm e^{i \sigma^z \Omega_0 t / 2}$, $|\psi^\mathrm{rot}(t)\rangle = V(t)|\psi_\mathrm{lab}(t)\rangle$, we find the effective Hamiltonian
	\begin{eqnarray}
	H^\mathrm{rot}(t) &=& \frac{\omega_q - \Omega_0}{2} \sigma^z + g \cos(\Omega_0 t+\phi) 
	\left[ \sigma^x \cos(\Omega_0 t) - \sigma^y \sin(\Omega_0 t) \right] 
	\nonumber\\
	&=& \frac{\omega_q - \Omega_0}{2} \sigma^z + \frac{g}{2} (\sigma^x \cos \phi + \sigma^y \sin \phi)
	+ \frac{g}{2} \left(\sigma^x \cos(2 \Omega_0 t+\phi)- \sigma^y \sin(2 \Omega_0 t+\phi) \right) ~.
	\label{eq:H_rot_qubit_realistic}
	\end{eqnarray}
	To more clearly demonstrate the Hamiltonians that are generally simulated in these systems, we parameterize the detuning $\delta \equiv \omega_q - \Omega_0$ and the drive strength $g$ as $g=-B \sin \theta$ and $\delta=-B \cos \theta$. Then keeping these values constant while taking the high frequency limit, $\Omega_0 \to \infty$, this model allows one to simulate arbitrary single-qubit Hamiltonians of the form $H = -{\bf B} \cdot {\boldsymbol \sigma} / 2$. It is precisely in this limit that Schroer et al.~\cite{schroer_14} measured topological transitions in a superconducting qubit using leading non-adiabatic corrections akin to Eq.~\eqref{eq:linear_respose_FNS}.
	
	However, at lower frequencies, the strong micromotion induced by the time-dependent (counter-rotating) third term in Eq.~\eqref{eq:H_rot_qubit_realistic}, will have a strong effect on the non-adiabatic corrections to the dynamics. Note that the rotating frame Hamiltonian is actually periodic with frequency $\Omega=2 \Omega_0$; therefore we rewrite the Hamiltonian as 
	\begin{equation}
	H^\mathrm{rot}(t) = -\frac{B}{2} \Big[  \cos \theta \sigma^z +  \sin\theta (\sigma^x \cos \phi + \sigma^y \sin \phi) + \sin \theta (\sigma^x \cos(\Omega t+\phi) - \sigma^y \sin(\Omega t+\phi) \Big]~.
	\label{eq:H_rewritten_qubit_realistic}
	\end{equation}
	Unlike the related model of a qubit in a circularly-polarised drive, which can be solved exactly by mapping it to a time-independent Hamiltonian, there exists no simple closed-form solution for the present model.
	
	As discussed earlier, the phase-averaged Berry curvature $\overline{F}^F_{\theta \phi}$ is measurable from Floquet linear response via
	\be
	\overline{M}_\phi = \overline{M}_\phi^{(0)} + \dot \theta \overline{F}^F_{\phi \theta} + \mathcal{O}(\dot \theta^2)~,
	\ee
	where $M_\phi=\langle -\partial_\phi H_{\rm lab} \rangle$ is the generalized force in the lab frame, related to the work done on the system by the periodic drive (see Eq.~\eqref{eq:partial_phi_H_lab} below), and $\overline{M}_\phi^{(0)}$ indicates its phase average in the Floquet ground state. This average is equal to zero by gauge invariance of the Floquet spectrum; according to Eq.~\eqref{eq:Floquet_gen_force}
	\[
	\overline{M}_\phi^{(0)}=-\partial_\phi \epsilon_0^F=0.
	\]
	We also note that $-\partial_\phi H_{\rm lab}$, which for static problems can be interpreted as a simple magnetization, is now a more complicated time-dependent observable.
	
	Consider now a ramp of $\theta$ in the time interval $[-t_R, t_R]$ such that $\theta(t=-t_R)=0$ and $\theta(t=t_R)=\pi$. For larger $t_R$, this ramp more adiabatically interpolates between $\theta=0$ and $\theta=\pi$. Then for fixed $\phi$, integrating the expectation value $\overline{M}_\phi$ along the ramp gives
	\be
	\int \overline{M}_\phi (t,\theta) \mathrm{d}t  \approx  \int \dot \theta \overline{F}^F_{\phi \theta}(\theta) \mathrm{d}t=  \int \overline{F}^F_{\phi \theta}(\theta) \mathrm{d} \theta= \frac{1}{2\pi}\int \overline{F}^F_{\phi \theta}(\theta) \mathrm{d} \theta \mathrm{d}\phi=C_1^F.
	\label{eq:M_phi_rot_frame}
	\ee
	Here we use the fact that since $\phi$ is just the driving phase in the lab frame, phase-averaged values of the Floquet Berry curvature are $\phi$-independent. Therefore one can extract the Floquet Chern number by simply averaging the experimentally-measurable generalized force over the angle $\theta$. Note that Eq.~\eqref{eq:M_phi_rot_frame} is completely general, relying solely on the validity of FAPT. Below we will show that this generalized force is also related to the work done on the system.  
	
	This procedure is carried out for the qubit model in Fig.~\ref{fig:qubit_chern}. Note that the long-time integration automatically averages over many cycles, so the phase averaging is done automatically by the slow ramp. In the high-frequency limit, as expected, the Chern number is found to be $C_1^F=1$ as in the $\Omega \to \infty$ case. However, this behavior continues down to much lower frequencies where the high-frequency limit is no longer valid. Even more interesting is the fact that, as the frequency is further lowered, the Floquet ground state ``inverts,'' as seen in Fig.~\ref{fig:qubit_chern}(b). This causes the Chern number to jump discontinuously to $C_1^F=-1$, i.e., the system undergoes a topological transition similar to those found in non-interacting Floquet topological insulators~\cite{mikami_15}. This is confirmed by numerical simulation in Fig.~\ref{fig:qubit_chern}(c). Thus we see that not only is the Floquet Chern number measurable, but we can actually get novel topological transitions in the low-frequency regime.
	
	While these ideas have been illustrated for the case of a qubit model, they are completely general. Thus situations such as cold atoms in flux lattices or Floquet topological insulators that have quantized Floquet invariants should in principle be susceptible to having these invariants measured by procedures analogous to that above. It bears mention that these techniques require one to measure $\partial_\lambda H(t)$, which in can be a highly oscillatory observable especially if the driven part of the Hamiltonian is directly coupled to $\lambda$. However, this is not an issue in the experimentally-relevant case of measuring the transverse response of a Floquet Chern insulator in cold atoms, where an effective electric field is created by introducing a static magnetic field gradient which effectively tilts the lattice~\cite{jotzu_14,aidelsburger_14,flaeschner_15}. The transverse generalized force is then the current along the direction perpendicular to the field gradient, a static observable. Note that while this transverse current operator is static, its expectation value is still oscillating in time~\cite{bukov_14_pra}, and thus appropriate averaging over the phase of the measurement must be done to obtain the topological response.

	\subsubsection{Floquet System as a Topological (Discrete) Energy Pump.}
	
	An interesting consequence of the fact that one of our parameters, $\phi$, was simply the phase of the drive in the lab frame, is that there is a deep connection between the topology measured above and energy absorption. Consider a generic Floquet Hamiltonian for which a closed manifold is defined by some parameter $\theta \in [0,\pi]$ and the driving phase $\phi$, such that $H_\mathrm{lab} = H_\mathrm{lab}(\theta,\Omega_0 t + \phi)$. By construction the Hamiltonian is a periodic function of $\phi$. Let us observe that the generalized force with respect to $\phi$ is
	\begin{equation}
	M_\phi = -\langle \partial_\phi H_\mathrm{lab} \rangle = -\frac{1}{\Omega_{0}}\langle\partial_{t}H_{\mathrm{lab}}\rangle = -\frac{1}{\Omega_{0}}\left\langle\frac{\mathrm{d}H_{\mathrm{lab}}}{\mathrm{d}t}\right\rangle+\frac{\dot{\theta}}{\Omega_{0}}\langle\partial_{\theta}H_{\mathrm{lab}}\rangle=-\frac{1}{\Omega_{0}}\frac{\mathrm{d}}{\mathrm{d}t}\langle H_{\mathrm{lab}}\rangle+\frac{\dot{\theta}}{\Omega_{0}}\langle\partial_{\theta}H_{\mathrm{lab}}\rangle.
	\label{eq:partial_phi_H_lab}
	\end{equation}
	At order $\dot \theta$, we can replace the last expression by its value in the Floquet ground state. Then performing the phase average and integrating over time from $-t_R$ to $t_R$ (see the protocol in the caption of Fig.~\ref{fig:qubit_chern}), we see that
	\begin{multline}
	C_1^F=\int \overline{M}_\phi \mathrm{d}t \approx -\frac{1}{\Omega_{0}}\int_{-t_R}^{t_R}  \mathrm{d}t \frac{\mathrm{d}}{\mathrm{d}t}\langle H_{\mathrm{lab}}\rangle + \frac{1}{\Omega_{0}}\int_0^\pi \mathrm{d}\theta \overline{\langle\partial_{\theta}H_{\mathrm{lab}}\rangle_0}
	\stackrel{(\ref{eq:Floquet_gen_force})}{=}
	-{W\over \Omega_0} +{1\over \Omega_0}\int_0^{\pi} \mathrm{d}\theta \partial_\theta \epsilon^F_0=\frac{W^F_{\rm ad} - W}{\Omega_0}~,
	\label{eq:chern_vs_energy}
	\end{multline}
	where $W=\Delta E$ is the phase-averaged work done {\em on} the system, or equivalently the energy  pumped into the system, and $W^F_\mathrm{ad}=\epsilon^F_0(\pi) - \epsilon^F_0(0)$ is the adiabatic Floquet work done on the system. Note that $W^F_\mathrm{ad}$ vanishes identically for any cyclic process, and in particular vanishes for the qubit system discussed above. Thus the Chern number is simply related to work done on the system during the adiabatic cycle:
	\[
	W=-C_1^F\Omega_0
	\]
	
	This result indicates that the work done on the system  during one adiabatic cycle is quantized in units of the driving frequency, opening the possibility of realizing a Floquet energy pump similar to the Thouless pump in equilibrium systems~\cite{thouless_83}. Physically this energy change amounts to generating (or removing) an integer number of photons from the driving field. For the particular example of the qubit one can check that if the angle $\theta$ keeps increasing from $\pi$ to $2\pi$ the total Chern number will be zero, and thus the system will not continuously absorb the energy. In order to realize the continuous energy pump in this system, during the second half of the cycle one can uncouple the qubit from the drive and reinitialize it in the ground state corresponding to $\theta=0$. Alternatively, one can apply the process to a sequence of qubits and do the $\pi$-rotation to each of them. We leave the detailed analysis of this interesting possibility, including the particularly intriguing situation where the system couples coherently to a photonic reservoir such that they cannot be treated as an external periodic drive, to future work.

	\section{\label{sec:examples_MB} Many-Body Examples.}
	
	We now analyse the applicability of FAPT by applying it to more complex, many-body systems. We first study the transverse-field Ising chain, a quintessential integrable many-body system. We show that an integrability-preserving drive of the transverse field results in obeying FAPT for driving frequencies above the single-particle bandwidth where photon absorption is only virtual (off-shell). Below the single-particle bandwidth, real (on-shell) photon absorption processes become important and the adiabaticity becomes only limited with a non-analytic power-law dependence of observables on the ramping rate. This comes from the fact that for such low driving frequencies even infinitesimal driving amplitude opens a gap in the Floquet spectrum. Therefore the whole setup becomes very similar to the Kibble-Zurek type scenario, where one starts the ramping protocol at a critical point (see e.g. Ref.~\cite{degrandi_13} for details). 
	
	We then introduce a longitudinal field which breaks integrability and makes the model generic. By measuring the  Floquet diagonal entropy $S^F_d$ (cf. Eq.~\eqref{eq:FAPT_entropy}), we show that even for this complicated model, a regime of validity exists for FAPT, at least for finite-size systems. At the same time, similarly to the Kapitza pendulum example, very slow ramps result in strong heating due to crossing many-body photon resonances.
	
	\subsection{\label{subsec:TFI} The Driven Transverse-Field Ising Model.}
	
	The transverse-field Ising model (TFIM) is the prototypical example to study quantum phase transitions~\cite{sachdev_book}. The Hamiltonian is given by
	\begin{eqnarray}
	\label{eq:static_TFIM}
	H = -J_0\sum_j \sigma^z_{j+1}\sigma^z_j - h^x_0\sum_j \sigma^x_j,
	\end{eqnarray} 
	with the nearest-neighbour Ising interaction $J_0$ and transverse magnetic field $h^x_0$. We consider periodic boundary conditions and restrict the discussion to chains with even total number of sites. It is well-known that this model exhibits a quantum phase transition at $J_0/h^x_0=1$ from an $x$-paramagnet to a $z$-ferromagnet~\cite{sachdev_book}. More importantly for our purposes, it is an exactly solvable many-body model that serves as a jumping off point to even more complicated cases.
	
	We now add a periodic modulation of the transverse field $h_1(t) = A_f\lambda\Omega\cos\Omega t$, so the total Hamiltonian of the system reads
	\begin{eqnarray}
	\label{eq:H_TFIM}
	H(t) = -J_0\sum_j \sigma^z_{j+1}\sigma^z_j - (h^x_0 + A_f\lambda\Omega\cos\Omega t)\sum_j \sigma^x_j.
	\end{eqnarray}  
	At fixed $\lambda$, this model was studied in Ref.~\cite{bastidas_12}, where it was shown that the ground state of the Floquet Hamiltonian still defines two different phases separated by a quantum critical point as in equilibrium. As we shall explain in the next few paragraphs, the critical magnetic field is controlled by $\lambda$ and can be made arbitrarily small if we tune the system to the dynamical localization regime where the effective spin-spin interaction becomes very small. The nonequilibrium physics in the presence of the drive near this critical point was extensively studied by Russomanno et al.~\cite{russomanno_15,russomanno_12,russomanno_15_epl,russomanno_16_epl,russomanno_15_jstat,russomanno_13_jstat}. In the discussion below, we carefully avoid crossing any critical points, as we want to focus on the perturbative non-adiabatic effects in Floquet systems which requires avoiding any Kibble-Zurek-type physics that would unnecessarily complicate the analysis.

	The TFIM can be solved exactly using the Jordan-Wigner transformation~\cite{sachdev_book}, which maps the Hamiltonian of spins to a quadratic Hamiltonian of spinless fermions $\{c_i,c^\dagger_j\}=\delta_{ij}$ that conserves particle number modulo two. The ground state is in the sector with an even number of particles, and thus we restrict ourselves to that sector. Because our Hamiltonian is translationally invariant it is also advantageous to Fourier transform to momentum space leading to:
	\begin{eqnarray}
	\label{eq:TFIM_driven}
	H(t) = \sum_{k\in\mathrm{BZ}} \Big[ 2(h^x_0+A_f\lambda\Omega\cos\Omega t -J_0\cos k )c^\dagger_kc_k+J_0\sin k(c_{-k}^\dagger c_k^\dagger + c_k c_{-k})\Big]~ ,
	\end{eqnarray}
	where $\mathrm{BZ} = [-\pi,\pi]$ is the first Brillouin zone. Since the driving amplitude scales with the driving frequency $\Omega$, we go to the rotating frame using the following transformation~\cite{bastidas_12}:
	\begin{equation}
	V(t)=\exp\left(-2iA_f\lambda\sin(\Omega t)\sum_{k\in\mathrm{BZ}}c_k^\dagger c_k\right)
	\end{equation}
	which leads to the following rotating-frame Hamiltonian:
	\begin{eqnarray}
	\label{eq:TFIM_driven_rot_frame}
	H^\mathrm{rot}(t) &=& H_\mathrm{ave} + \sum_{k\in\mathrm{BZ}} J_0\sin k\left[\sum_{\substack{\ell=-\infty \\ \ell\neq 0}}^\infty \mathcal{J}_\ell(4A_f\lambda)(\mathrm e^{i\ell\Omega t}c_{-k}^\dagger c_k^\dagger + \mathrm e^{-i\ell\Omega t}c_k c_{-k})\right],\nonumber\\
	H_\mathrm{ave} &=& \sum_{k\in\mathrm{BZ}} \Big[ 2(h^x_0-J_0\cos k )c^\dagger_kc_k+J_\mathrm{ave}(\lambda)\sin k (c_{-k}^\dagger c_k^\dagger + c_k c_{-k}) \Big] ~.
	\end{eqnarray}
	We separated the time average explicitly, taking into account the renormalisation of the model parameters by the drive: $J_\mathrm{ave}(\lambda) = J_0\mathcal{J}_0(4A_f\lambda)$, where $\mathcal{J}_\ell$ is the $\ell^{\text{th}}$ Bessel function of the first kind. Interestingly, by tuning the combination $4 A_f\lambda$ to the zero of the Bessel function $\mathcal J_0$ it is possible to completely suppress superconducting term in $H_\mathrm{ave}$ and effectively map the model to the XY chain~\cite{sachdev_book}. The single-particle dispersion relation of the time-averaged Hamiltonian has the two bands
	\begin{eqnarray}
	\label{eq:TFIM_disp_ave}
	E_k = \pm\sqrt{(h^x_0 - J_0\cos k)^2 + J_\mathrm{ave}^2(\lambda)\sin^2 k}.
	\end{eqnarray}
	
	The drive in the rotating frame couples to the two-particle excitation operator, which suggests that one can excite two particles with a single drive quantum $\Omega$. Thus, whenever the driving frequency becomes smaller than \emph{twice} the single-particle bandwidth of the time-averaged Hamiltonian, $\Omega \leq 2 W_\mathrm{ave}$, with $W_\mathrm{ave} = 2\mathrm{max}_k |E_k| =  2|h^x_0+J_\mathrm{ave}|$, a resonance occurs. The situation is similar to the parametric resonance observed in the periodically-driven weakly-interacting Bose-Hubbard model, as seen perturbatively in Refs.~\cite{bukov_15_prl,lellouch_16}. Based on this argument, we expect that  FAPT fails for this model if $\Omega\leq 2W_\mathrm{ave}$. On the other hand, our previous results suggest that FAPT should reproduce the correct behaviour of the system at small ramp speeds for $\Omega>2W_\mathrm{ave}$.
	
	\begin{figure}[h!]
		\centering
		\includegraphics[width=0.5\columnwidth]{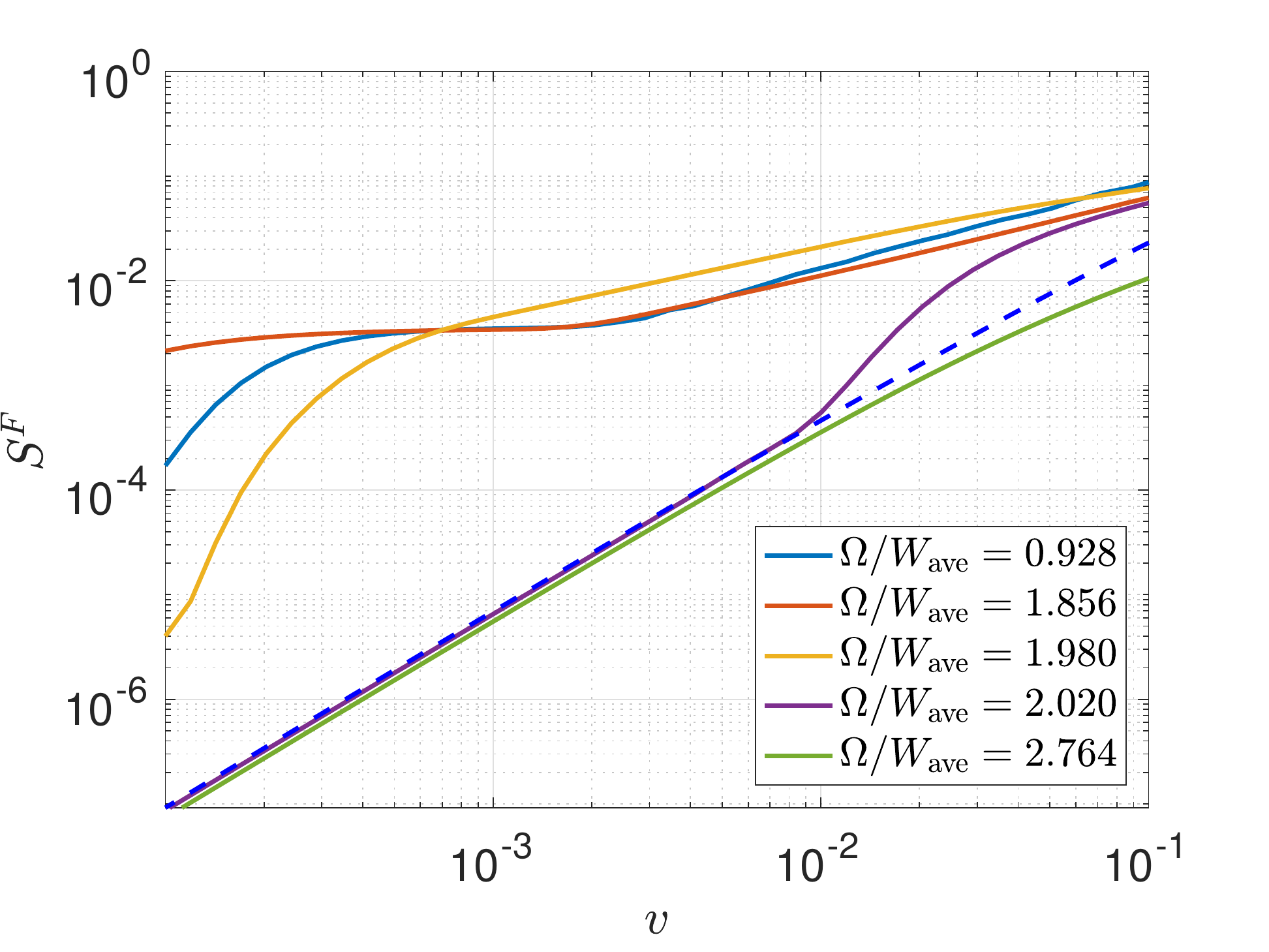}
		\caption{\label{fig:TFIM_nint}(Color online)[Driven transverse-field Ising model]. Floquet diagonal entropy of the driven TFI chain, showing that FAPT applies for $\Omega>2 W_\mathrm{ave}$ (the effective single-particle bandwidth, see text). The dashed line shows a $v^2$ power law. Model parameters are $A_f = 0.5$, $h^x_0/J_0=0.809$, $2W_\mathrm{ave}/J_0 = 7.235$.}
	\end{figure}

	To test these predictions, we prepare the system in the ferromagnetic GS of the Hamiltonian $H_0(\lambda=0) = H_\mathrm{ave}(\lambda=0)$ and ramp up the amplitude of the drive slowly according to the protocol $\lambda(t)=\left[(t-t_i)/|t_i|\right]^2$ from zero to unity. We put the system on a ring of $L=200$ sites, and ensure that the results do not change if we further increase the system size. As a measure of adiabaticity, we choose the Floquet diagonal entropy $S^F_d$, cf.~Eq.~\eqref{eq:FAPT_entropy}, to avoid the difficulties associated with identifying the adiabatically-connected state. Figure~\ref{fig:TFIM_nint} clearly shows that for $\Omega> 2W_\mathrm{ave}$ FAPT applies and the non-adiabatic excitations are captured by the leading-order correction. On the other hand, for $\Omega<2 W_\mathrm{ave}$ FAPT breaks down and the system is excited much stronger than in the high frequency regime. Nevertheless, a certain notion of limited adiabaticity still holds in a sense that slower rates result in fewer excitations of the system. This behavior can be traced back to the equilibrium-like Kibble-Zurek physics resulting in a non-analytic scaling of various observables with the ramp rate $v$ (see e.g.~Ref.~\cite{degrandi_13}), associated with the emergence of a degeneracy analogous to a quantum critical point in the Floquet system. We leave the details of this interesting story to a future work, as this setup is not sufficiently generic. Let us only point out that the existence of the singular point is intuitively clear by noting that at zero driving amplitude all Floquet levels are completely decoupled while an infinitesimal driving amplitude immediately opens a gap in the resonantly coupled states, which always exist for $\Omega<2 W_{\rm res}$. This gap opening is similar to the ordinary second-order phase transition in the Floquet Hamiltonian and drives this Kibble-Zurek physics. Therefore, increasing the driving amplitude from zero is similar to starting at the quantum critical point leading to the Kibble-Zurek scenario. If one starts the ramp from a finite driving amplitude, FAPT is expected to be restored in this system even in this low-frequency regime if one appropriately defines the adiabatic limit.
	
	The conclusions drawn from this model apply to arbitrary periodically-driven non-interacting band systems. In particular, our results are readily extensible to capture the dynamics of free bosons and fermions in various lattice geometries with periodic boundary conditions. Hence, it proves useful to study the applicability of FAPT to such non-interacting quantum many-body systems before we consider the fully interacting case in the next section. In Refs.~\cite{privitera_15,dalessio_14_Chern} the adiabatic loading in the ground state of the Floquet Haldane model was investigated with special emphasis put on crossing the topological critical point as the drive is gradually turned on. The failure of adiabaticity was caused by crossing intra-band resonances~\cite{privitera_15}. We note in passing that the results presented above pertain directly to recent cold atoms experiments realising dynamical localisation, artificial gauge fields, and topological bands with non-interacting particles.
	
	\begin{figure}[h!]
		\centering
		\includegraphics[width=0.8\columnwidth]{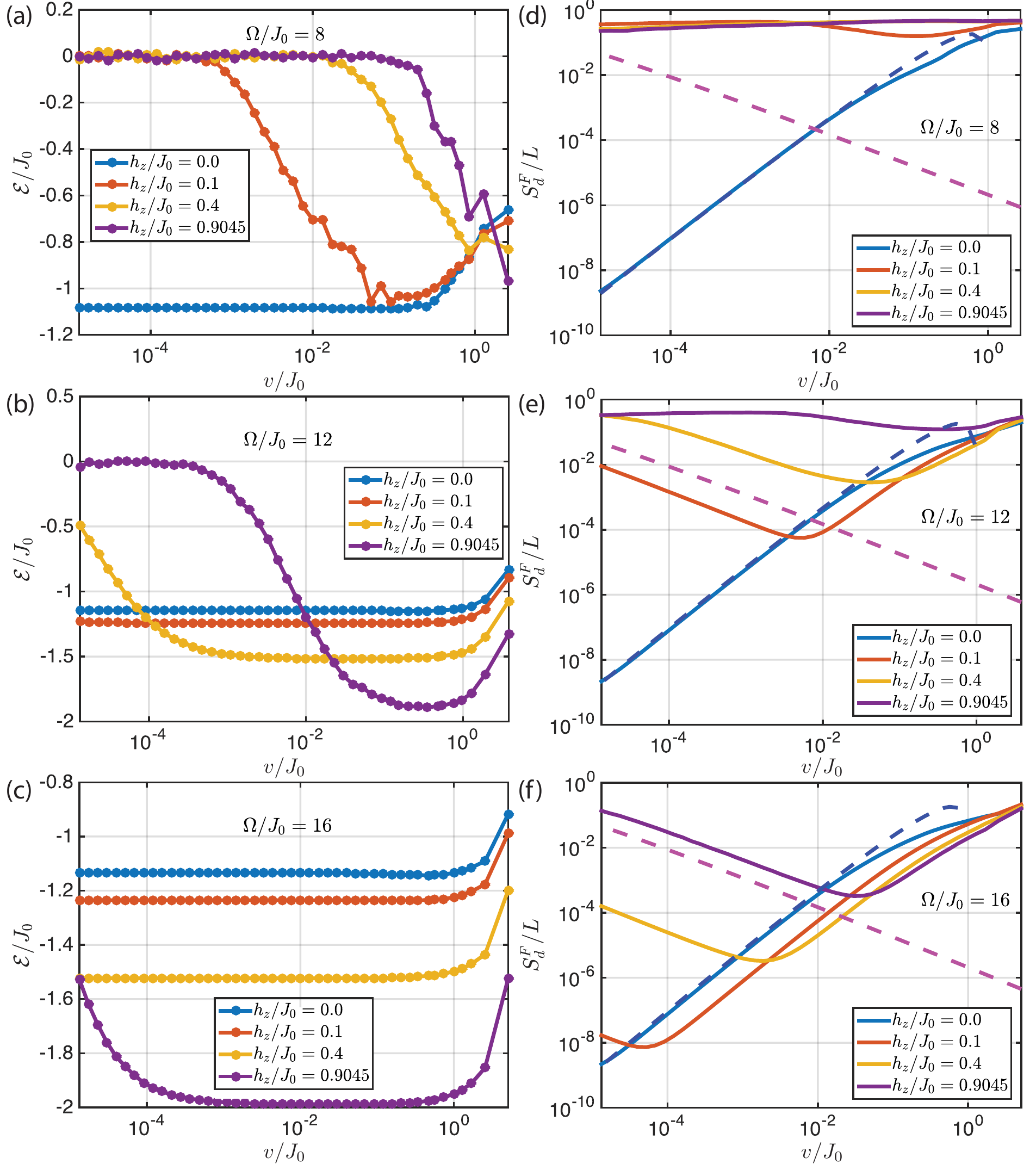}
		\caption{\label{fig:TFIM_Edensity}(Color online)[Driven non-integrable transverse-field Ising model]. Energy density $\mathcal{E}=\langle \psi(t=0)|H(\lambda=0)|\psi(t=0)\rangle/L$  (a-c) and Floquet diagonal entropy density $S_d^F/L$ (d-f) as a function of $v$, showing heating due to resonances at low velocities and for large $h_z$. The dashed lines are the same as in Fig.~\eqref{fig:Kapitza_entropy}. The model parameters are $A_f = 0.5$, and $h^x_0/J_0=0.809$. The system size is $L=18$ with a many-body bandwidth: $W^\mathrm{MB}_{\rm ave}/J_0 =  42.19$, $43.80$, $48.90$, $58.16$ for $h^z/J_0=0$, $0.1$, $0.4$, $0.9045$ respectively.}
	\end{figure}

	\subsection{\label{subsec:TFI2} The Driven Transverse-Field Ising Model in a Longitudinal Magnetic Field.}

	Let us generalise the TFIM from the previous section by switching on an additional static magnetic field $h^z$ along the $z$-direction. The driven Hamiltonian now reads
	\begin{eqnarray}
	\label{eq:H_TFIM_nonintegrable}
	H(t) = -J_0\sum_j \sigma^z_{j+1}\sigma^z_j - (h^x_0 + A_f\lambda\Omega\cos\Omega t)\sum_j \sigma^x_j - h^z\sum_j \sigma^z_{j}.
	\end{eqnarray}
	The non-driven version is no longer analytically solvable. Its spectrum exhibits Wigner-Dyson statistics which suggests that this model is quantum chaotic~\cite{kim_13}. Hence, it represents a generic interacting periodically-driven quantum system.

	\begin{figure}[h!]
		\centering
		\includegraphics[width=0.8\columnwidth]{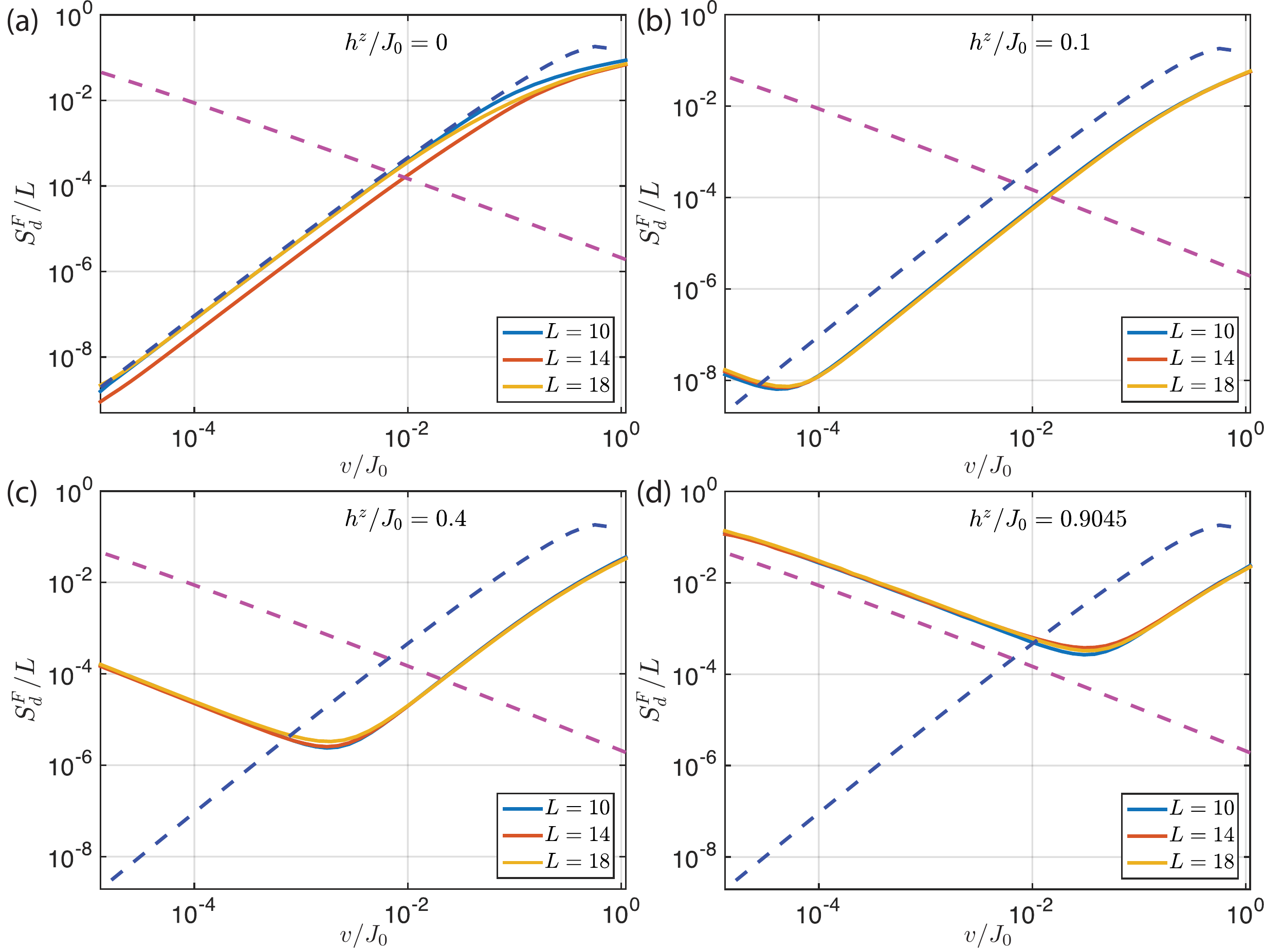}
		\caption{\label{fig:TFIM_nonint_L}(Color online)[Driven non-integrable transverse-field Ising model]. Diagonal entropy density in the Floquet eigenbasis versus the ramp speed $v$ for different system sizes, confirming the presence of a regime of applicability for FAPT. The dashed lines are the same as in Fig.~\eqref{fig:Kapitza_entropy}. The many-body bandwidth at $\lambda=0$ and $h^z/J_0=0.9045$ as a function of the system size $L$ reads: $W^\mathrm{MB}_{\rm ave}/J_0 = 32.34$ for $L=10$; $W^\mathrm{MB}_{\rm ave}/J_0 = 45.24$ for $L=14$, and $W^\mathrm{MB}_{\rm ave}/J_0 = 58.16$ for $L=18$. The model parameters are $A_f = 0.5$, $h^x_0/J_0=0.809$, and $\Omega/J_0=16$.}
	\end{figure}
	
	Studying periodically-driven many-body systems in the thermodynamic limit requires a certain degree of caution, in order not to be mislead by finite-size effects. As these systems have unbounded spectra for $L\to\infty$, it is necessary to clearly define the meaning of the thermodynamic limit $L\to\infty$ and the infinite-frequency limit $\Omega\to\infty$, none of which are strictly speaking accessible either experimentally or numerically. Here, we consider the situation where we first send $L\to\infty$ and only then are allowed to take $\Omega\to\infty$. We thus choose a driving frequency much less than the many-body bandwidth $W^{\rm MB}_{\rm ave}$, while still larger than twice the single-particle bandwidth: $W^{\rm MB}_{\rm ave} \gg \Omega > 2W_\mathrm{ave}$. This condition is crucial to allow for the appearance of Floquet many-body resonances~\cite{russomanno_15,bukov_15_erg} which, as we already demonstrated, represent the limiting factor in the applicability of FAPT, while preventing single-particle resonances that, as we already encountered, lead to the trivial breakdown of FAPT. For a given driving frequency, we test the largest values $L$ possible in an attempt to push towards the thermodynamic limit. All simulations in this section were performed in the zero (total) momentum sector with positive spatial parity.

	Similarly to the TFIM, we start from the GS of the non-driven Hamiltonian and slowly ramp-up the periodic drive. We measure the Floquet diagonal entropy as a function of the ramp speed at the end of the ramp. Figure~\ref{fig:TFIM_Edensity},~\ref{fig:TFIM_nonint_L} show the existence of large velocity windows at intermediate frequencies for which FAPT holds, so long as one does not cross any phase transitions~\cite{russomanno_15}. We find that this window shrinks down as a function of the integrability breaking parameter $h^z$, as expected, but remains clearly visible even at moderate-to-strong longitudinal fields.  When the ramp speed $v$ becomes too small, however, as with the quantum Kapitza pendulum, photon absorption resonances become important and the system eventually heats up. This can be easily detected numerically by looking at the expectation value of the non-driven Hamiltonian $H_0$ at the measurement point, i.e., the physical energy of the system, in Fig.~\ref{fig:TFIM_Edensity}. While there is a clear plateau which holds over several decades along the $v$-axis, energy absorption is eventually enabled by the strong hybridisation of the Floquet many-body levels in the vicinity of the photon absorption avoided crossings. 
	
	Even though we cannot conclude from the available system sizes what the fate of this window is in the thermodynamic limit, we observe that the region of validity of FAPT does not show any severe system-size dependence, see Fig.~\ref{fig:TFIM_nonint_L}. Let us point out that, at high frequencies in the thermodynamic limit, in the \emph{absence} of a ramp, energy absorption in spin and fermion systems can happen at most exponentially slowly in the driving frequency due to the exponentially suppressed matrix elements responsible for the appearance of many-body resonances~\cite{mori_15,abanin_15,kuwahara_15,abanin_15_2}. Therefore, for a ramped system at high-frequencies, the photon absorption gaps leading to the appearance of many-body resonances and consequently to non-adiabatic heating, become invisible to the system at these small but finite ramp rates. Hence, we expect that the large window where FAPT is valid will be present in experimentally-relevant setups with Floquet many-body Hamiltonians.
	
	It has been recently shown that the onset of heating in driven nonintegrable many-body systems can be traced back to the proliferation of many-body resonances~\cite{bukov_15_erg}. Here, we have identified the same resonances as the origin of breakdown of FAPT. This explains the observation that the window of adiabaticity shrinks as we lower the driving frequency, see Fig.~\ref{fig:TFIM_Edensity}. To analytically explore this phenomenon of adiabaticity breaking and understand its origin in a greater detail, we shall return to the simpler example of the Kapitza pendulum in the next section and apply the inverse-frequency expansion.

	\section{\label{sec:HFE} Floquet Adiabatic Perturbation Theory and the Inverse-Frequency Expansion.}
	
	With the exception of special Floquet integrable situations such as a two-level system in the presence of circularly polarized time-periodic fields or the driven harmonic oscillator discussed in Sec.~\ref{subsec:HO}, it is hard or even impossible to compute the Floquet Hamiltonian analytically. An important tool for studying periodically driven systems at high frequencies is the inverse-frequency expansion~\cite{rahav_03,rahav_03_pra,goldman_14,bukov_14,mikami_15,eckardt_15} for the Floquet Hamiltonian, which is the cornerstone of present-day Floquet engineering. Specifically, one uses strongly or resonantly-coupled periodic drives~\cite{bukov_14,goldman_14_res,bukov_15_SW} to generate nontrivial properties in the Floquet Hamiltonian in the high-frequency limit. It is, therefore, important to analyse the regimes of applicability of FAPT within the inverse-frequency expansion.
	
	Below we demonstrate that the photon absorption resonances in the Floquet spectrum that lead to the failure of FAPT also carry far-reaching consequences for the convergence properties of the inverse-frequency expansion, variants of which, such as the Floquet-Magnus expansion~\cite{blanes_09,bukov_14}, the van Vleck expansion~\cite{rahav_03,rahav_03_pra,goldman_14,bukov_14,eckardt_15} or the Brillouin-Wigner expansion~\cite{mikami_15}, are widely used to study Floquet problems. FAPT was recently combined with the van Vleck inverse-frequency expansion in a unified description~\cite{novicenko_16}, but see also Ref.~\cite{ho_16}. In systems with unbounded spectra, such as many-particle systems in the thermodynamic limit, or generic single-particle systems with an unbounded kinetic energy, the convergence of the inverse-frequency expansion cannot be easily justified in mathematical sense. Nevertheless, if the driving frequency is much higher than the natural frequencies in the system, one can anticipate that this expansion should give a very good approximation to the exact Floquet Hamiltonian. In Refs.~\cite{mori_14,kuwahara_15,mori_15}, it was indeed proven that the inverse frequency expansion is at least asymptotic, which implies that the corresponding approximate Floquet Hamiltonian is guaranteed to describe the dynamics up to exponentially long (in the driving frequency) times. For single-particle systems like the Kapitza pendulum, the very notion of a natural frequency depends on the energy: for example, at small energies close to the stable minimum of the potential, the system is nearly a harmonic oscillator with a well-defined oscillation frequency. At high energies, the Kapitza pendulum becomes a free rotor weakly perturbed by the driving field. Because the energy gaps between consecutive levels of the free rotor are linearly spaced, even at high driving frequencies one can always find high-energy states, which are nearly resonant with the drive, and thus the validity of the inverse-frequency expansion should depend on the energy. To investigate all these and other issues, we give a detailed comparison between the exact Floquet spectrum and eigenstates, and those obtained within the inverse-frequency expansion for the Kapitza pendulum.

	The inverse-frequency expansion~\cite{rahav_03_pra,rahav_03,goldman_14,bukov_14,eckardt_15,mikami_15} relies on Floquet's theorem. To introduce this concept, notice that Eq.~\ref{eq:Floquet_thm_general} can be rewritten as
	\begin{eqnarray}
	U(t,t_0) =  \mathrm e^{-iK(t)} \mathrm e^{-i(t-t_0)H_F}\mathrm e^{iK(t_0)},
	\end{eqnarray}
	where the periodic Kick operator $K(t)=K(t+T)$ is Hermitian because the micromotion operator $P(t)=\mathrm e^{-iK(t)}$ is unitary. As we discussed earlier, both the Floquet Hamiltonian and the kick operator depend on the choice of the initial time $t_0$ (or equivalently the phase of the drive)\footnote{This is known as the Floquet gauge~\cite{bukov_14}.}. Our previous choice of stroboscopic gauge corresponds to the boundary condition $K(t_0)=0$, and consequently $U(t_0+T,t_0)=\exp[-i H_F[t_0] T]$. This choice is, however, not always convenient for calculating the spectrum perturbatively as it explicitly breaks the gauge symmetry with respect to the choice of $t_0$~\cite{rahav_03_pra,eckardt_15}. While the full Floquet spectrum is gauge independent, it is desirable that this invariance is also respected at each order in the inverse-frequency expansion. To achieve this, one needs to impose a different boundary condition on the kick operator, namely $\int_0^T\mathrm{d}tK_{\rm eff}(t)=0$, for which choice the Floquet Hamiltonian is denoted $H_{\rm eff}$. This is related to our previous stroboscopic gauge choice by the static $t_0$-dependent unitary transformation: $H_F[t_0]=\exp[-i K_{\rm eff}(t_0)] H_{\rm eff}\exp[i K_{\rm eff}(t_0)]$.
	
	This gauge choice leads to the van Vleck inverse-frequency expansion\cite{rahav_03_pra,rahav_03,goldman_14,bukov_14,mikami_15,eckardt_15} for the Floquet Hamiltonian $H_{\rm eff}$ and the kick operator $K_{\rm eff}(t)$, where $H_{\rm eff}$ is manifestly independent of the driving phase~\cite{rahav_03_pra,rahav_03}. All information about the phase is, therefore, contained in the kick operator $K_{\rm eff} (t)$. Assuming that the driving frequency is the largest single-particle energy scale in the problem, one can expand $H_\mathrm{eff}$ and $K_\mathrm{eff}(t)$ in powers of the inverse frequency\footnote{Since the series might be asymptotic, i.e.~in particular not convergent, one must be careful in interpreting the equal sign in Eq.~\ref{eq:HFE_ansatz}.}
	\begin{eqnarray}
	H_\mathrm{eff} = H_\mathrm{eff}^{(0)}+H_\mathrm{eff}^{(1)}+H_\mathrm{eff}^{(2)}+\dots,\ \ \ \ \ \ \ \  K_\mathrm{eff}(t) =K_\mathrm{eff}^{(0)}(t)+K_\mathrm{eff}^{(1)}(t)+K_\mathrm{eff}^{(2)}(t)+\dots,
	\label{eq:HFE_ansatz}
	\end{eqnarray}
	where $H_\mathrm{eff}^{(n)} = \mathcal{O}(\Omega^{-n})$, and similarly for the kick operator. The first few terms are given by 
	\begin{eqnarray}
	H_\text{eff}^{(0)} &=& H_0 = \frac{1}{T}\int_0^T\mathrm{d}t\,H(t),\nonumber\\
	H_\text{eff}^{(1)} &=& \frac{1}{\Omega}\sum_{\ell=1}^\infty \frac{1}{\ell} [H_\ell,H_{-\ell}] = \frac{1}{2Ti}\int_{0}^{T}\mathrm{d}t_1\int_{0}^{t_1}\mathrm{d}t_2\, f(t_1-t_2) [H(t_1),H(t_2)],\nonumber\\
	K_\text{eff}^{(0)}(t) &=& \bm{0},\nonumber\\
	K_\text{eff}^{(1)}(t) &=& \frac{1}{i\Omega}\sum_{\ell\neq 0}\frac{\mathrm{e}^{i\ell\Omega t}}{\ell} H_\ell = \frac{1}{2}\int_{t}^{T+t}\mathrm{d}t'H(t')f(t-t'),
	\label{eq:kick_operator_HFE}
	\end{eqnarray}
	where we Fourier-decomposed the Hamiltonian as $H(t) = \sum_{\ell=-\infty}^\infty H_\ell \mathrm e^{i\ell\Omega t}$ with operator-valued coefficients $H_\ell$, and the weight function $f(x) = (1-2x/T) = f(x+T)$, $x\in[0,T]$ in the integrands is defined to be periodic with period $T$~\cite{eckardt_15}. Higher-order terms up to third order of the van Vleck high-frequency expansion (vV HFE) can be found in Ref.~\cite{mikami_15}.

	A peculiarity of the inverse-frequency expansion is that it gives rise to an unfolded Floquet spectrum. This is intuitively clear for a many-body system, since each of the subsequent orders of the series produces static many-body Hamiltonians, the spectra of which all scale with the system size. For many-body systems this follows from the extensivity of the harmonics $H_\ell$ and their commutators. For single-particle systems with unbounded spectra, one can easily convince oneself that at any order in the inverse-frequency expansion one also obtains an unbounded effective Hamiltonian. If we now recall that the Floquet spectrum is defined modulo $\Omega$, and fold it artificially (i.e.~ad hoc), we find that the approximate van Vleck Floquet Hamiltonian necessarily features a multitude of unavoided level crossings. The level crossings are unavoided because there are no matrix elements in the approximate Floquet Hamiltonian to couple the resonant states~\cite{bukov_15_erg}, since all these Hamiltonians are local operators\footnote{In this sense, any approximate Floquet Hamiltonian obtained via the inverse-frequency expansion is ``Floquet integrable'', although the exact Floquet Hamiltonian may not be.}. However, this is in contradiction with the appearance of avoided photon absorption level crossings, which as we argued earlier, are ultimately responsible for the breakdown of FAPT. From this simple argument we can conclude that the Floquet spectrum close to such absorption resonances \emph{cannot} be reproduced in any finite order of the inverse frequency expansion. Recalling that the approximate Floquet Hamiltonian is a well-defined static many-body Hamiltonian, which satisfies conventional adiabatic perturbation theory (APT, see Sec.~\ref{subsec:APT}), from the argument above we can also conclude that the violation of adiabaticity we observed in the models discussed so far is a non-perturbative effect, which cannot be captured by the inverse-frequency expansion.
	
\begin{figure}[ht]
		\includegraphics[width=\columnwidth]{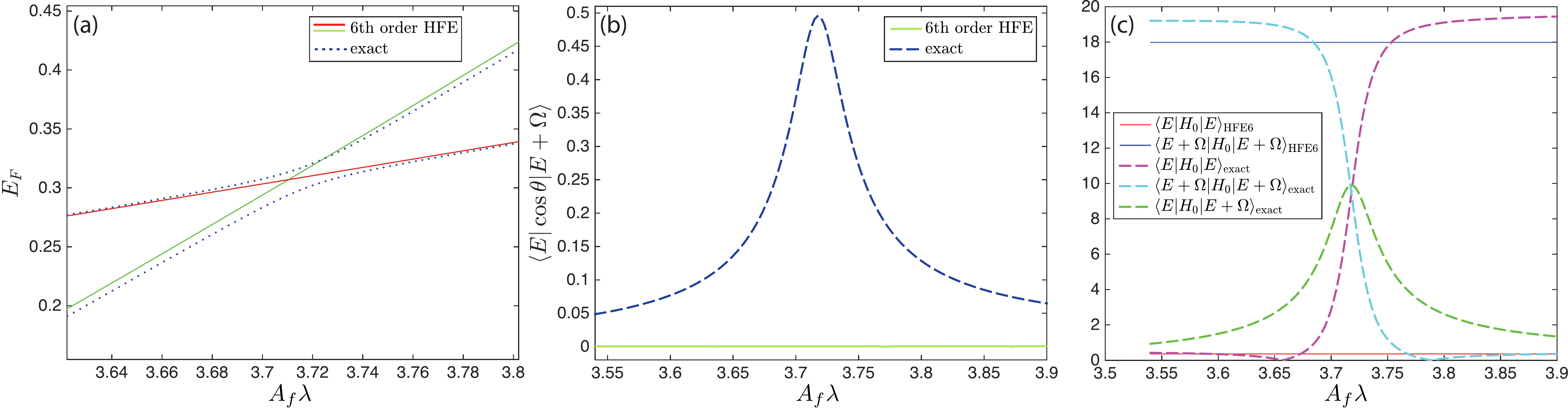}
		\caption{\label{fig:crossing_HFE_vs_exact}(Color online)[Kapitza pendulum]. High-frequency expansions and resonance effects on a characteristic photon absorption avoided crossing between the Floquet GS and a state one drive quantum higher in energy. (a) Photon absorption level crossing in the exact quasienergy spectrum of the Kapitza pendulum (blue dots), which is not present in the approximate spectrum calculated with the sixth order vV HFE [the approximate Floquet GS (red line) and the excited state (green line)]. (b) Transition matrix element of the dipole operator $d=\cos\theta$ between the two exact Floquet states show a resonance, which is not captured by the approximate states. (c) Matrix elements of the non-driven Hamiltonian $H_0=H(t,\lambda=0)$ in the exact Floquet states (dashed lines) and the approximate vV HFE states (solid lines). The model parameters are $m\omega_0 = 1$, $A_f = 2$, and $\Omega/\omega_0 = 20$, and the data shown corresponds to a single-photon resonance. }
	\end{figure}

	To confirm these qualitative considerations we come back to the Kapitza pendulum, a single-particle Hamiltonian with an unbounded spectrum, whose photon absorption avoided crossings were studied in Sec.~\ref{subsec:kapitza} (cf. Refs.~\cite{kapitza_51,dalessio_13,bukov_14} for more details on this model). Let us recall that the Hamiltonian for the Kapitza pendulum (cf.~Eq.~\eqref{eq:H_Kapitza}) is
	\begin{eqnarray}
	H(t) = \frac{p_\theta^2}{2m}  - m\omega_0^2\cos\theta -m A_f\lambda(t) \Omega\cos\Omega t\cos\theta~.
	\label{eq:kapitza_lab}
	\end{eqnarray}
	Since the driving amplitude scales linearly with the driving frequency~\cite{bukov_14_pra,bukov_14}, before we apply the inverse-frequency expansion it is advantageous to bring the Hamiltonian to the rotating frame
	\begin{eqnarray}
	H_\text{rot}(t) &=& V^\dagger(t)H(t)V(t) - iV^\dagger(t)\partial_t V(t),\nonumber\\
	V(t) &=& \exp\left[-i m A_f\lambda\sin\Omega t\cos\theta\right],
	\label{eq:kapitza_V(t)}
	\end{eqnarray}
	where $V(t)$ is a periodic unitary transformation which obeys $V(lT)={\bf 1}$. It is straightforward to check that the rotating frame Hamiltonian reads
	\begin{eqnarray}
	H_\text{rot}(t) &=& \frac{p_\theta^2}{2m}  - m\omega_0^2\cos\theta + \frac{m\lambda^2A_f^2\sin^2\Omega t}{2}\sin^2\theta - \frac{\lambda A_f\sin\Omega t}{2}\{\sin\theta,p_\theta\}_+ = \sum_{\ell=-2}^2 H_\ell \mathrm e^{i\ell\Omega t}, \nonumber\\
	H_0 &=& \frac{p_\theta^2}{2m}  - m\omega_0^2\cos\theta + \frac{m\lambda^2A_f^2}{4}\sin^2\theta,\nonumber\\
	H_{1} &=& i\frac{\lambda A_f}{4} \{\sin\theta,p_\theta\}_+ = -H_{-1}, \quad H_{2} = -\frac{m\lambda^2A_f^2}{8}\sin^2\theta = H_{-2},
	\label{eq:kapitza_H(t)_rot_frame}
	\end{eqnarray}
	where $\{\cdot,\cdot\}_+$ denotes the anti-commutator and all parameters remain finite as $\Omega\to\infty$. We then compare the exact Floquet spectrum obtained numerically to the spectrum produced by the vV HFE, including terms up to sixth order $H_\mathrm{eff}^{[n_\mathrm{HFE}=6]}$ [all odd orders $H_\mathrm{eff}^{(2n+1)}=0$ vanish for this model] for a frequency of $\Omega/\omega_0=20$. In doing so we make sure we eliminate all dependence on the spectral cut-off $M$ from the discussion; see Sec.~\ref{subsec:kapitza} for the precise definition of the cut-off parameter $M$. Since we have identified the Floquet ground state, it is straightforward to find a \emph{photon absorption} avoided crossing in the quasienergy spectrum: all we need to do is locate a state which crosses the Floquet ground state (a.k.a.~the adiabatically-connected state) coming from below on the quasienergy axis (see green curve in Fig.~\ref{fig:crossing_HFE_vs_exact}a). Moreover, for a reliable comparison with the vV HFE, the candidate Floquet states have to be well-approximated by the eigenstates of $H_F^{[n_\mathrm{HFE}]}$\footnote{Let us remark that since we work with the van Vleck expansion, the eigenstates of the non-stroboscopic $H_\mathrm{eff}^{[n_\mathrm{HFE}=6]}$ need to be rotated back to the basis of stroboscopic Floquet-Magnus Hamiltonian $H_F[0]$ with the help of the kick operator. This is because numerically it is straightforward to diagonalize the unitary evolution operator over the period yielding the eigenstates of the stroboscopic Floquet-Magnus Hamiltonian $H_F[0]$.}. In~\ref{subsec:HFE_convergence} we show a detailed analysis of the logarithmic inverse participation ratio (log-IPR) of the exact Floquet eigenstates in the basis of approximate Hamiltonian $H_F^{[n_\mathrm{HFE}]}$ for different orders of the vV HFE. Interestingly, close to resonances increasing the order of the expansion barely affects the log-IPR (see Fig.~\ref{fig:IPR}(c)), while increasing the frequency makes the log-IPR visibly closer to unity.
	
	Figure~\ref{fig:crossing_HFE_vs_exact} (a) shows a photon absorption avoided crossing of the Floquet ground state and a higher-energy state. As anticipated above, we see that the vV HFE up to sixth order does not know about the avoided crossing. We stress that the matrix elements in the effective Hamiltonian that would induce a direct transition between the two states participating in the crossing are already enabled by the high number of nested commutators in the sixth order of the expansion\footnote{Figure~\ref{fig:crossing_HFE_vs_exact} (a) actually holds to seventh order since all odd orders of the effective Hamiltonian vanish. This is not true for the kick operator, though.}. From the point of view of adiabaticity, this means that within the vV HFE, FAPT can fail only due to standard avoided crossings, whose physics is unrelated to that of the photon absorption crossings as seen in Sec.~\ref{subsec:kapitza}. Thus, a certain amount of caution needs to be exerted when one uses the vV HFE in such problems. 
	
	To bring out the nature of the avoided crossing we define the dipole operator $d=\cos\theta$, which changes the angular momentum of a basis state by one quantum, $\langle l|\cos\theta|l'\rangle \sim \delta_{l,l'\pm 1}$. The matrix element of the dipole operator between the two states that undergo the crossing is shown in Fig.~\ref{fig:crossing_HFE_vs_exact} (b). Clearly, due to the large angular momentum difference of the approximate states in the vV HFE, the matrix element not surprisingly vanishes identically, while the exact Floquet states produce a nice resonance owing to the strong hybridisation present in the vicinity of the crossing. Finally, Fig.~\ref{fig:crossing_HFE_vs_exact} (c) shows the matrix elements of the non-driven Hamiltonian $H_0 = H(t,\lambda=0)$ in the exact and approximate Floquet states. Once again, this is a measure of the hybridisation between the two Floquet states involved in the crossing, which is completely absent in their approximate counterparts.

	The failure of the vV HFE to capture resonance effects can also be observed in simpler systems. For instance, consider a harmonic oscillator with periodically stretched confining potential, which models the dynamics of a child on a swing. This simple system exhibits the phenomenon of parametric resonance: whenever the driving frequency hits twice the natural frequency of the oscillator, $\Omega = 2\omega_0$, all physical observables feature exponential growth in time. The quantum version of this driven oscillator is even more intriguing, as one has to accommodate this exponential growth in the wave functions of the system which are expected to be normalised. Strikingly, it was shown that, on resonance, the Floquet states are non-normalisable and the Floquet spectrum becomes continuous~\cite{weigert_02,popov_70}. In a more complicated system, the spectrum of the non-driven model might not be commensurate, but the appearance of resonances is guaranteed, as we have seen in various examples throughout this work. Recently, it was also shown that in a spin chain the second derivative of the Floquet ground state quasienergy with respect to the driving frequency features divergences~\cite{russomanno_15}, which was used as a precursor of frequency-induced topological phase transitions~\cite{mikami_15,russomanno_15}, and hints towards a non-analytic (in frequency) behavior. The vV HFE, on the other hand, is by assumption/construction analytic in the inverse frequency and, as we have seen using the example of the Kapitza pendulum, it is not sensitive to such resonances. If we decompose the interactions of the system's degrees of freedom with the drive into real (on shell) and virtual (off-shell), the vV HFE only captures the virtual excitations, similarly to the Schrieffer-Wolff transformation in equilibrium systems~\cite{bukov_15_SW}.

	\section{\label{sec:outro}Discussion and Outlook.}
	
	In this review, we extensively discussed Floquet adiabatic perturbation theory in various single-particle and many-body models, focusing on non-adiabatic corrections due to a smooth change of parameters in a periodically driven Hamiltonian. We first analysed those ``Floquet integrable" systems in which the Floquet adiabatic limit is well defined. Examples of such systems include the driven harmonic oscillator and the driven one-dimensional transverse field Ising spin chain. We identified the leading non-adiabatic corrections to various observables generalising the quantum adiabatic theorem to Floquet systems. We showed that non-adiabatic response in general is determined by both the slowly changing Floquet Hamiltonian and the slowly changing kick operator. For the phase-averaged observables (i.e., observables averaged over the period or equivalently over the driving phase), these corrections take a remarkably simple form, similar to non-driven systems. In particular, we showed that the leading linear non-adiabatic response to generalised forces is proportional to the Floquet Berry curvature even in interacting systems, suggesting the possibility of measuring it together with the associated Chern numbers in experiments. Based on these results, we showed how one can realise an energy Thouless pump, where by adiabatically changing some parameter in the Floquet Hamiltonian one can increase or decrease the energy of the system in discrete units of the driving frequency with the energy quantum determined by the Floquet Chern number. Physically, this corresponds to adding or removing an integer number of photons from the drive in each cycle. Our results also imply that the Floquet Chern numbers for band insulators~\cite{kitagawa_11,oka_09,aidelsburger_14,jotzu_14,dehghani_15,dehghani_15_2} are measurable through the standard Hall-type response if one averages the signal over the driving period.
	
	For generic Floquet systems, whose non-driven part features an unbounded spectrum, the adiabatic limit strictly speaking does not exist, as infinitesimally slow ramping leads to infinite heating through a cascade of photon-absorption resonances. Nevertheless, we demonstrated that at high driving frequencies FAPT still works very well in a broad window of ramp speeds such that the ramp rates are sufficiently fast for the resonant level crossings to be passed diabatically, while all standard avoided crossings (i.e.~the level crossings of the approximate unfolded Floquet Hamiltonian) are passed adiabatically, as illustrated in Fig.~\ref{fig:avoided_crossings}. As the driving frequency is decreased, this adiabaticity window shrinks down and, at sufficiently slow frequencies, it completely disappears. We carefully analysed these photon absorption resonances using the example of the quantum Kapitza pendulum and presented a strong evidence that they are not captured at any finite order of the inverse frequency expansion. This suggests that the non-adiabatic effects associated with the Floquet resonances might have a non-perturbative dependence on the inverse frequency.

	In some cases, Floquet engineering makes use of resonant drives, where the driving frequency is locked to an integer multiple of some (mostly single-particle) energy scale in the non-driven problem. If possible, for such high-frequency resonant driving, one can significantly simplify the theoretical analysis by first going to a rotating frame with respect to both the driving protocol and the resonant term~\cite{goldman_14_res,bukov_15_SW}. Effectively, this leads to strong hybridisation between the resonant levels. One can then apply FAPT in this rotating frame~\footnote{This is analogous to our analysis of the TFIM, see Sec.~\ref{subsec:TFI}.}. However, if one starts the resonant driving from a zero driving amplitude, then FAPT is expected to break down due to Kibble-Zurek type physics. One can understand this as follows: the drive amplitude sets the degree of hybridisation between the resonant levels. Starting from zero amplitude and slowly ramping it up, one opens up a gap between the quasienergies and the physics of the problem is essentially that of the Kibble-Zurek problem starting from the critical point. Therefore, it could be advantageous to design a protocol which avoids this scenario by, e.g., first quenching to some finite value of the drive amplitude, and then sweeping it gradually to the desired final value (see Ref.~\cite{tsuji_11} and the supplemental material to Ref.~\cite{bukov_15_SW} for specific examples). In doing so, the quench helps avoid the initial non-analyticities in the Floquet spectrum as $\lambda\to 0$ leading to complications associated with Kibble-Zurek physics. This combination of an initial quench followed up by an adiabatic sweep has not yet been studied systematically but offers an intriguing alternative to the conventional ramp-through approach, to be analysed in future studies.
	
	The breakdown of FAPT in generic Floquet systems is intimately related to heating. At a fixed drive coupling, recent numerical studies suggest that energy absorption can be either completely absent for finite-size systems~\cite{dalessio_13,prosen_98a,prosen_98b,prosen_99,prosen_02,bukov_15_erg} or, if present in the thermodynamic limit, it is typically exponentially suppressed with increasing the drive frequency $\Omega$. This  provides a justification for the existence of long-lived Floquet steady states in certain parameter regimes at high frequencies~\cite{abanin_15,kuwahara_15,mori_15,abanin_15_2,canovi_15,bukov_15_prl}. As we have demonstrated in this chapter, the situation becomes even more complex once a model parameter is being ramped up in the presence of the drive. In particular, during ramps one necessarily crosses photon-absorption resonances and, when the ramp speeds are infinitesimal, this leads to heating of the system even in the high-frequency regime. These results are strikingly similar to those recently predicted in single-particle disordered systems~\cite{khemani_15}, and they most likely also apply to MBL systems. 
	
	The notion of Floquet adiabaticity discussed in this review is intriguing and can naturally be extended and applied to a variety of novel physical situations. Our conclusions tell us about possible issues with any low-frequency linear response theory applied to isolated Floquet systems, which is very relevant as a method for measuring these systems and has been discussed extensively in the recent literature~\cite{dehghani_15,dehghani_15_2,iadecola_14_bath,seetharam_15,rigolin_14}. We have derived FAPT by looking into properties of the gauge potential $\mathcal A_F$, which is an operator whose diagonal components give generalizations of the Berry connections. However, as an operator, this actually gives us access to many more properties such as off-diagonal connections. Furthermore, it readily gives access to non-Abelian Floquet Berry connections in the presence of degeneracies, as has recently been explored in non-driven systems~\cite{rigolin_10,rigolin_12,rigolin_14,kolodrubetz_16_Ch2}. This is particularly important as experiments involving Floquet systems are quite active in the creation of topologically non-trivial states, the most interesting of which have non-trivial non-Abelian Berry phases. The generalization of these techniques to cases with degeneracies will, therefore, lead to a proposal for a measurement protocol for these non-Abelian effects, and will provide valuable insight on differences in topological physics between driven and non-driven systems. Moreover, the methods we discussed are expected to naturally extend to mixed states, and an important open question is how they are modified by weak interactions with the environment~\cite{avron_12,avron_12_jstat,xu_14,albert_15}. At the same time, the absence of a well-defined adiabatic limit in open periodically-driven systems poses interesting fundamental questions regarding the stability of Floquet phases which, in equilibrium systems, is based on the maximum entropy principle and the resulting equivalence of adiabatic generalized forces. In the absence of adiabaticity, one must re-examine fundamental questions, such as the existence of the equations of state, and many other statements which we take for granted in equilibrium thermodynamics.
	
	\section{Acknowledgments}
	The authors would like to thank M.~Aidelsburger, M.~Atala, E.~Dalla~Torre, N.~Goldman, M.~Heyl, D.~Huse, G.~Jotzu, C. Kennedy, M.~Lohse, T.~Mori, L.~Pollet, M.~Rudner, A.~Russomanno, and C.~Schweizer for fruitful discussions. This work was supported by AFOSR FA9550-13-1-0039, NSF DMR-1506340, ARO W911NF1410540, and the Hungarian research grant OTKA  Nos.~K101244, K105149. M.~K.~was supported by Laboratory Directed Research and Development (LDRD) funding from Berkeley Lab, provided by the Director, Office of Science, of the U.S.~Department of Energy under Contract No.~DE-AC02-05CH11231. The authors are pleased to acknowledge that the computational work reported on in this paper was performed on the Shared Computing Cluster which is administered by \href{www.bu.edu/tech/support/research/}{Boston University's Research Computing Services}. The authors also acknowledge the Research Computing Services group for providing consulting support which has contributed to the results reported within this paper. The study of the driven non-integrable transverse-field Ising model was carried out using QuSpin~\cite{quspin} -- an open-source state-of-the-art Python package for dynamics and exact diagonalisation of quantum many-body systems, available to download \href{https://github.com/weinbe58/QuSpin}{here}.

	\appendix

	\section{\label{app:general_thy}Details in the Derivation of Floquet Adiabatic Perturbation Theory}
	
	In this Appendix, we give the details of the derivation of Eq.~\eqref{eq:c_n_FAPT} from the main text. To this end, let us write down again Eq.~\eqref{eq:FAPT_exact}:
	\begin{equation}
	i\dot c_n'=-\dot\lambda\sum_{m\neq n} \langle n_F(\lambda) | \mathcal A^F_\lambda(\lambda,t) |m_F(\lambda)\rangle \mathrm e^{i (\Phi_n^F(t)-\Phi_m^F(t))} c_m',
	\label{eq:FAPT_exact2}
	\end{equation} 
	which is an exact Schr\"odinger equation describing Floquet systems. If $\lambda(t)$ is a slowly-varying, monotonic function of time, and the ramp is adiabatic: $\dot\lambda\to 0$, the RHS vanishes identically and the system remains in the instantaneous adiabatically-connected state. However, any finite ramp speed results in excitations, the magnitude of which is governed by the instantaneous velocity $\dot\lambda(t)$. In order to understand these excitations, our strategy is to formally integrate Eq.~\eqref{eq:FAPT_exact2} perturbatively, using the ramp speed as a small parameter. 
	
	Let us assume that at time $t_i$ we start the evolution in the ground state of the non-driven Hamiltonian, i.e.~$c_n(t_i)=\delta_{n0}$. Straightforward integration then leads us to the first correction to the amplitude for making a transition to the $n$-th excited state ($n\neq 0$):
	\begin{eqnarray}
	c'_n(\lambda(t),t)&=&  i\int_{t_0}^{t}\mathrm{d} t'\dot{\lambda}(t')\mathrm e^{i(\Phi^F_{n}(t')-\Phi^F_{0}(t'))}A^F_{n0}(\lambda(t'),t')+\mathcal{O}(\dot{\lambda}^2)
	\label{eq:cancelphase},
	\end{eqnarray}
	where we define $A^F_{n0}(\lambda,t)= \langle n_F(\lambda) | \mathcal A^F_\lambda(\lambda,t)| 0_F(\lambda)\rangle$ to simplify notation. Next  we need to evaluate the rapidly oscillating integral in Eq.~\eqref{eq:cancelphase}. To do this, we cast the Floquet gauge potential at fixed $\lambda$  in Fourier space 
	\begin{equation}
	\mathcal{A}^F(\lambda,t)=\sum_{\ell=-\infty}^\infty \mathcal{A}^{F,\ell}(\lambda)\mathrm e^{i\Omega \ell t}.
	\end{equation}
	Thus, Eq.~\eqref{eq:cancelphase} assumes the form:
	\begin{equation}
	c'_n(\lambda(t),t)= i\sum_\ell \int_{t_0}^{t}\mathrm{d} t'\dot{\lambda}(t')\mathrm e^{i(\Phi^F_{n}(t')-\Phi^F_{0}(t')+\Omega \ell t')}A_{n0}^{F,\ell}(\lambda(t'))+\mathcal{O}(\dot{\lambda}^2).
	\end{equation}
	We are faced with a rapidly oscillating phase, multiplied by a slow function in the integral so we can use standard techniques for evaluating it approximately. For finite values of $\lambda(t_i)$ and $\lambda(t_f)$, this can be done by integration by parts, assuming $\dot{\lambda}(t_i)=0$ and staying to linear order in $\dot{\lambda}(t)$:
	\begin{eqnarray}
	i\sum_\ell\int_{t_0}^{t}\mathrm{d} t'\dot{\lambda}(t')\mathrm e^{i(\Phi^F_{n}(t')-\Phi^F_{0}(t')+\Omega \ell t')}\mathcal A_{n0}^{F,\ell}(\lambda(t'))
	&=&\sum_\ell\int_{t_0}^{t}\mathrm{d} t'\frac{\dot{\lambda}(t')\mathcal A_{n0}^{F,\ell}(\lambda(t'))}{\epsilon_{n}(\lambda(t'))-\epsilon_{0}(\lambda(t'))+\ell\Omega}\frac{\mathrm{d}}{\mathrm{d} t'}\mathrm e^{i(\Phi^F_{n}(t')-\Phi^F_{0}(t')+\Omega \ell t')}\nonumber\\
	&=&\sum_\ell\frac{\dot{\lambda}(t)\mathcal A_{n0}^{F,\ell}(\lambda(t))\mathrm e^{i(\Phi^F_{n}(t)-\Phi^F_{0}(t)+\Omega \ell t)}}{\epsilon_{n}(\lambda(t))-\epsilon_{0}(\lambda(t))+\ell\Omega}+\mathcal{O}(\dot{\lambda}^2,\ddot{\lambda}).
	\end{eqnarray}
	The first equality comes from noticing that $d\Phi^F_n(t)/dt=\epsilon_n(\lambda(t))$, while for the second equality we have integrated by parts and used the assumption that $\dot{\lambda}(t_0)=0$. To simplify the notation, it is useful to suppress some of the $t$-dependence. Going back to the original $c_n(t)$ frame we obtain:
	\begin{equation}
	c_n(t)= \mathrm e^{-i\Phi_0^F(t)} \dot \lambda(t) \sum_{\ell=-\infty}^\infty \mathrm e^{i \ell \Omega t} {\langle n_F(\lambda)|\mathcal A^{F,\ell}_\lambda|0_F(\lambda)\rangle\over \epsilon^F_n-\epsilon^F_0+\ell\Omega}+\mathcal{O}\left(\ddot \lambda, \dot\lambda^2\right).
	\label{eq:amplitude_1}
	\end{equation}

	\section{\label{app:HO_thy}Exact Solution to the Harmonic Oscillator with Periodically Displaced Potential.}
	
	In this appendix we outline the exact solution to the driven Harmonic oscillator from Sec.~\ref{subsec:HO}, as well as derivations of various formulas from the main text. Once again, throughout the derivations below we set $\hbar=1$. The general solution is outlined in a set of notes by Peter H\"anggi on periodically driven systems and can be found online~\cite{haenggi_notes}. As we follow the derivation of Ref.~\cite{haenggi_notes}, we use the notion of rotating frames to be consistent with the discussion in the main text. 
	
	The Hamiltonian we would like to solve reads as:
	\begin{equation}
	H(t)=\frac{p^2}{2m}+\frac{1}{2}m\omega_0^2 x^2-f(t)x=H_\mathrm{HO}-f(t)x
	\end{equation}
	By going to a rotating frame using the \emph{consecutive} transformations 
	\begin{equation}
	\label{eq:app_V1V2}
	V_1(t)=\mathrm e^{-i\eta(t) p};\ \ \  V_2(t)=\mathrm e^{im\dot{\eta}(t) x},
	\end{equation}
	one finds that transformed Hamiltonian:
	\begin{gather}
	|\psi_\mathrm{rot}(t)\rangle = V^\dagger_2(t)V^\dagger_1(t)|\psi(t)\rangle\\
	H^\mathrm{rot}(t)=H_\mathrm{HO}+x\left[m\ddot{\eta}(t)+m\omega_0^2\eta(t)-f(t)\right]-L(\eta,\dot{\eta},t),\label{eq:app_H_HO_rot}
	\end{gather}
	where $L$ is the classical Lagrangian for the lab frame $H(t)$:
	\begin{equation}
	L(\eta,\dot{\eta},t)=\frac{1}{2}m\dot{\eta}^2(t)-\frac{1}{2}m\omega_0^2\eta^2(t)+\eta(t)f(t).
	\end{equation}
	Throughout, we adopt the notation $\dot{(\cdot)} = \mathrm{d}/\mathrm{d}t = \partial_t + \dot\lambda \partial_\lambda$. 
	
	We see that we can remove the linear term in $x$ from the Hamiltonian~\eqref{eq:app_H_HO_rot}, if $\eta(t)$ satisfies the classical equation of motion: 
	\begin{equation}
	\label{eq:app_EOM}
	m\ddot{\eta}+m\omega_0^2\eta=f(t)
	\end{equation}
	Finally, by doing another unitary transformation one can also remove the Lagrangian for the classical variable $L(\eta,\dot{\eta},t)$, leaving just $H_\mathrm{HO}$. In this frame the exact solution to the time-dependent Schr\"odinger equation is simply given in terms of the eigenstates and eigenenergies of $H_\mathrm{HO}$, $|n\rangle$ and $E_n=\omega_0(n+1/2)$:
	\begin{equation}
	|\psi^\mathrm{rot}(t)\rangle=\sum_n c_n \mathrm e^{-iE_nt}|n\rangle
	\end{equation}
	where $c_n$ are \emph{time independent}. Transforming back to the original lab frame, the exact solution can be written as:
	\begin{gather}
	|\psi(t)\rangle=\sum_n c_n |\chi_n(t)\rangle\\\label{eq:HO_gen_sol}
	|\chi_n(t)\rangle=\mathrm e^{i\varphi_n(\eta(t),\dot{\eta}(t),t)}\mathrm e^{-i\eta(t)\hat{p}}\mathrm e^{im\dot{\eta}(t)\hat{x}}|n\rangle
	\end{gather}
	where $\varphi_n(\eta,\dot{\eta},t)=-E_n t+\int_{t_i}^t \mathrm{d}t'L(\eta,\dot{\eta},t')$. Here we explicitly put hats on $\hat{x}$ and $\hat{p}$ to emphasize that they act as operators on the state $|n\rangle$. 
	
	In the following, it will prove useful to distinguish between two classical solutions: (i) $\eta(t) = \xi(t)$ is defined as the exact solution to the ramped classical problem~\eqref{eq:app_EOM}, i.e.~for $f(t)=\lambda(t)A_f\Omega^2\cos(\Omega t+\varphi_0)$. We shall see below that in this case we need not require that $\lambda(t)$ changes slowly. (ii) we denote by $\eta(t)=\zeta(t)$ the classical trajectory for the Floquet solution at a fixed $\lambda$, i.e.~for~$f(t)=\lambda A_f\Omega^2\cos(\Omega t+\varphi_0)$ at a fixed $\lambda$.

	\subsection{Exact Floquet Solution}
	
	\emph{Floquet Solution.---}In this subsection we discuss the application of the solution outlined above to the Harmonic oscillator in the presence of a periodically displaced potential. We then use that solution to calculate various quantities in FAPT. In this subsection, we consider $\lambda = const(t)$.
	
	From the previous discussion, it is clear that no assumptions are made about the initial conditions for the classical trajectory in the definition of the solution. Even though the initial conditions do not matter for the solution itself, the initial conditions affect the basis $|\chi_n(t)\rangle$. Therefore, in order to make use of the general solution to find the Floquet Hamiltonian and micromotion operator, the initial conditions of $\zeta(t)$ must be chosen such that $|\chi_n(t)\rangle$ manifestly satisfy Floquet's theorem. This can be accomplished if $\zeta(t)$ is the \emph{periodic} solution of: 
	\begin{equation*}
	m\ddot{\zeta}+m\omega_0^2\zeta=\lambda A_f\Omega^2\cos(\Omega t+\varphi_0),
	\end{equation*}
	which given by~\cite{haenggi_notes}:
	\begin{equation}
	\zeta(t)=\frac{A_f\Omega^2\lambda\cos(\Omega t+\varphi_0)}{m(\omega_0^2-\Omega^2)}.
	\end{equation}
	Note that the notation $V_1$ and $V_2$ for the rotators in Eq.~\eqref{eq:app_V1V2} was not an accident. Indeed, if we take the $\Omega\rightarrow\infty$ limit of $\zeta$, we see that $V_1$ and $V_2$ become the same transformations mentioned in the main text, see Eqs.~\eqref{eq:HO_rot_frame}. To get the Floquet quasi-energies one must separate the periodic and non-periodic parts of $\varphi_n(t)$~\cite{haenggi_notes}:
	\begin{gather}
	\epsilon_n^F(\lambda)=\omega_0\left(n+\frac{1}{2}\right)-\frac{A_f^2\Omega^4\lambda^2}{4m(\omega_0^2-\Omega^2)}\\
	|n^F(\lambda,t)\rangle=\mathrm e^{i\varphi(\zeta,\partial_t{\zeta},t)}\mathrm e^{-i\zeta\hat{p}}\mathrm e^{im\partial_t{\zeta}\hat{x}}|n\rangle\label{eq:HO_F_ket}
	\end{gather}
	where the $\lambda$ dependent constant part of $\epsilon_n^F$ comes from the period average of $L$, and  $\varphi(\zeta,\partial_t{\zeta},t)=\varphi_n(\zeta,\partial_t{\zeta},t)-\epsilon_n^F t$ is a periodic function with frequency $\Omega$. 
	
	For the rest of the calculation it will convenient to use the instantaneous Floquet basis $|n^F(\lambda,t)\rangle=P(\lambda,t)|n_F(\lambda)\rangle$ because of its simple form. It should be noted, that the form above suggests that the non-stroboscopic Floquet Hamiltonian $H_\mathrm{eff}$ and kick operator $K_\mathrm{eff}(t)$, cf.~Ref.~\cite{bukov_14}, are given by:
	\begin{gather}
	H_\mathrm{eff}=H_\mathrm{HO}-\frac{A_f^2\Omega^4\lambda^2}{4m(\omega_0^2-\Omega^2)}\\
	\mathrm e^{-iK_\mathrm{eff}(t)}=\mathrm e^{i\varphi(\zeta,\partial_t{\zeta},t)}\mathrm e^{-i\zeta\hat{p}}\mathrm e^{im\partial_t{\zeta}\hat{x}}
	\end{gather}
	By doing simple manipulations on the moving frame basis, one can find the stroboscopic micromotion operator $P(t)$, and the Floquet eigenstates which can be used to determine the stroboscopic $H_F[t_i]$\footnote{Note that while the stroboscopic Floquet Hamiltonian $H_F[t_i]$ depends explicitly on the initial time, the non-stroboscopic $H_\mathrm{eff}$ is manifestly $t_i$--independent.}:
	\begin{gather}
	H_F[t_i]=\frac{1}{2m}\left(p-m\partial_t{\zeta}(t_i)\right)^2+\frac{1}{2}m\omega_0^2\left(x-\zeta(t_i)\right)^2 - \frac{A_f^2\Omega^4\lambda^2}{4m(\omega_0^2-\Omega^2)}\\
	P(t)=\mathrm e^{i\varphi(\zeta,\partial_t{\zeta},t)}\mathrm e^{-i(\zeta(t)-\zeta(t_i))p}\mathrm e^{im(\partial_t{\zeta}(t)-\partial_t{\zeta}(t_i))(x+\zeta(t_i))}
	\end{gather}
	where $t_i$ defines the Floquet gauge~\cite{bukov_14}. Note that we don't show the explicit $\lambda$--dependence in $P(t)$, but it indeed comes in through the classical solution $\zeta$ which depends on the driving amplitude. Just as we mentioned in the main text, the Floquet spectrum only depends on $\lambda$ via a constant shift. Also, note that even at finite $\Omega$, all of the Floquet eigenstates are smooth functions of $\lambda$ and, therefore, satisfy the assumptions of FAPT.

	\emph{FAPT calculations.---}Now that we know the Floquet solution we can use it to calculate physical quantities to test FAPT. In order to do this, we first need the Floquet gauge potential $\mathcal{A}^F_{\lambda,m,n}=\langle m^F(\lambda,t)|i\partial_\lambda|n^F(\lambda,t)\rangle$, see Sec.~\ref{subsec:FAPT}. This calculation is very simple if one uses Eq.~\eqref{eq:HO_F_ket}, which leads to:
	\begin{gather}
	\mathcal{A}^F_{\lambda,m,n}=\mathcal{A}^F_{\lambda,0}\delta_{m,n}+\mathcal{A}^F_{\lambda,n,n+1}\delta_{m,n+1}+\mathcal{A}^F_{\lambda,n,n-1}\delta_{m,n-1}\\
	\mathcal{A}^F_{\lambda,0}=\left(m\partial_t{\zeta}\partial_\lambda\zeta-\partial_\lambda\varphi\right)\\
	\mathcal{A}^F_{\lambda,n,n+1}=\sqrt{\frac{n+1}{2m\omega_0}}\frac{A_f\lambda\Omega^2}{(\omega_0^2-\Omega^2)}\left[\Omega\sin(\Omega t+\varphi_0)+i\omega_0\cos(\Omega t+\varphi_0)\right]\\\label{eq:HO_F_guage}
	\mathcal{A}^F_{\lambda,n,n-1}=\sqrt{\frac{n}{2m\omega_0}}\frac{A_f\lambda\Omega^2}{(\omega_0^2-\Omega^2)}\left[\Omega\sin(\Omega t+\varphi_0)-i\omega_0\cos(\Omega t+\varphi_0)\right]
	\end{gather}
	An important observation here is that the off-diagonal elements of $\mathcal{A}^F$ only have two harmonics: $\ell = \pm 1$. From this, we conclude that only the micromotion operator $P(t)$ plays a role for the non-adiabatic corrections, because the Floquet Hamiltonian only appears in the FAPT expression via the $\ell = 0$ harmonic, cf.~Eq.~\eqref{eq:c_n_FAPT}. Another important observation is that we can always find a driving phase $\varphi_0$ which makes $\mathcal{A}^F_{\lambda,10}$ imaginary. As a result, observables which have real matrix elements in the moving Floquet basis have no $\dot{\lambda}$ corrections in FAPT. This implies that if one has precise control of the driving phase and ramp time, it should be possible to systematically reduce excitations at the measurement time.
	
	Now that we have the Floquet gauge potential we can calculate the leading non-adiabatic corrections  to the probability amplitudes of our wave function (see Eq.~\eqref{eq:c_n_FAPT}):
	\begin{equation}
	|\psi(t)\rangle=\sum_n c_n(t)|n^F(\lambda(t),t)\rangle
	\end{equation}
	with initial conditions: $c_n(t_i)=\delta_{n,0}$. Using Eq.~\eqref{eq:HO_F_guage} and the Floquet quasi-energies we find that $c_n(t)$ is given by:
	\begin{equation}
	c_n(t)\approx i\dot{\lambda}(t) \mathrm e^{-i\Phi^F_0(t)}\sqrt{\frac{1}{2m\omega_0}}\frac{A_f\lambda\Omega^2}{2}\left(\frac{\mathrm e^{i(\Omega t +\varphi_0)}}{(\omega_0+\Omega)^2}+\frac{\mathrm e^{-i(\Omega t+\varphi_0)}}{(\omega_0-\Omega)^2}\right)\delta_{n,1}+\mathcal{O}(\ddot{\lambda},\dot{\lambda}^2).
	\end{equation}
	As a result, the general expression for the expectation values for observables reads:
	\begin{equation}
	\langle O(\lambda(t),t)\rangle\approx \langle 0^F(\lambda(t),t)|O|0^F(\lambda(t),t)\rangle
	+\sum_{n>0}\left(\mathrm e^{i\Phi^F_0(t)}c_n(t)\langle 0^F(\lambda(t),t)|O|n^F(\lambda(t),t)\rangle + c.c.\right)+\mathcal{O}(\ddot{\lambda},\dot{\lambda}^2)
	\end{equation}
	Calculating matrix elements of the form: $\langle 0^F(\lambda(t),t)|O|n^F(\lambda(t),t)\rangle$ is straightforward using Eq.~\eqref{eq:HO_F_ket}. We show here the final results for $\langle p^2\rangle$ and $\langle x\rangle$:
	\begin{gather}
	\langle p^2(t)\rangle\approx m^2[\partial_t{\zeta}(t)]^2+\frac{m\omega_0}{2}+\dot{\lambda}(t)m\partial_t{\zeta}(t)\frac{4A_f\Omega^2(\Omega^2+\omega_0^2)}{(\omega_0^2-\Omega^2)^2}\cos(\Omega t+\varphi_0)+\mathcal{O}(\ddot{\lambda},\dot{\lambda}^2)\\
	\langle x(t)\rangle\approx \zeta(t)+\dot{\lambda}(t)\frac{2A_f\Omega^3}{m(\omega_0^2-\Omega^2)^2}\sin(\Omega t+\varphi_0)+\mathcal{O}(\ddot{\lambda},\dot{\lambda}^2),
	\end{gather}
	which are plotted in Fig.~\ref{fig:HO_obs} and discussed in Sec.~\ref{subsec:HO}. One can also calculate a closed expression for the probabilities of occupying a given Floquet state $|n^F(\lambda(t),t)\rangle$, which is given by $|c_n(t)|^2$:
	
	\begin{equation}
	p^F_n(t)\approx\frac{A_f^2\Omega^4 \dot{\lambda}(t)^2}{2m\omega_0}\frac{\left(4\omega_0^2\Omega^2\sin^2(\Omega t +\varphi_0)+(\omega_0^2+\Omega^2)^2\cos^2(\Omega t+\varphi_0)\right)}{(\omega_0^2-\Omega^2)^4}\delta_{n,1}+\mathcal{O}(\ddot{\lambda},\dot{\lambda}^3)
	\label{eq:HO_FAPT_Pex}
	\end{equation}
	which can be used to calculate the log-fidelity $f_d$ and the Floquet diagonal entropy $S_d^F$, cf.~Sec.~\ref{subsec:FAPT}.

	\subsection{Exact Solution to the Ramped Problem}

	In this subsection, we discuss the exact solution of the ramped problem. As mentioned above, the solution entails solving the classical equations of motion: 
	\begin{equation*}
	m\ddot{\xi}(t)+m\omega_0^2\xi(t)=f(t);\ \ \ \ f(t)=\lambda(t)A_f\Omega^2\cos(\Omega t +\varphi_0)
	\end{equation*}
	Note that in this subsection we have a time-dependent ramp function $\lambda=\lambda(t)$, and thus $\dot{\xi} = \partial_t \xi + \dot\lambda\partial_\lambda\xi$ is the full time derivative. 
	
	Any initial conditions for $\xi(t)$ gives a valid solution, but one can simplify the solution greatly by picking the proper initial condition: as we start in the ground state of $H_\mathrm{HO}$ initially, a natural choice for the initial condition on $\xi(t)$ is: $\xi(t_i)=\dot{\xi}(t_i)=0$. With this choice the basis $|\chi_n(t)\rangle$ (see Eq.~\eqref{eq:HO_gen_sol}) at $t=t_i$ is simply the eigenbasis of $H_\mathrm{HO}$, $\{|n\rangle\}$, and so we find that the solution has a very simple form:
	
	\begin{equation}
	|\psi(t)\rangle=\mathrm e^{i\varphi_0(\xi,\dot{\xi},t)}\mathrm e^{-i\xi\hat{p}}\mathrm e^{im\dot{\xi}\hat{x}}|0\rangle\label{eq:HO_R_ket}
	\end{equation}
	This has a simple physical interpretation: the wave function is simply a Gaussian wave packet following the classical trajectory. Using the initial conditions given above, $\xi(t)$ can be calculated for any $f(t)$ as:
	\begin{equation}
	\xi(t)=\sin(\omega_0 t)\int_{t_i}^t \mathrm{d}t' \cos(\omega_0 t')\frac{f(t')}{m}+\cos(\omega_0 t)\int_{t_i}^t \mathrm{d}s \sin(\omega_0 t')\frac{f(t')}{m}
	\end{equation}
	If we pick $\lambda(t)$ to have a simple enough form (e.g.~a power law in $t$), these integrals can be evaluated exactly in terms of simple functions, although the expressions are too long to give here. Using the quadratic form in Eq.~\eqref{eq:HO_R_ket}, it is a simple to evaluate the expectation value of any analytic function of $\hat{x}$ and $\hat{p}$:
	\begin{equation}
	\langle\psi(t)|g(\hat{x},\hat{p})|\psi(t)\rangle=\langle 0|g(\hat{x}+\xi(t),\hat{p}+m\dot{\xi}(t))|0\rangle
	\end{equation}

	Finally, we also calculate is the overlap with the instantaneous Floquet basis which define the exact $c_n(t)$:
	\begin{equation}
	c_n(t)=\langle n^F(\lambda(t),t)|\psi(t)\rangle.
	\end{equation}
	This can also be evaluated exactly by rewriting this overlap as an integral and then using the generating function of the Hermite Polynomials. Nevertheless, the result is quite long and not very useful. Fortunately, the exact probabilities $p_n^F$ can be defined in terms of a function of the difference between the classical trajectories $\zeta$ and $\xi$. To this end, we define
	\begin{equation}
	\Theta(t)=\frac{m}{2\omega_0}\left[\left[\partial_t {\zeta}(t,\lambda(t))-\dot{\xi}(t)\right]^2+\omega^2_0\left[\zeta(t,\lambda(t))-\xi(t)\right]^2\right].
	\end{equation}
	In terms of this function the probabilities, log-fidelity and Floquet diagonal entropy are given as:
	\begin{gather}
	p_n^F(t)=\mathrm e^{-\Theta(t)}\frac{\Theta(t)^n}{n!}\\
	f_d(t)=\Theta(t)\\
	S_d^F(t)=\Theta(t)(1-\log(\Theta(t)))+\mathrm e^{-\Theta(t)}\sum_{k=0}^\infty\frac{\Theta(t)^n\log(n!)}{n!}
	\end{gather}
	One can also check that if one expands the exact solutions above to leading order in $\dot{\lambda}$, one recovers all the expressions from FAPT, as expected.

	\section{\label{subsec:HFE_convergence}Numerical Results on the Convergence of the Inverse-Frequency Expansion in Systems with Unbounded Spectra.}

\begin{figure}[ht]
		\includegraphics[width=\columnwidth]{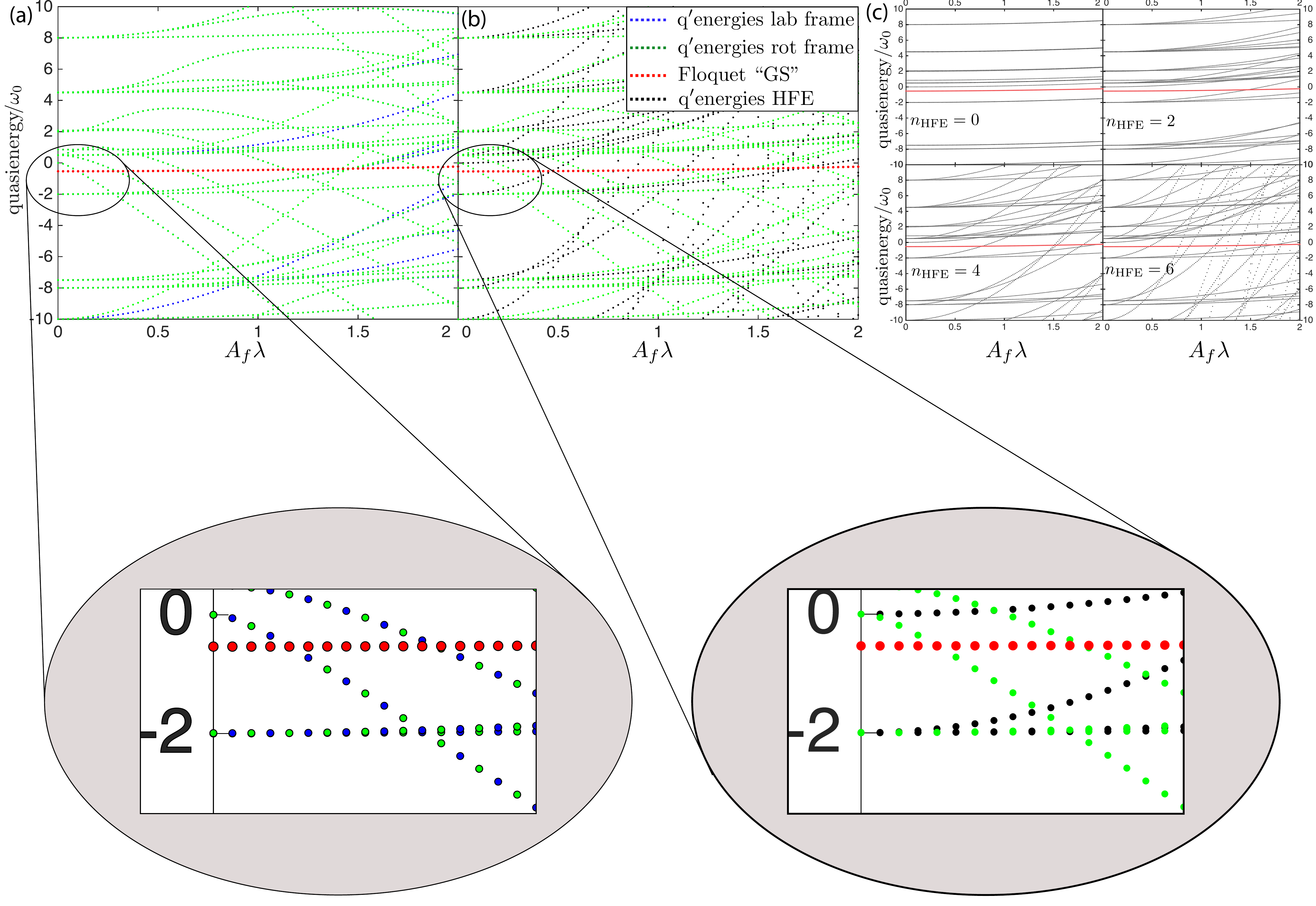}
		\caption{\label{fig:qspectra_comparison}(Color online)[Kapitza pendulum]. (a) Comparison between the exact quasienergy spectrum of the Floquet Hamiltonians obtained numerically using the lab frame Hamiltonian $H(t)$ (blue dots) and the rot frame Hamiltonian $H^\mathrm{rot}(t)$ (green dots) for a finite cutoff $M=30$. These spectra coincide in the infinite cutoff limit but in general differ at finite cutoff. Identifying nearly identical states allows us to numerically identify cutoff independent Floquet eigenstates (see text for details).  (b) Comparison between the exact Floquet quasienergy spectrum in the rot frame (green dots) and the approximate spectrum of the vV HFE to order six (black dots). (c) Spectrum of the vV HFE at different orders $n_\mathrm{HFE}$ of the vV HFE. The approximate spectra (black dots) are calculated for $M=100$, but only the lowest $31$ states are kept; hence all approximate spectra are cut-of independent. In all panels the Floquet ground state is denoted in red. The model parameters are $m\omega_0 = 1$, $A_f = 2$, and $\Omega=20\omega_0$.}
	\end{figure}

	Let us now ask the more general question whether the vV HFE for the Kapitza pendulum converges to the Floquet Hamiltonian in the regions away from the detrimental resonances. If we want to compare the behaviour of the exact Floquet eigenstates and quasienergies in the limit $M\to\infty$ to those of the vV HFE, we first need to make sure our results are independent of the cut-off $M$, which is expected to affect the states at the top end of the spectrum. Usually, when one discusses a cut-off dependence in static systems, one makes sure the results remain unchanged with increasing the value of the cut-off itself. However, in Floquet systems a larger cut-off amounts to a larger number of states, all folded within the same Floquet zone. Therefore, this procedure quickly makes interpreting quasienergy plots difficult. Unfortunately, there is no straightforward algorithm to cut off the high-energy states of an exact Floquet spectrum obtained numerically, since the latter always comes out folded.  Thus, we follow a slightly different route: we do the numerical calculation of the exact quasispectrum both in the lab frame and in the rotating frame. In the limit of sending the cut-off $M\to\infty$, both calculations trivially result in the same Floquet spectrum. However, since the transformation to the rotating frame in Eq.~\eqref{eq:kapitza_V(t)} couples differently to the cut-off dependent states, this allows us to immediately identify all cut-off dependencies upon comparing the two spectra. Figure~\ref{fig:qspectra_comparison} (a) shows the exact quasienergy spectrum of the Kapitza pendulum for $\Omega/\omega_0 = 20$ obtained in the lab (blue dots) and the rotating frame (green dots), while the Floquet ground state is shown in red. Let us fix a value of $\lambda$, and focus on a particular quasienergy level, which corresponds to a single dot in the plot. Then, if the rot-frame and lab-frame data coincide, the quasienergy is identified as cut-off independent. It is interesting to note that (i) for the Kapitza pendulum, all cut-off dependent states necessarily have large physical energies, and are thus close to the top of the spectrum, and (ii), where present, the cut-off dependence becomes more pronounced at larger driving amplitude $\lambda$. 
	
	Once we have identified the cut-off dependence in the exact Floquet spectrum, we have to do so for the approximate spectrum obtained within the vV HFE. However, since the latter produces an unfolded spectrum, this is easily done with standard methods: we first calculate the approximate spectrum for a larger value of the cut-off $M_\mathrm{up}$, and \emph{after diagonalisation} we artificially keep only a desired small number $M$ of the energy states $M<M_\mathrm{up}$ from the bottom of the approximate spectrum. Finally, we make sure the chosen $M$ states do not depend on the choice of $M_\mathrm{up}$. Figure~\ref{fig:qspectra_comparison} (b) shows the quasienergy spectra of the exact Floquet Hamiltonian $H_\mathrm{eff}$ in the rot frame, and the approximate one, $H_\mathrm{eff}^{[6]}$. 
	
	Note first that all states which are relatively flat as a function of $\lambda$ display a nice agreement. Not surprisingly, these are the low-energy states, which are not resonantly coupled to the drive. Second, the entire fan of states which bend down and cross the Floquet GS (red dots) in the exact quasienergy spectrum is absent in the vV HFE spectrum. Moreover, this entire fan of exact Floquet eigenstates is cut-off independent, as can be seen from Fig.~\ref{fig:qspectra_comparison} (a). These are precisely the states that lead to the breakdown of FAPT, as discussed in Sec.~\ref{subsec:kapitza}. This already hints towards a serious problem with the convergence of the vV HFE for this model, although the mechanism behind it is expected to be fairly general. Figure~\ref{fig:qspectra_comparison} (c) shows the cut-off independent folded spectra of the vV HFE depending on the order of approximation $n_\mathrm{HFE}$. Observe how, while for the flat quasienergy levels with $E\lesssim\Omega$ the convergence seems quite reasonable, the high-energy levels with $E\gtrsim\Omega$ obviously diverge as we increase the order of the expansion. Note that the frequency used, $\Omega/\omega_0=20$, is already an order of magnitude larger than the single-particle parameters, and the vV HFE was naively eexpected to converge for such frequencies. 

\begin{figure}
	\includegraphics[width=\columnwidth]{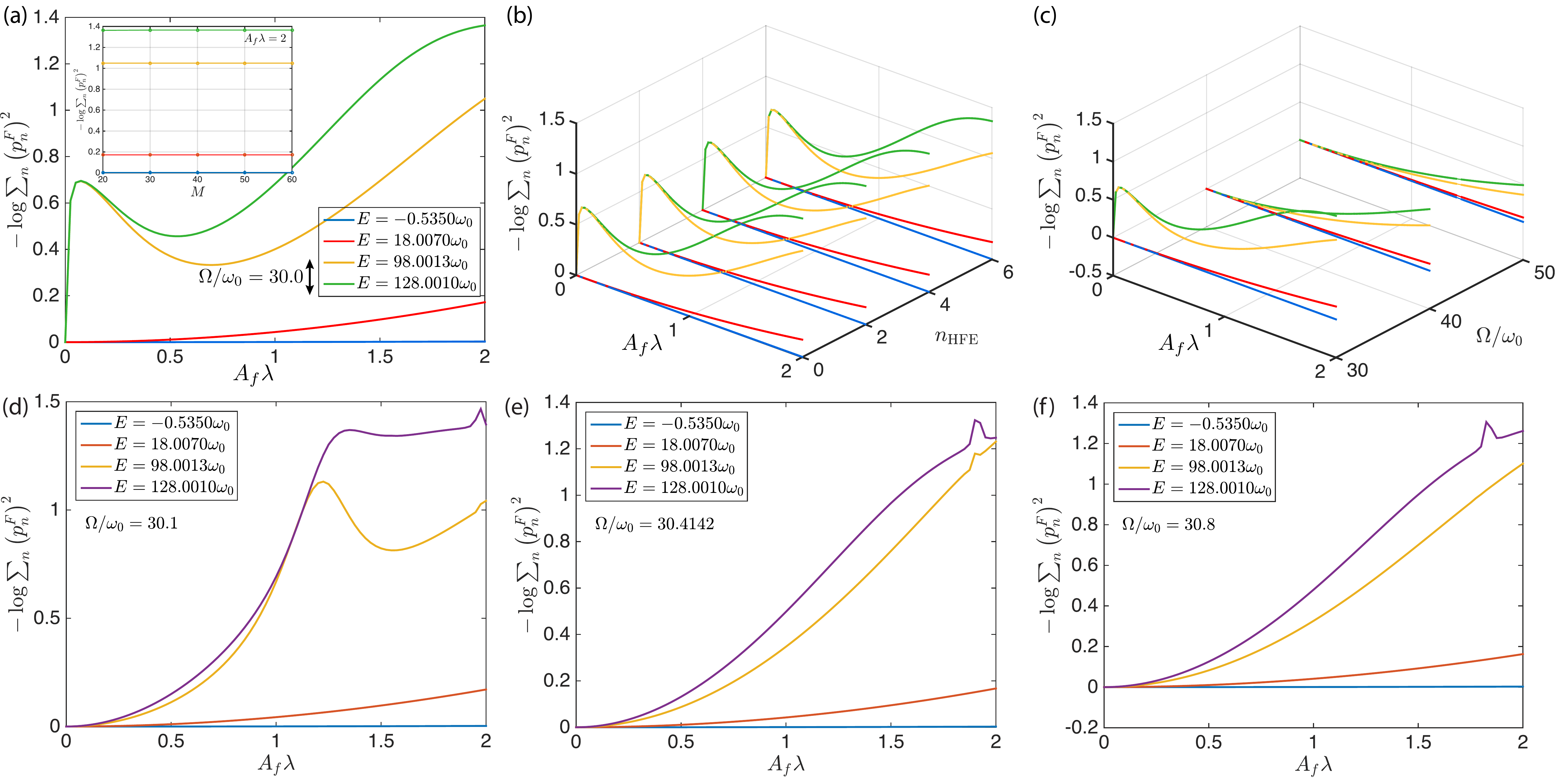}
	\caption{\label{fig:IPR}(Color online)[Kapitza pendulum] Inverse participation ratio (log-IPR) of four distinguished approximate states in the exact Floquet spectrum as a function of the amplitude $A_f\lambda$. Inset: system size-dependence of the log-IPR for $A_f\lambda=2$. $\Omega/\omega_0 = 30$. (b) The dependence of the log-IPR on the order of the vV HFE is negligible compared to the curvature of the curves for $\Omega/\omega_0 = 30$. (c) Frequency-dependence of the log-IPR as a function of $\lambda$ for $n_\mathrm{HFE}=6$. (d-f) same as (a) but for the off-resonant frequencies $\Omega/\omega_0 = 30.1$, $\Omega/\omega_0 = 30.41$ and $\Omega/\omega_0 = 30.8$, respectively. The model parameters are $m\omega_0 = 1$, $A_f = 2$. }
\end{figure}

	In the following, we focus only on states which do not exhibit any cut-off dependence. To better quantify the convergence of the vV HFE for such states to their exact Floquet counterparts, we define the log inverse participation ratio (log-IPR or collision entropy) of the approximate eigenstates $|\nu\rangle$ of the vV HFE in the exact Floquet eigenbasis $|n\rangle$ as
	\begin{eqnarray}
	R = -\log\sum_n \left(p_n\right)^2~,~p_n = |\langle n|\nu\rangle|^2
	\label{eq:inv_part_ratio}
	\end{eqnarray}
	If an approximate state matches an exact Floquet state, then $R=0$, while whenever $R>0$ the log-IPR measures the deviation between the two. We now focus on four representative approximate states and calculate their participation ratios in the exact Floquet spectrum; these are the Floquet ground state, a state with $E\gtrsim\Omega$, and two high-energy states whose energy differ by approximately $\Omega$. We label these four states by their physical energy $E$. Figure~\ref{fig:IPR} (a) shows the inverse participation ratios of these states within the sixth order vV HFE with the inset confirming that the data used is cut-off independent. As expected from examining the quasienergy spectra discussed earlier, the log-IPR increases with the physical energy. Interestingly, the participation ratios of the low-energy states grow as $R\sim \lambda^2$ for $\lambda\to 0$.  Remarkably, $R$ can also exhibit a non-analytic behaviour at small $\lambda$ for high-energy states. This can be explained as follows: whenever two states of the non-driven model at $\lambda=0$ have energies close to resonance with the driving frequency, any weak coupling strongly hybridises them, leading to the opening of a photon avoided crossing at infinitesimally small $\lambda$. This is the worst-case scenario for FAPT, since it means that one can never do a successful adiabatic ramp-up of the drive, because if the ramp starts smoothly with a vanishing velocity and acceleration in the region of $\lambda\to 0$, the system will immediately absorb energy due to the avoided crossing at $\lambda\to 0$. In such cases, it is possible that quench-starting the drive will produce less excess energy than slowly turning it up. However, this non-analytic behaviour is only present when the drive frequency matches precisely the energy difference of two bare levels, as confirmed by Fig.~\ref{fig:IPR}(d-e) for frequencies close but not equal to the resonant one. This suggest that the width of such resonances can indeed be very small. Figure~\ref{fig:IPR} (b) shows the inverse participation ratios as a function of the order in the vV HFE. The results clearly suggest that the vV HFE does not converge to the exact Floquet Hamiltonian. Interestingly, however, at fixed $\lambda$ the log-IPR curves of these states do seem to converge, but it is not clear what the limit is and whether it carries a physical meaning. Last, Fig.~\ref{fig:IPR} (c) displays the frequency-dependence of the log-IPR $R$. Unlike increasing the order in the vV HFE, increasing $\Omega$ reduces the participation ratio, as expected, since the vV HFE becomes exact as $\Omega\to\infty$.
	
	The results shown in this section put in doubt the convergence of the inverse-frequency expansion to the exact Floquet Hamiltonian for systems with unbounded spectra.  It has become clear that the origin of divergence of the vV HFE can be traced back to the unbounded spectrum due to the operator $p^2_\theta$, which enables the appearance of photon absorption crossings. Hence, if one further eliminates this unboundedness by going to yet another rotating frame, which amounts to applying the generalised Schrieffer-Wolff (SW) transformation~\cite{bukov_15_SW}, the convergence properties of the expansion improve significantly. The intuition behind this is that this way of `folding' the unbounded part of the spectrum is equivalent to a re-summation of an infinite vV HFE subseries~\cite{bukov_14}. Hence, if the vV HFE is divergent due to the presence of a non-analytic in $1/\Omega$ term in the exact Floquet Hamiltonian, this re-summation circumvents the expansion of the non-analytic piece. There exists evidence that the generalised SW transformation, unlike the bare vV HFE, captures these photon absorption avoided crossings; however, this is beyond the scope of the present paper.





\bibliography{Floquet_bib}
\bibliographystyle{apsrev4-1} 







\end{document}